\documentclass[11pt,a4paper]{article}

\usepackage[utf8]{inputenc}
\usepackage{a4wide}

\usepackage[colorlinks=true,citecolor=blue]{hyperref}

\usepackage{graphicx}
\usepackage{float}
\usepackage{epstopdf}
\usepackage{floatflt} 
\usepackage{subcaption}

\usepackage{tikz}
\newlength\fwidth
\usetikzlibrary{arrows}
\usetikzlibrary{decorations.pathreplacing}
\usetikzlibrary{plotmarks}
\usepackage{pgfplots} 
\usepackage{pgfgantt}
\usepackage{pdflscape}
\pgfplotsset{compat=newest} 
\pgfplotsset{plot coordinates/math parser=false}
\usepackage{varwidth}
\usepgfplotslibrary{fillbetween}
\pgfplotsset{compat = 1.3}
\usepgfplotslibrary{external}
\tikzsetexternalprefix{figures/}

\usepackage{amsmath}
\usepackage{amsthm}
\usepackage{amssymb}
\usepackage{breqn} 
\usepackage{empheq} 
\usepackage{cite} 
\usepackage{nicefrac} 

\providecommand{\abs}[1]{\lvert#1\rvert}
 
\DeclareMathOperator*{\argmax}{argmax}
\providecommand{\minimum}[1]{\min \{ #1 \}}
\providecommand{\colvector}[1]{\left( \begin{array}{c} #1 \end{array} \right)}
\providecommand{\einhalb}[0]{\frac{1}{2}}
\providecommand{\eindrittel}[0]{\frac{1}{3}}
\providecommand{\vmax}[0]{v^{\max}}

\providecommand{\rhomax}[0]{\rho^{\max}}
\newcommand\fmax{f^{\max}}

\DeclareMathOperator*{\incoming}{in}

\providecommand{\Nxmyroad}[1]{N_{x_{#1}}}
\newcommand\Nt{N_t}

\newcommand\km{\text{km}}
\newcommand\kmh{\frac{\text{km}}{\text{h}}}
\newcommand\ckm{\frac{\text{cars}}{\text{km}}}
\newcommand\ch{\frac{\text{cars}}{\text{h}}}

\newcommand\intx{\mathrm{d}x}
\newcommand\intt{\mathrm{d}t}

\DeclareMathOperator*{\LWR}{LWR}

\DeclareMathOperator*{\ALWR}{ALWR}
\DeclareMathOperator*{\onramp}{or}
\DeclareMathOperator*{\limcase}{lim}

\providecommand{\onrampinflow}{f_{\onramp}^{\incoming}}

\providecommand{\dx}[0]{\Delta x}
\providecommand{\dt}[0]{\Delta t}
\providecommand{\LWRvelocity}{V}
\providecommand{\geqzero}{{\geq 0}}
\providecommand{\road}{e}
\providecommand{\junction}{k}
\providecommand{\graph}{G}
\providecommand{\setofjunctions}{\mathcal{V}}
\providecommand{\setofroads}{\mathcal{E}}
\providecommand{\Demand}[0]{D}
\providecommand{\Supply}[0]{S}
\providecommand{\LWRdemand}[0]{D^{\LWR}}
\providecommand{\LWRsupply}[0]{S^{\LWR}}
\providecommand{\fluxmax}[0]{\sigma}
\providecommand{\Leftroad}[0]{1}
\providecommand{\Rightroad}[0]{2}
\providecommand{\queuelength}{l}
\providecommand{\testfunction}{\phi}
\providecommand{\intx}{\, dx}
\providecommand{\intt}{\, dt}
\providecommand{\Kruzhkov}{Kruzhkov }
\DeclareMathOperator{\sign}{sign}
\providecommand{\massflux}{q}
\providecommand{\momflux}{r}
\providecommand{\set}[1]{\lbrace #1 \rbrace}
\providecommand{\relaxationtime}{\tau}
\providecommand{\pressure}{p}
\providecommand{\eigenvalue}{\lambda}
\providecommand{\pressexp}{\gamma}
\providecommand{\ARZdemand}[0]{D^{\ARZ}}
\providecommand{\ARZsupply}[0]{S^{\ARZ}}
\DeclareMathOperator*{\ARZ}{ARZ}
\providecommand{\staterhow}{U}
\providecommand{\ALWRsupply}[0]{S^{\ALWR}}
\providecommand{\demandonramp}{\Demand^{\onramp}}
\providecommand{\Godunov}{\mathcal{G}}
\providecommand{\Cell}[1]{C_{#1}}
\DeclareMathOperator*{\niceeinhalb}{\nicefrac{1}{2}}
\providecommand{\setofjunctions}{\mathcal{V}}
\providecommand{\setofroads}{\mathcal{E}}
\providecommand{\Nxroad}{N_{x_{\road}}}
\providecommand{\lengthofroad}{L}
\providecommand{\dtbydx}[0]{\frac{\dt}{\dx}}
\providecommand{\control}[0]{u}
\providecommand{\constrafficstate}{Q}	
\providecommand{\staterhow}{U}
\providecommand{\RHSARZ}{g}
\providecommand{\sumofincomingdemands}[0]{\mathcal{Z}}
\providecommand{\rholim}{\rho_{\limcase}}
\providecommand{\wlim}{w_{\limcase}}
\providecommand{\sigmalim}{\sigma_{\limcase}}
\providecommand{\tilderhoequalssigmal}{\Sigma}
\providecommand{\curveofmaxima}{\mathcal{M}}
\providecommand{\curveofequalsupplyatmaxarz}{\mathcal{W}}
\providecommand{\gcapdrop}{\overline{g}}

\newtheorem{Def}[]{Definition}[]

\newtheorem{Remark}[]{Remark}[]

\usepackage{booktabs}
\usepackage{multirow}
\usepackage{multicol}

\usepackage{color}

\definecolor{myblack}{rgb}{0.0,0.0,0.0} 
\definecolor{myblue}{rgb}{0.2,0.2,0.6} 
\definecolor{myred}{rgb}{0.8,0.0,0.0} 
\definecolor{mygray}{rgb}{0.52,0.52,0.51} 
\definecolor{mygreen}{rgb}{0.0,0.42,0.24} 

\begin{document}
	\title{A combined first and second order model for a junction with ramp buffer}
	\author{Jennifer Weißen\footnotemark[1], \; Oliver Kolb\footnotemark[1], \; Simone G\"ottlich\footnotemark[1]}
	
	\footnotetext[1]{University of Mannheim, Department of Mathematics, 68131 Mannheim, Germany (jennifer.weissen@uni-mannheim.de, \{kolb,goettlich\}@uni-mannheim.de).}

	\date{\today}

	\maketitle
	
	
	\definecolor{darkgreen}{rgb}{0,0.5,0}
	\definecolor{darkblue}{rgb}{0,0,0.5}
	\definecolor{gold}{rgb}{1, 0.84, 0}
	\definecolor{light-red}{rgb}{1, 0.59, 0.59}
	\definecolor{hintergrund}{rgb}{0.94, 1.0, 0.94}
	\definecolor{hintergrund2}{rgb}{1, 1, 0.7}
	\definecolor{blond}{rgb}{0.98, 0.94, 0.75}
	\definecolor{floralwhite}{rgb}{1.0, 0.98, 0.94}
	\definecolor{hellgray}{rgb}{0.9, 0.9, 0.9}
	\definecolor{orange}{rgb}{1.0, 0.75, 0.0}
	\definecolor{uniblau}{RGB}{0,19,122}
	\definecolor{rwthblau}{RGB}{0,85,170}
	\definecolor{petrol}{rgb}{0.24,0.67,0.52}
	\definecolor{hellblau}{rgb}{0.75,0.83,0.89}
	
	\begin{abstract}
Second order macroscopic traffic flow models are able to reproduce the so-called capacity drop effect, i.e., the phenomenon that the outflow of a congested region is substantially lower than the maximum achievable flow. Within this work, we propose a first order model for a junction with ramp buffer that is solely modified at the intersection so that the capacity drop is captured. 
Theoretical investigations motivate the new choice of coupling conditions and illustrate the difference to purely first and second order models. 	
The numerical example considering the optimal control of the onramp merging into a main road highlights that the combined model generates similar results as the second order model.
	\end{abstract}

{\bf AMS subject classifications:} 65M08, 90C30

\smallskip

{\bf Keywords:} traffic flow, numerical analysis, ramp metering control
	

\section{Introduction}

Starting with the pioneering works on the Lighthill-Whitham-Richards (LWR) model~\cite{LigWhi1955,Ric1956}, traffic flow models based on scalar conservation laws, i.e., macroscopic traffic models, have received considerable attention in the academic literature over the past decades. Macroscopic models are mainly distinguished in first order models, which consist of a scalar conservation law for the traffic density, and second order models, using a system of two conservation or balance laws with an additional equation for the mean traffic speed. The first order LWR model describes the time evolution of the traffic density $\rho$. The main drawback of the LWR model is the direct link of the velocity and flux to the traffic density, which does not allow for a correct description of traffic instabilities. Second order models have been developed aiming to improve traffic descriptions. Aw, Rascle and Zhang ~\cite{AwRas2000, Zha2002} introduced the second order ARZ model which is capable to capture traffic instabilities and additionally overcame the drawbacks of earlier developed second order models, see~\cite{Dag1995}. They introduced a pressure function $\pressure(\rho)$ in the dynamics. The pressure function represents an anticipation factor, which takes into account the reaction of drivers to the traffic in front of them~\cite{Goa2006}. It defines the relation between the density $\rho$ and the speed $v$ in the fundamental diagram via a Lagrangian marker $w$. In comparison to the LWR model, the second order model enables to portray more important traffic phenomena~\cite{TreKes2013}. Greenberg~\cite{Gre2001} extended the ARZ model with a relaxation term towards a preferred velocity. This extension in turn was further generalized by~\cite{SieMau2006}. Since then, the extension to traffic flow on road networks has been investigated intensively, see for example~\cite{BreCanGar2014, CocGarPic2005, HauBas2007, HerMouRas2006, HerRas2006, PicGar2006,GoaGoeKol2016,KolGoeGoa2017,KolCosGoa2018}. The crucial point at intersections is the definition of suitable coupling conditions, i.e., conservation of mass and possibly momentum in the case of second order models.

Many interesting effects within traffic dynamics related to congestion and traffic jams have been investigated for microscopic, mesoscopic and macroscopic traffic models, see~\cite{Hel2001}. Among these effects, the so-called capacity drop phenomenon is of interest~\cite{HauBasChi2005,ParBui2012,HauBas2007,KolGoeGoa2017,trafficandgranularflow_lebacque}. The capacity drop is the phenomenon that the outflow of a congested region is significantly lower than the maximum achievable flow in this region. A capacity drop is observed when the traffic volume upstream of a bottleneck increases, but the discharge flow leaving the bottleneck decreases in comparison to the flow measured prior to the increase in traffic volume~\cite{ChuRudCas2007}. Studies conducted on freeways showed that the discharge flow diminishes with the onset of upstream queues~\cite{HalAgy1991,Ban1991,CasBer1999}. This results in a flow lower than the maximum one. The traffic jams arising from the capacity drop at bottlenecks can even become permanent~\cite{HauBasChi2005}.

Specifically, in case of an onramp at a freeway, the capacity drop is explained by the fact that drivers mutually impede each other, when too many vehicles try to access the main road~\cite{ParBui2012}. Additionally, drivers at the incoming road upstream from the onramp entry are impacted and might break and slow down. While vehicles can decelerate fast, the acceleration process takes substantially longer~\cite{Leb2003}. 

The capacity drop has major influence on the traffic flow and is especially important in the consideration of traffic control. In general, the observational findings indicate that approaches to control the traffic density are promising for increasing bottleneck discharge flows~\cite{ChuRudCas2007}.  

%
%
%
%
Here, we completely stick to the LWR model to describe the dynamics on the roads. To mirror the capacity drop effect, we put special emphasis on the intersections, where the velocities and Lagrangian markers are incorporated to describe the dynamics. The capacity drop is achieved by a flow reduction derived from second order dynamics rather than by an adaption of the fundamental diagram. We therefore consider a first order macroscopic model on the roads and we merely apply adapted coupling conditions of a second order macroscopic model at the junctions. This combination then allows to model the capacity drop effect. 
A similar approach has been considered in~\cite{MouHer2009}, where the LWR model was coupled to a kinetic model at the junctions. Comparable to~\cite{MouHer2009}, the combined model is also able to cover the capacity drop phenomenon. As in~\cite{SanDonPel2019}, our model is a first order model with a point constraint at the junction. In both models, the supply function of the outgoing road incorporates information of the upstream traffic conditions. The junction constraint in this work is motivated by the second order ARZ model and combines the first order approach with some second order information. Using the combined model, we apply a discrete optimization approach for the control of the onramp, see e.g.~\cite{GoaGoeKol2016, KolGoeGoa2017, ReiSamDel2015}. The combined model serves as a suitable substitute model for a second order model when investigating optimal control strategies for ramp metering.

The paper is organized as follows: In Section~\ref{sec:trafficnetworks}, we briefly recall the LWR and ARZ model. Section~\ref{sec:capdrop_litreview} discusses the capacity drop. We construct our combined model in Section~\ref{sec:constructionofALWRmodel} with specific emphasis on the respective demand and supply functions, which determine the flux at the junction. The supply function is analyzed intensively in Section~\ref{sec:augmentedsupply}. In Section~\ref{sec:NumericalTreatment},  we present a numerical comparison for several Riemann problems in the LWR and the combined model based on the discretization in Section~\ref{sec:discretization_firstorder} and Section~\ref{sec:RP_ALWRvsLWR}. The ability to capture the capacity drop is illustrated in Table~\ref{tab:fluxvalues_ex}. Section~\ref{sec:comparison_ALWR_ARZ} presents a comparison of the combined model with second order models. The numerical discretization for second order models is introduced in Section~\ref{sec:discretization_secondorder} and comparisons of the combined model with second order models are drawn in Section~\ref{sec:RP_ARZvsALWR}. The paper concludes with numerical results for an optimal control problem for ramp metering in Section~\ref{sec:RampMetering_ALWR}.


\section{Models and notations for traffic flow} \label{sec:trafficnetworks}

We consider a directed graph $\graph = (\setofjunctions,\setofroads)$
with a finite number of roads $e \in E$ and junctions $\junction \in \setofjunctions$. The roads $e$ are modeled by intervals $(a_\road,b_\road)$ with (possibly infinite) lengths $\lengthofroad_\road = b_\road - a_\road$. Vehicles are treated as a continuum with density $\rho$ and velocity $v$. We shortly introduce two well-known first and second order macroscopic approaches for traffic flow on networks and explain how they model a junction with ramp buffer.

\subsection{The LWR model} \label{sec:basics_LWRmodel}

Given an initial density $\rho_\road(x,0)$ and denoting $\rho_\road = \rho_\road(x,t)$, the dynamics of the LWR model are given by
\begin{equation} \label{eq:LWR_network}
	\partial_t \rho_\road + \partial_x \left( \rho_\road \LWRvelocity(\rho_\road)\right) = 0, \qquad (x,t) \in (a_\road, b_\road) \times \mathbb{R}_\geqzero,
\end{equation}
on each road $e$. Here, we choose for each road $\road$ the linear velocity function
\begin{equation} \label{eq:LWR_Velocity}
	\LWRvelocity(\rho_\road) = \vmax \left( 1 - \frac{\rho_\road}{\rhomax}\right),
\end{equation}
with maximum velocity $\vmax$ and maximum density $\rhomax$ as for example described in~\cite{PicGar2006}. We denote the concave flux function $f(\rho_\road) = \rho_\road \LWRvelocity(\rho_\road) $. It holds that $f(0) = f(\rhomax) = 0$ and the density $\nicefrac{\rhomax}{2}$ maximizes the flux in the interval $[0, \rhomax]$. Traffic flow $f$ and velocity $V$ are directly coupled to the traffic density $\rho$.

\begin{figure}[tbhp]
	\centering 
	\tikzsetnextfilename{OnetoonewithOR} 
	\begin{tikzpicture}[]  
		\node (A) at (0,0) [circle,draw,minimum size=0.8cm, align=center, text centered] { \small \emph{in}};
		\node (B) at (3,0) [circle,draw,minimum size=0.8cm,align=center, text centered] {\small  \emph{~ramp} };
		\node (C) at (6,0) [circle,draw,minimum size=0.8cm,align=center, text centered] {\small  \emph{out}};
		\node (O) at (3,1.5)[circle] {};
		\draw[->, very thick, dashed] (O) to node [right, text width = 3cm] {  $\queuelength(t)$, $(1- \beta$)} (B);
		\draw[->, very thick] (A) to node [below,text width = 1.5cm, align=center] {   $\road =\Leftroad$, $\beta$} (B);
		\draw[->, very thick] (B) to node [below,text width = 1.5cm, align=center] { $\road = \Rightroad$}(C);
	\end{tikzpicture}
	\caption{1-to-1 junction with an onramp}
	\label{img:onetoonewithOR}
\end{figure}

In this work, we are especially interested in the dynamics at a junction with ramp buffer. Figure~\ref{img:onetoonewithOR} illustrates the 1-to-1 junction with an onramp at the node \emph{ramp}, which is examined in the following. For simplicity, we assume as prescribed in~\eqref{eq:LWR_network} that the roads $\road =\Leftroad, \Rightroad$ have the same parameters $\rhomax$, $\vmax$. At the onramp traffic is allowed to enter the outgoing road $\road =\Rightroad$ and the parameter $\beta$ describes the mixture of incoming traffic from road $\road = \Leftroad$ and from the onramp. The quantity $\queuelength(t)$ describes the length of the queue at the ramp buffer and is modeled by the following ordinary differential equation (ODE):
\begin{align} \label{eq:evolutionqueue}
	\frac{\mathrm{d} \queuelength (t)}{\mathrm{d} t} = \onrampinflow(t) - q_\onramp(t), \qquad t \in \mathbb{R}_\geqzero,
\end{align}
where $\onrampinflow$ denotes a (possibly) time--dependent flux that enters the onramp from outside of the network and $q_\onramp$ is the flux that exits from the onramp to the outgoing road $\road = \Rightroad$.

To define an initial value problem for this particular network structure, we consider Riemann data on the incoming and outgoing road, i.e., $\rho_\road(x,0) = \rho_\road^0$, $\road=\Leftroad, \Rightroad$ and a fixed inflow $\onrampinflow > 0$ from the onramp. The end of the incoming road is set to $b_\Leftroad = 0$ and the beginning of the outgoing road to $a_\Rightroad = 0$, such that the junction is located at $x=0$. Then, the (half-) Riemann problem for the roads and the evolution of the queue length  at the junction reads
\begin{align} \label{eq:LWR_RP_11withOR}
	\begin{cases}
		\partial_t \rho_\road + \partial_x \left(\rho_\road \LWRvelocity(\rho_\road) \right) = 0, &\qquad (x,t) \in (a_\road, b_\road) \times \mathbb{R}_\geqzero,~\road = 1,2, \\
		\rho_\road(x,0) = \begin{pmatrix} \rho_\road^+ & \mbox{ for } x>0 \\ \rho_\road^- & \mbox{ for } x \leq 0. \end{pmatrix}, &\qquad x \in (a_\road, b_\road),~\road =1,2,\\
		\frac{\mathrm{d} \queuelength(t)}{\mathrm{d} t} = \onrampinflow - q_\onramp(t), &\qquad t \in \mathbb{R}_\geqzero, \\
		\queuelength(0) =\queuelength^0. &
	\end{cases}
\end{align}
Here, $\queuelength^0 \geq 0$ indicates the initial buffer load. Depending on whether the road is incoming or outgoing, only one of the Riemann data $\rho_\road^-, \rho_\road^+$ is defined for $t=0$. In particular, for road $\road = \Leftroad$ entering the junction $\rho_\Leftroad^- = \rho_\Leftroad^0$, and for the road $\road = \Rightroad$ exiting the junction $\rho_\Rightroad^+ = \rho_\Rightroad^0$. The other datum as well as the flux $q_\onramp(t)$ exiting the onramp is defined by the solution through some suitable coupling conditions at the junction. This will be the focus of this work. We fix a priority parameter $\beta$ and define the mixture rule:
\paragraph{(M1)}
In case that not all cars can enter the outgoing road, let $\massflux$ be the flux that can enter the outgoing road. Then $\beta \massflux$ comes from the incoming road $\road = \Leftroad$ and $(1- \beta) \massflux$ comes from the onramp. In case that the amounts $\beta \massflux$ (resp. $(1-\beta) \massflux$) cannot be served from the incoming road  (resp. onramp), we allow that the missing amount is filled up by the onramp (resp. incoming road).

We define solutions to~\eqref{eq:LWR_RP_11withOR} in the following, compare also~\cite{ReiDelSam2014}.
\begin{Def} \label{def:ALWR_networksolution}
	A triple $(\rho_\Leftroad, \rho_\Rightroad, \queuelength)$ is called an admissible weak solution to~\eqref{eq:LWR_RP_11withOR} if the following holds:
	\begin{itemize}
		\item The densities $\rho_\road \colon  (a_\road, b_\road) \times [0, \infty) \rightarrow [0, \rhomax]$, $\road= \Leftroad, \Rightroad$ are weak solutions such that
		\begin{align*}
			\int_{0}^{\infty} \int_{a_\road}^{b_\road} \left(\rho_\road \partial_t \testfunction_\road + f(\rho_\road) \partial_x \testfunction_\road \right) \intx \intt = 0, \qquad \road = \Leftroad, \Rightroad,
		\end{align*}
		for every test function $\testfunction_\road: (a_\road, b_\road) \times \mathbb{R}_\geqzero \rightarrow \mathbb{R}$ with compact support.
		\item The densities $\rho_\road$, $\road = \Leftroad, \Rightroad$ satisfy the \Kruzhkov entropy condition~\cite{Kru1970}. I.e., for $z \in \mathbb{R}$ and $\testfunction_\road \colon (a_\road, b_\road) \times \mathbb{R}_\geqzero \rightarrow \mathbb{R}$ smooth, positive with compact support, we have that
		\begin{align}  \label{eq:LWR_weakentropysolution}
			\begin{split}
			\int_0^\infty \int_{a_\road}^{b_\road} \big( \abs{\rho_\road - z}\partial_t \testfunction_\road + \sign(\rho_\road - z) &\left(f(\rho_\road) - f(z)\right) \partial_x \testfunction_\road \big) \intx \intt \\
			~&+ \int_{a_\road}^{b_\road} \abs{\rho_\road^0 - z} \testfunction_\road(x,0) \intx \geq 0, \qquad \road= \Leftroad, \Rightroad.
			\end{split}
		\end{align} 
		\item The Rankine Hugoniot condition at the junction, ensuring conservation of mass, is fulfilled
		\begin{align*} 
			\underbrace{f\left(\rho_\Leftroad(0,t)\right)}_{:=\massflux_\Leftroad} + \massflux_\onramp(t)= \underbrace{f\left(\rho_\Rightroad(0,t)\right)}_{:=\massflux_\Rightroad}.
		\end{align*}
		\item The flux entering the outgoing road is maximal subject to the mixture rule (M1).
		\item The queue length $\queuelength$ is a solution of~\eqref{eq:evolutionqueue} for almost every $t \in \mathbb{R}_\geqzero$.
	\end{itemize}
\end{Def}

Let $\massflux_\Leftroad$, $\massflux_\onramp$ denote the initially unknown mass flux coming from the incoming road and the onramp. The total flux has to exit into the outgoing road such that $\massflux_\Rightroad = \massflux_\Leftroad + \massflux_\onramp$. In agreement with the literature~\cite{CocGarPic2005,trafficandgranularflow_lebacque}, we employ the widely used demand and supply formulation for admissible flux values. The fluxes at the junction are bounded by demand and supply 
\begin{align*}
	0 \leq \massflux_\Leftroad \leq \LWRdemand(\rho_\Leftroad) , \qquad	0 \leq \massflux_\onramp \leq \demandonramp(l), \qquad 0 \leq \massflux_\Rightroad \leq \LWRsupply(\rho_\Rightroad).
\end{align*}
These are used to determine the flux at the junction and allow for an appropriate description of the coupling and boundary conditions~\cite{trafficandgranularflow_lebacque}. The demand $\LWRdemand(\rho_\Leftroad)$ of the incoming road and the supply $\LWRsupply(\rho_\Rightroad)$ of the outgoing road are defined as follows:
\begin{align} \label{eq:LWR_demandsupply}
	\LWRdemand(\rho_\Leftroad) =\begin{cases}
		\rho_\Leftroad \LWRvelocity(\rho_\Leftroad) &\text{ if } \rho_\Leftroad \leq \frac{\rhomax}{2}, \\
		\fmax  &\text{ if } \rho_\Leftroad > \frac{\rhomax}{2},
	\end{cases} \qquad
	\LWRsupply(\rho_\Rightroad)=\begin{cases}
		\fmax &\text{ if } \rho_\Rightroad \leq \frac{\rhomax}{2}, \\
		\rho_\Rightroad \LWRvelocity(\rho_\Rightroad)  &\text{ if } \rho_\Rightroad > \frac{\rhomax}{2},
	\end{cases}
\end{align}
where $\fmax= f(\nicefrac{\rhomax}{2})$. Moreover, the demand of the onramp is modeled by
\begin{align} \label{eq:demand_onramp}
	\demandonramp(\queuelength) = \begin{cases}
		\fmax_\onramp	&\text{ if } \queuelength > 0, \\
		\min \set{\onrampinflow(t), \fmax_\onramp} &\text{ otherwise,}
	\end{cases}
\end{align}
where $\fmax_\onramp$ denotes the maximum flux which is allowed to enter from the onramp into the outgoing road, i.e., a physical restriction on the number of cars per time unit that can exit from the onramp. Collecting together equations~\eqref{eq:LWR_demandsupply}-\eqref{eq:demand_onramp} and the mixture rule (M1), we can formulate the mass fluxes at the 1-to-1 junction with onramp in the compact form
\begin{align*}
	\massflux_\Leftroad &= \min \lbrace \LWRdemand(\rho_\Leftroad), \max \{ \beta \LWRsupply(\rho_\Rightroad), \LWRsupply(\rho_\Rightroad) - \demandonramp(l) \} \rbrace, \\
	\massflux_\onramp &= \min \{ \demandonramp(\queuelength), \max \lbrace (1-\beta) \LWRsupply( \rho_\Rightroad), \LWRsupply( \rho_\Rightroad) - \LWRdemand( \rho_\Leftroad)\} \rbrace, \\
	\massflux_\Rightroad &= \massflux_\Leftroad + \massflux_\onramp,
\end{align*} 
see also~\cite{GoaGoeKol2016}. The above expression clarifies that the fluxes at the junction are obtained from the densities on incoming and outgoing road as well as the queue length at the onramp. It is a first order approach where velocity and flux at the junction are directly linked to the traffic densities $\rho_\Leftroad, \rho_\Rightroad$. The demand is obtained purely from the density on the incoming road $\rho_\Leftroad$ and the supply purely from the density on the outgoing road $\rho_\Rightroad$. All in all, it is a simple and robust model to describe traffic flow~\cite{trafficandgranularflow_lebacque}. Although the LWR model can predict some traffic situations rather well, the model exhibits various deficits~\cite{Dag1995}. In general, the model cannot portray traffic instabilities, which include growing traffic waves and so-called capacity drops~\cite{TreKes2013}.

\subsection{The ARZ and the Greenberg model}  \label{sec:basics_ARZmodel}

In the second order ARZ model, the traffic density $\rho_\road = \rho_\road(x,t)$ and velocity $v_\road = v_\road(x,t)$ are not directly coupled, but evolve according to a system of conservation laws
\begin{equation}\label{eq:ARZ_network}
	\partial_t \colvector{\rho_\road \\ \rho_\road w_\road} + \partial_x \colvector{\rho_\road v_\road \\ \rho_\road v_\road w_\road } = 0,  \qquad (x,t) \in (a_\road, b_\road) \times \mathbb{R}_\geqzero.
\end{equation}
The first equation describes the conservation of mass and the second equation includes the evolution of the vehicle speed. The Lagrangian marker $w_\road$ for each road in~\eqref{eq:ARZ_network} is defined by the relation 
\begin{equation}\label{eq:Lagrangianmarker}
	w_\road = v_\road + \pressure (\rho_\road),
\end{equation}
where we assume for simplicity that the pressure function $\pressure(\rho_\road)$ is identical on all roads. The ARZ model is hyperbolic with eigenvalues $\eigenvalue_1 = v_\road- \rho_\road p'(\rho_\road) \leq \eigenvalue_2 = v_\road$ and Riemann invariants $w_\road$ and $v_\road$, see~\cite{AwRas2000}. In this case, the Lagrangian marker $w_\road$ travels with the velocity $v_\road$ of the cars. It can be seen as a driver attribute describing the empty road velocity of drivers.

The pressure function $p(\rho_\road) \sim \rho_\road^\pressexp$ with $\pressexp = 2$ is the prototype function in traffic flow~\cite{AwRas2000,GarPic2006} from which the Lagrangian marker~\eqref{eq:Lagrangianmarker} is computed. Here, we consider the pressure function from~\cite{KolGoeGoa2017,ParBui2012, HauBas2007} given by
\begin{equation}\label{eq:pressurefunction}
	\pressure(\rho_\road) = \frac{\vmax}{2} \left( \frac{\rho_\road}{\rhomax} \right)^{2}, 
\end{equation}
which was derived from microscopic considerations~\cite{AwKlaMat2002}. The pressure function satisfies $p'(\rho_\road) > 0$ and $\rho_\road p''(\rho_\road) + 2p'(\rho_\road)>0$ for all $\rho_\road$. The former property ensures that the pressure function is strictly increasing and the latter ensures the existence of a point $\fluxmax(w_\road)$ that maximizes the flux $\rho_\road v_\road = \rho_\road (w_\road-\pressure(\rho_\road))$ for a given value of $w_\road$. The inverse function of the pressure function is given by 
\begin{equation} \label{invpressure}
	p^{-1} (y) = \rho^{\max} \left( \frac{2 y}{\vmax} \right)^{\frac{1}{2}}, 
\end{equation}
and $\fluxmax(w_\road) $ can also be computed explicitly
\begin{equation} \label{eq:sonicpoint}
	\fluxmax(w_\road)  = \argmax_{\rho_\road \in [0, \rhomax]} \rho_\road(w_\road-\pressure(\rho_\road)) = \rhomax \left( \frac{2 w_\road }{3 \vmax} \right)^{\frac{1}{2}}. 
\end{equation}

The (half-)Riemann problem for the ARZ model at the 1-to-1 junction with onramp located at $x=0$ reads:
\begin{align} \label{eq:ARZ_RP_11withOR}
	\begin{cases}
		\partial_t \colvector{\rho_\road \\ \rho_\road w_\road } + \partial_x \colvector{ \rho_\road v_\road \\ \rho_\road w_\road v_\road } = 0, &\qquad (x,t) \in (a_\road, b_\road) \times \mathbb{R}_\geqzero$,~$\road = 1,2,\\
		(\rho_\road, w_\road) = \begin{pmatrix} (\rho_\road^+,w_\road^+) & \mbox{ for } x>0 \\ (\rho_\road^-, w_\road^-) & \mbox{ for } x \leq 0 \end{pmatrix}, &\qquad x \in (a_\road, b_\road),~\road =1,2,\\
		\frac{\mathrm{d} \queuelength(t)}{\intt} = \onrampinflow - q_\onramp(t), &\qquad t \in \mathbb{R}_\geqzero, \\
		\queuelength(0) =\queuelength^0. &
	\end{cases}
\end{align}
As for the first order model, we formulate the admissible fluxes at the junction through a demand and supply framework. Following~\cite{HerRas2006, KolGoeGoa2017}, the demand and supply functions for the ARZ model are given by
\begin{align}
	\ARZdemand(\rho, w) =\begin{cases}
		\rho (w - \pressure(\rho)) &\text{ if } \rho \leq \fluxmax(w), \\
		\fluxmax(w)(w- \pressure(\fluxmax(w)))  &\text{ if } \rho > \fluxmax(w), 
	\end{cases} \label{eq:ARZDemand}
	\\
	\ARZsupply(\rho, w) =\begin{cases}
		\fluxmax(w) (w - \pressure(\fluxmax(w))) &\text{ if } \rho \leq \fluxmax(w), \\
		\rho (w- \pressure(\rho))  &\text{ if } \rho > \fluxmax(w).
	\end{cases} \label{eq:ARZSupplyimplicit}
\end{align}
Without entering the discussion in detail, by knowledge on the Riemann solution of the hyperbolic second order model, see e.g.~\cite{Lax_maththeoryofshockwaves,bressanhyperbolic}, we know that the Riemann problem for~\eqref{eq:ARZ_network} on the whole line $x \in \mathbb{R}$
with traffic state $\staterhow_L=(\rho_L, \rho_L w_L)$ on the left ($x\leq 0$) and $\staterhow_R = (\rho_R,\rho_R w_R)$ on the right ($x>0$), is solved by a 1-wave connecting $\staterhow_L, \tilde{\staterhow}$ and a 2-contact-discontinuity connecting $\tilde{\staterhow}, \staterhow_R$, see also~\cite{AwRas2000}. The intermediate state is given by $\tilde{\staterhow}  = (\tilde{\rho}, \tilde{\rho} w_L)$ and satisfies $v(\tilde{\staterhow}) = v_R$, i.e., the second order velocity is given by the velocity of the density on the right.

This can be transferred to define solutions to the (half-) Riemann problem in~\eqref{eq:ARZ_RP_11withOR}. We follow the approach of~\cite{KolGoeGoa2017} to define the Riemann solver at the 1-to-1 junction with onramp: We choose to rewrite the supply function~\eqref{eq:ARZSupplyimplicit} at the 1-to-1 junction with onramp depending on the traffic states left and right from the junction
\begin{equation}
	\ARZsupply(\rho_\Leftroad, \rho_\Rightroad, w_\Leftroad, v_\Rightroad) =\begin{cases}
		\fluxmax( w_\Leftroad) (w_\Leftroad - \pressure(\fluxmax(w_\Leftroad))) &\text{ if } \tilde{\rho} \leq \fluxmax(w_\Leftroad), \\
		\tilde{\rho} (w_\Leftroad- \pressure(\tilde{\rho}))  &\text{ if } \tilde{\rho} > \fluxmax(w_\Leftroad).
	\end{cases} \label{eq:ARZSupplyexplicit}
\end{equation}
The mass flux at the interface $x=0$ is determined by the intermediate density value $\tilde{\rho}$ where $\tilde{\rho} = \tilde{\rho}(\rho_\Leftroad, \rho_\Rightroad, w_\Leftroad, v_\Rightroad)$ is the density of the state $\tilde{\staterhow}$ such that $w(\tilde{\staterhow}) = v_\Leftroad + p(\rho_\Leftroad)$ and $v(\tilde{\staterhow}) = v_\Rightroad$. By assumption on the pressure function, there exists at most one  state with $\tilde{\rho} \neq 0$ satisfying these conditions\footnote{We remark that the approach in~\cite{KolGoeGoa2017} works with a fixed pressure function $\pressure$. There are also Riemann solvers considering a change in the pressure function after merge type junctions~\cite{HerRas2006, HerMouRas2006}. }. The mass fluxes for the 1-to-1 junction with onramp are then given by
\begin{align} 
	\begin{split}\label{eq:ARZ_11withonramp}
		\massflux_\Leftroad &= \min \bigg \lbrace \ARZdemand(\rho_\Leftroad, w_\Leftroad), \max \Big \{ \beta \ARZsupply, \ARZsupply - \demandonramp(l) \Big \} \bigg \rbrace, \\
		\massflux_\onramp &= \min \bigg \{ \demandonramp(\queuelength), \max \Big \lbrace (1-\beta) \ARZsupply, \ARZsupply
		- \ARZdemand( \rho_\Leftroad, w_\Leftroad) \Big \}  \bigg \rbrace, \\
		\massflux_\Rightroad &= \massflux_\Leftroad + \massflux_\onramp,
	\end{split}
\end{align} 
where $\ARZsupply = \ARZsupply(\rho_\Leftroad, \rho_\Rightroad, w_\Leftroad, v_\Rightroad)$. One weakness of the ARZ model as pointed out by Greenberg~\cite{Gre2001} is that drivers travelling at speed $v_\road$ in traffic with local density $\rho_\road$ never try to adjust their speed to a maximum allowable speed $\LWRvelocity(\rho_\road)$. The  ARZ model was extended by Greenberg~\cite{Gre2001} with a relaxation term on the right-hand side of the second equation to correct this weakness. Given initial densities and velocities, the traffic dynamics are governed by the equations
\begin{align} \label{eq:Greenberg_network}
	\partial_t \colvector{\rho_\road \\ \rho_\road w_\road} + \partial_x \colvector{\rho_\road v_\road \\ \rho_\road w_\road v_\road} = \colvector{0 \\ - \rho_\road \frac{v_\road - \LWRvelocity(\rho_\road)}{\relaxationtime}}, \qquad (x,t) \in (a_\road, b_\road) \times \mathbb{R}_\geqzero.
\end{align}
The relaxation term on the right-hand side includes that drivers tend to adopt an equilibrium (or preferential) speed $V(\rho_\road)$. The factor $\relaxationtime > 0$ is interpreted as relaxation time towards an equilibrium speed $\LWRvelocity(\rho_\road)$. In the numerical examples, we use the Greenberg model with the LWR velocity given by~\eqref{eq:LWR_Velocity} as equilibrium velocity. The model therefore possibly contains a relaxation towards the LWR model. For $\relaxationtime = \infty$, we obtain the original ARZ model. For any value $\relaxationtime \in (0, \infty)$, we obtain the relaxed model to which we will refer to as "Greenberg model". The Greenberg model was mathematically derived from car following models in~\cite{AwKlaMat2002}. A generalization with a more flexible source term where the relaxation depends on the density and velocity is considered in~\cite{SieMau2006, SieMauMou2009}.  For an analytical convergence analysis of the Greenberg model for $\tau \rightarrow 0$, we refer to~\cite{Ras2002, AwKlaMat2002}. Moreover, some numerical experiments comparing the Greenberg model with the LWR model are discussed in~\cite{KolGoeGoa2017}.
\begin{Remark}
	The source term in the Greenberg model~\eqref{eq:Greenberg_network} is relevant for the road dynamics but is neglected when considering Riemann problems at intersections~\cite{SieMauMou2009}. Coupling conditions for the Greenberg model and the 1-to-1 junction with ramp buffer are therefore also given by~\eqref{eq:ARZ_11withonramp}.
\end{Remark}
	

\section{Capacity drop and introduction of the combined model}\label{sec:capdropandALWRmodel}

First, we introduce the capacity drop. Then, we establish a new modeling approach, which combines first and second order traffic models at junction points leading to new coupling conditions.
\subsection{The capacity drop} \label{sec:capdrop_litreview}

Several traffic network models which equip the first order LWR model with coupling conditions at the junction~\cite{CocGarPic2005,HerKla2003,HolRis1995} do not reproduce the capacity drop, see also the discussion in~\cite{HauBasChi2005}. Lebacque~\cite{trafficandgranularflow_lebacque} introduced an additional new state variable for the junction in the LWR model. Using this modification, the model is able to represent the capacity drop. Haut et al.~\cite{HauBasChi2005} triggered a capacity drop by considering a $\gcapdrop$\emph{-capacity drop function} $\gcapdrop(\sumofincomingdemands)$ at the junction. If the sum of incoming demands $\sumofincomingdemands$ is above the maximum flux $\fmax$ of the outgoing road, the discharge flow at the junction is reduced. The shape of the flow reduction is prescribed by the $\gcapdrop$-capacity drop function. Nonlocal point constraints limiting the discharging flow at the junction by averaging the density conditions upstream are considered in~\cite{AndDonRaz2015}. The constraint function at the junction $x = 0$  depends on the solution itself in an upstream neighborhood $x < 0$. They analyzed the capacity drop in the model and further extended the nonlocal point constraint to the ARZ model~\cite{AndDonRos2016}.

Within the second order traffic network models, the capacity drop is included more naturally. The junction model by Haut and Bastin~\cite{HauBas2007} is able to reproduce the capacity drop without the introduction of any new parameters compared to those of the single road model~\eqref{eq:ARZ_network}. Parzani and Buisson~\cite{ParBui2012} study the model from~\cite{HauBas2007} in a traffic scenario where the bottleneck is caused by an onramp. Furthermore, the Riemann solver in~\cite{KolGoeGoa2017} for the junction with onramp, which we recalled in Section~\ref{sec:trafficnetworks}, shows that the Greenberg model is able to portray the capacity drop and can be applied within traffic control strategies.

In general, traffic control strategies aim to reduce the total time drivers spend in a traffic network. Drivers enter highways at onramps on several locations and controlling these entry points by ramp metering approaches can decrease the total travel time in comparison to the uncontrolled case~\cite{Ban1991}. For example, a control strategy for ramp metering that can manage to sustain high flows at the sources of the network leads to a decrease of the total time spent in the network~\cite{Pap2002}. A model covering the capacity drop is therefore an essential requirement for the development of traffic control strategies.

\subsection{Construction of the combined model} \label{sec:constructionofALWRmodel}

We begin to develop a junction model for the 1-to-1 junction with ramp buffer covering the capacity drop. The junction model~\eqref{eq:ARZ_11withonramp} of the Greenberg model includes the capacity drop, see~\cite{KolCosGoa2018}. However considering the second order dynamics, it comes at increased computational cost in comparison to the LWR model. The sharp analysis of second order models by Daganzo~\cite{Dag1995} states that the traffic for low densities is free and similar to the dynamics of the LWR model. We therefore represent the traffic on the incoming and outgoing road by the LWR model, which gives a good description for low densities. Our idea is now to customize the coupling conditions of the LWR model in case of increasing densities upstream of the junction. We denote the new model as augmented LWR model, or shortly ALWR.

By including second order quantities, we seek to achieve a capacity drop similar to the second order models. However, the velocity  $\LWRvelocity_\road = \LWRvelocity(\rho_\road)$ stays coupled to the density $\rho_\road$ and the derived $w$-value for each road is always given by $w(\rho_\road) = \pressure(\rho_\road) +\LWRvelocity_\road$ and thus, also coupled to the density. At the junction point, our coupling conditions do not only consider the actual density $\rho_\Rightroad$, but additionally the velocity $\LWRvelocity_\Rightroad$ and the upstream $w$-value $w_\Leftroad = w(\rho_\Leftroad)$. The upstream $w$-value induces a nonlocal point constraint, similar to~\cite{SanDonPel2019} at the junction point, manipulating the flow through the junction such that a capacity drop is achieved. As explained above, our coupling conditions are motivated by the second order coupling conditions for the ARZ model and we combine them with the standard LWR coupling conditions by means of the augmented supply function
\begin{align} \label{eq:ALWRsupply}
	\ALWRsupply(\rho_\Leftroad, \rho_\Rightroad,\queuelength) &= \begin{cases}
		\LWRsupply(\rho_\Rightroad) \qquad \text{if } 0 \leq \LWRdemand(\rho_\Leftroad) + \demandonramp(\queuelength) \leq \fmax, \\[2pt]
		\min \lbrace \LWRsupply(\rho_\Rightroad),\ARZsupply \left(\rho_\Leftroad, \rho_\Rightroad, w_\Leftroad, \LWRvelocity_\Rightroad \right) \rbrace \qquad \text{otherwise.}
	\end{cases}
\end{align}
To determine the demands, the LWR demand function~\eqref{eq:LWR_demandsupply} and the onramp demand~\eqref{eq:onrampdemand} is applied. The value $\sumofincomingdemands = \LWRdemand(\rho_\Leftroad) + \demandonramp(\queuelength)$ indicates the cumulated desired inflow into the outgoing road.  It was observed in experiments that the threshold of traffic density at which a capacity drop is observed is merely constant across the observation days~\cite{ChuRudCas2007}. When the demand is at the level of the road capacity, breakdown and congestion occur~\cite{ParBui2012}. Therefore, the exceedance of the road capacity $\fmax$ in~\eqref{eq:ALWRsupply} is meant to trigger the formation of a jam upstream the junction. 

Analogously to~\cite{HauBasChi2005}, the flux value $\sumofincomingdemands$ works as a switch that activates a reduction of the flow through the junction. Contrary to their approach, the reduction of the flow here does not only consider the level of the accumulated demand $\sumofincomingdemands$, but also the interplay of (derived) traffic states up- and downstream the junction in $\ARZsupply$ using~\eqref{eq:ARZSupplyexplicit}. In comparison with~\cite{AndDonRaz2015}, the nonlocality is restricted only to the junction interface $x=0-$. 

In case that the cumulated desired inflow does not exceed the maximum flow $\fmax$ of the outgoing road, the supply function of the LWR model is applied. In contrast, to evaluate the supply at a junction when the desired inflow exceeds the maximum flow, the minimum of the LWR and ARZ supply is used.  Meaning that the incoming flow drops below the inflow predicted by the LWR model, if the application of the ARZ model to the situation with derived second order quantity $w_\Leftroad$ would lead to a lower inflow. In case of a flow reduction in comparison to the LWR model, we determine the flow by the ARZ supply~\eqref{eq:ARZSupplyexplicit} using the derived quantities $w_\Leftroad$ and  $\LWRvelocity_\Rightroad$ in the augmented supply function. For the coupling conditions using the augmented supply function, we introduce a quite useful partition of the phase plane, which we will explain in detail in the following.

\subsection{Discussion of the augmented supply function} \label{sec:augmentedsupply}

This section focuses on the investigation of the supply function in the combined model to show that the ALWR model reduces the flux at the junction in case of congestion. We analyze the problem~\eqref{eq:LWR_RP_11withOR} and examine the augmented supply function~\eqref{eq:ALWRsupply} in case that the desired inflow $\sumofincomingdemands$ exceeds the capacity of the outgoing road $\fmax$. 

Using the LWR velocity~\eqref{eq:LWR_Velocity} and the pressure function~\eqref{eq:pressurefunction}, the derived $w$-value reads 
\begin{equation} \label{calc:functionwofrho}
	w(\rho_\road) = \LWRvelocity(\rho_\road) + \pressure(\rho_\road) = \vmax \left( 1 - \frac{\rho_\road}{\rhomax}\right) + \frac{\vmax}{2} \left(\frac{\rho_\road}{\rhomax}\right)^2, 
\end{equation}
and is plotted along with the velocity function in Figure~\ref{img:w_valuesandvelocities}.

\begin{figure}[tbhp]
	\centering	
	\tikzsetnextfilename{wvalueandvelocity}
%
%
\begin{tikzpicture}
\setlength\fwidth{0.4\textwidth}
\begin{axis}[%
width=0.951\fwidth,
height=0.75\fwidth,
at={(0\fwidth,0\fwidth)},
scale only axis,
xmin=0,
xmax=600,
xtick={0, 100, 350, 500, 600},
xticklabels={0, $\rho_a$, $\rho_b$, $\rho_c$, $\rhomax$},
xlabel style={font=\color{white!15!black}},
xlabel={},
ymin=0,
ymax=100,
ytick={51.38, 58.68,84.72, 100},
yticklabels={ \small $w(\rho_c)$, \small $ w(\rho_b)$, \small $w(\rho_a)$, \small$\vmax$},
ylabel style={font=\color{white!12!black},anchor= near yticklabel},
ylabel={},
axis background/.style={fill=white},
legend style={ font=\small, anchor=north east, legend cell align=left, align=left, draw=white!15!black}
]
\addplot [color=black, line width=1.0pt]
  table[row sep=crcr]{%
0	100\\
16.7	97.2554013888889\\
33.4	94.5882722222223\\
50.1	91.9986125\\
66.8	89.4864222222222\\
83.5	87.0517013888889\\
100.1	84.7083347222223\\
116.7	82.4415125\\
133.3	80.2512347222222\\
149.9	78.1375013888889\\
166.5	76.1003125\\
183.1	74.1396680555556\\
199.7	72.2555680555556\\
216.3	70.4480125\\
232.9	68.7170013888889\\
249.5	67.0625347222223\\
266.1	65.4846125\\
282.7	63.9832347222223\\
299.3	62.5584013888889\\
315.9	61.2101125\\
332.5	59.9383680555555\\
349.1	58.7431680555555\\
365.7	57.6245125\\
382.3	56.5824013888889\\
398.9	55.6168347222223\\
415.5	54.7278125\\
432.1	53.9153347222223\\
448.7	53.1794013888889\\
465.3	52.5200125\\
481.9	51.9371680555556\\
498.5	51.4308680555556\\
515.1	51.0011125\\
531.7	50.6479013888888\\
548.3	50.3712347222222\\
564.9	50.1711125\\
581.5	50.0475347222222\\
598.1	50.0005013888889\\
600	50\\
};
\addlegendentry{$w(\rho_\road)$}

\addplot [color=black, dashdotted, line width=1.0pt]
  table[row sep=crcr]{%
0	100\\
600	0\\
};
\addlegendentry{$\LWRvelocity(\rho_\road)$}

\end{axis}
\end{tikzpicture}%
	\caption{Derived $w$-values and velocities.}
	\label{img:w_valuesandvelocities}
\end{figure}

To shorten the notation, we define the derived ARZ supply by
\begin{equation} \label{eq:derivedARZsupply}
	\ARZsupply(\rho_\Leftroad, \rho_\Rightroad) =  \ARZsupply(\rho_\Leftroad, \rho_\Rightroad,w_\Leftroad, \LWRvelocity_\Rightroad) = \begin{cases}
		\fluxmax_\Leftroad (w_\Leftroad - p(\fluxmax_\Leftroad))  &\text{if } \tilde{\rho} \leq \fluxmax_\Leftroad, \\
		\tilde{\rho} (w_\Leftroad - p(\tilde{\rho}))  &\text{otherwise,} 
	\end{cases}
\end{equation}
in the analysis below. Here, $\fluxmax_\Leftroad = \fluxmax(w_\Leftroad)$ is the sonic point given by~\eqref{eq:sonicpoint}. Note that we are able to do this because we assume that the dynamics on the roads are described by the LWR model.

The value $w_\Leftroad$ scales the supply function and also determines the position of the sonic point $\fluxmax_\Leftroad$. If we interpret the term $(w_\Leftroad - p(\rho_\Leftroad))$ as the velocity velocity function, then $w_\Leftroad$ is the empty road velocity. In the ARZ model, $w_\Leftroad$ is a driver attribute describing the empty road velocity of different drivers and travels with the speed of car. In the LWR model, the  empty road velocity is $\vmax$, which is prescribed by the velocity function~\eqref{eq:LWR_Velocity} and identical for all drivers. Using the derived marker $w_\Leftroad$ in~\eqref{eq:derivedARZsupply} can be seen as a correction to the empty road velocity depending on the density. For $\rho = 0$, we obtain $w(0) = \vmax$. For $\rho = \rhomax$, we obtain $w(\rhomax) = \vmax /2$ with the specific choices~\eqref{eq:LWR_Velocity} and~\eqref{eq:pressurefunction}, see also Figure~\ref{img:w_valuesandvelocities}. The higher the density, the lower the maximum speed, drivers can reach. Since a reduction in $w_\Leftroad$ reduces the flux, we can understand the above correction in the $w$-value as a flux reduction for high density values, see Figure~\ref{img:LWRvsARSupply}. 
The value $\tilde{\rho}$ in~\eqref{eq:derivedARZsupply} determines the flux at the junction in the second order model. This intermediate density, which is used to evaluate the supply, is derived from the combination of $\rho_\Leftroad$ and $\rho_\Rightroad$. Greenberg~\cite{Gre2001} refers to this property as the anticipatory nature of the ARZ model, i.e., that the density $\rho$ and velocity $v$ behind a contact are also determined by $\rho,v$ ahead of it. Traditionally, in the coupling at the junction of the LWR model, no connection is made between the states of the incoming and outgoing road in the supply. We therefore introduced $\tilde{\rho}$ motivated from the second order model to combine information from both roads in the supply.
Here, the derived value of $\tilde{\rho}$ decides which branch of the supply function~\eqref{eq:derivedARZsupply} is used to evaluate the flux at the junction. It holds that $\tilde{\rho}$ is either the intersection of the curves $\lbrace w(U)= w_\Leftroad \rbrace$ and $\lbrace v(U) = \LWRvelocity_\Rightroad \rbrace$ or $\tilde{\rho} = 0$, see~\cite{AwRas2000}. We can express $\tilde{\rho}$ explicitly by
\begin{equation} \label{eq:explicittilderho}
	\tilde{\rho} = \pressure^{-1}(\max \{w_\Leftroad -\LWRvelocity_\Rightroad,0\}), \qquad \text{ where }
	w_\Leftroad = \LWRvelocity_\Leftroad + \pressure(\rho_\Leftroad).
\end{equation}
If we plug $\tilde{\rho}$ in the augmented supply function~\eqref{eq:ALWRsupply}, then the augmented supply can either take the value of the LWR supply or the derived ARZ supply. In situations where the LWR supply is lower than the derived ARZ supply, the ALWR model is equal to the LWR model in its description of traffic dynamics. On the other hand, in situations where the derived ARZ supply is lower than the LWR supply, traffic flow at the junction is reduced and the model differs from the LWR model. Note that we aim to approximate the fluxes at the junction in the ARZ model, but not the traffic state or the solution structure since we keep the LWR model on the roads and only approximate the flux at the junction point.  

\begin{figure}
	\centering
	\subfloat[Derived ARZ and LWR supply function.]{ 
		\tikzsetnextfilename{LWRvsARSupply}
%
%
\begin{tikzpicture}
\setlength\fwidth{0.4\textwidth}
\begin{axis}[%
scaled ticks=false,
width=0.951\fwidth,
height=0.75\fwidth,
at={(0\fwidth,0\fwidth)},
scale only axis,
xmin=0,
xmax=600,
xtick={0,300, 375.2777, 469.9291,600},
xticklabels={0,$\frac{\rhomax}{2}$, $\fluxmax_b$, $\fluxmax_a$, $\rhomax$},
xlabel style={font=\color{white!15!black}},
xlabel={},
ymin=0,
ymax=40000,
ytick={15000},
yticklabels={\small $\fmax$},
ylabel style={font=\color{white!15!black}},
ylabel={},
axis background/.style={fill=white},
legend style={at={(0.03,0.7)}, font=\small, anchor=south west, legend cell align=left, align=left, draw=white!15!black},
legend columns ={2},
]
\addplot [color=black, line width=1.0pt]
  table[row sep=crcr]{%
0	15000\\
301	14999.8333333333\\
302.1	14999.265\\
303.299999999999	14998.185\\
304.6	14996.4733333333\\
306	14994\\
307.5	14990.625\\
309.200000000001	14985.8933333333\\
311	14979.8333333333\\
313	14971.8333333333\\
315.1	14961.9983333333\\
317.4	14949.54\\
319.9	14933.9983333333\\
322.5	14915.625\\
325.299999999999	14893.3183333333\\
328.299999999999	14866.5183333333\\
331.4	14835.6733333333\\
334.700000000001	14799.3183333333\\
338.200000000001	14756.7933333333\\
341.799999999999	14708.7933333333\\
345.6	14653.44\\
349.6	14589.9733333333\\
353.700000000001	14519.385\\
358	14439.3333333333\\
362.5	14348.9583333333\\
367.1	14249.5983333333\\
371.9	14138.3983333333\\
376.9	14014.3983333333\\
382	13879.3333333333\\
387.299999999999	13729.785\\
392.799999999999	13564.6933333333\\
398.4	13386.24\\
404.200000000001	13190.3933333333\\
410.200000000001	12975.9933333333\\
416.299999999999	12745.7183333333\\
422.6	12494.8733333333\\
429.1	12222.1983333333\\
435.799999999999	11926.3933333333\\
442.6	11610.8733333333\\
449.6	11269.9733333333\\
456.799999999999	10902.2933333333\\
464.1	10511.865\\
471.6	10092.24\\
479.299999999999	9641.91833333333\\
487.1	9165.59833333333\\
495.1	8655.99833333333\\
503.299999999999	8111.51833333334\\
511.700000000001	7530.51833333333\\
520.200000000001	6918.66\\
528.9	6267.465\\
537.799999999999	5575.19333333334\\
546.799999999999	4848.29333333334\\
556	4077.33333333334\\
565.4	3260.47333333334\\
574.9	2404.99833333334\\
584.6	1500.47333333333\\
594.5	544.958333333332\\
600	0\\
};
\addlegendentry{$\LWRsupply(\rho)$}

\addplot [color=black, dashdotted, line width=1.0pt, forget plot]
  table[row sep=crcr]{%
0	25468.9106409441\\
451.799999999999	25468.76669\\
452.799999999999	25468.2491733333\\
453.900000000001	25467.2440529167\\
455.099999999999	25465.6255345833\\
456.400000000001	25463.2558133333\\
457.799999999999	25459.9849233333\\
459.400000000001	25455.33103\\
461.099999999999	25449.3123429167\\
463	25441.27125\\
465	25431.3020833333\\
467.200000000001	25418.5454933333\\
469.599999999999	25402.4797866667\\
472.099999999999	25383.34758875\\
474.799999999999	25359.92264\\
477.599999999999	25332.5832533333\\
480.599999999999	25299.82797\\
483.700000000001	25262.19385375\\
487	25217.8745833333\\
490.5	25166.03921875\\
494.099999999999	25107.49866375\\
497.900000000001	25039.9158695833\\
501.799999999999	24964.32419\\
505.900000000001	24877.9971695833\\
510.099999999999	24782.2159304167\\
514.5	24673.8300520833\\
519	24554.3945833333\\
523.700000000001	24420.2990204167\\
528.5	24273.3994270833\\
533.5	24109.5874479167\\
538.599999999999	23931.04827\\
543.799999999999	23736.9878233333\\
549.200000000001	23522.4922933333\\
554.700000000001	23290.3052329167\\
560.400000000001	23034.9182133333\\
566.200000000001	22759.4753433333\\
572.099999999999	22462.9971720833\\
578.200000000001	22139.01781\\
584.400000000001	21791.3539466667\\
590.799999999999	21412.83704\\
597.299999999999	21007.7635670833\\
600	20833.3333333333\\
};
\addlegendimage{no markers, dashdotted,line width=1.0pt, draw=black};
\addlegendentry{$\ARZsupply(\rho_a, \rho)$};

\addplot [color=black, dashed, line width=1.0pt, forget plot]
  table[row sep=crcr]{%
0	14681.0016366853\\
376.299999999999	14680.8380629167\\
377.4	14680.2959966667\\
378.6	14679.2706033333\\
379.9	14677.6470279167\\
381.299999999999	14675.3001670833\\
382.799999999999	14672.0945066667\\
384.5	14667.59359375\\
386.299999999999	14661.81852125\\
388.299999999999	14654.1782795833\\
390.4	14644.76288\\
392.700000000001	14632.8043079167\\
395.1	14618.47995125\\
397.700000000001	14600.8214120833\\
400.4	14580.1122133333\\
403.299999999999	14555.1591754167\\
406.4	14525.3591466667\\
409.6	14491.1837866667\\
413	14451.0420833333\\
416.5	14405.56359375\\
420.200000000001	14352.8624433333\\
424	14293.7466666667\\
428	14226.0066666667\\
432.1	14150.6502554167\\
436.4	14065.1138133333\\
440.799999999999	13970.62704\\
445.4	13864.2404633333\\
450.1	13747.4286804167\\
454.9	13619.5790070833\\
459.9	13477.11336125\\
465	13321.9270833333\\
470.299999999999	13149.9740379167\\
475.700000000001	12963.4533204167\\
481.200000000001	12761.5950933333\\
486.9	12539.59487375\\
492.700000000001	12300.1672245833\\
498.6	12042.43927\\
504.700000000001	11760.76544125\\
510.9	11458.4361070833\\
517.200000000001	11134.4654933333\\
523.700000000001	10782.2781870833\\
530.299999999999	10405.80512125\\
537	10003.9370833333\\
543.799999999999	9575.52949\\
550.799999999999	9112.63104\\
557.9	8620.14311958334\\
565.1	8096.78340958333\\
572.4	7541.23008\\
579.9	6943.97952791667\\
587.5	6311.03515625001\\
595.200000000001	5640.92202666666\\
600	5208.33333333333\\
};

\addlegendimage{no markers, dashed,line width=1.0pt, draw=black};
\addlegendentry{$\ARZsupply(\rho_b, \rho)$};

\addplot [color=black, dotted, line width=1.0pt, forget plot]
  table[row sep=crcr]{%
0	12031.456446159\\
352.200000000001	12031.3065766667\\
353.4	12030.7392633333\\
354.6	12029.74787\\
355.9	12028.1936279167\\
357.299999999999	12025.95923375\\
358.799999999999	12022.9175733333\\
360.5	12018.6569270833\\
362.299999999999	12013.1992545833\\
364.299999999999	12005.9876795833\\
366.4	11997.10848\\
368.700000000001	11985.8383745833\\
371.1	11972.3452179167\\
373.700000000001	11955.7171454167\\
376.5	11935.4552604167\\
379.4	11911.8763633333\\
382.5	11883.73046875\\
385.700000000001	11851.4633620833\\
389.1	11813.5726429167\\
392.6	11770.6518366667\\
396.299999999999	11720.9198129167\\
400.1	11665.1372220833\\
404.1	11601.21653875\\
408.200000000001	11530.1056433333\\
412.5	11449.3815104167\\
416.9	11360.2001654167\\
421.5	11259.7724479167\\
426.200000000001	11149.4843433333\\
431	11028.75125\\
436	10894.1866666667\\
441.1	10747.57242625\\
446.4	10585.07648\\
451.799999999999	10408.76669\\
457.299999999999	10217.90548375\\
463	10007.9379166667\\
468.799999999999	9781.41824\\
474.700000000001	9537.50823291667\\
480.799999999999	9270.84970666667\\
487	8984.54125\\
493.299999999999	8677.63413375\\
499.700000000001	8349.14792041667\\
506.299999999999	7992.52985458334\\
513	7611.70875000001\\
519.799999999999	7205.58022333334\\
526.700000000001	6773.00428291667\\
533.799999999999	6306.07299\\
541	5809.66375\\
548.299999999999	5282.49811291667\\
555.700000000001	4723.25782041667\\
563.200000000001	4130.58389333333\\
570.9	3494.63669041667\\
578.700000000001	2821.66966625\\
586.6	2110.16640333334\\
594.6	1358.56520333334\\
600	833.333333333336\\
};
\addlegendimage{no markers, dotted,line width=1.0pt, draw=black};
\addlegendentry{$\ARZsupply(\rho_c, \rho)$};
\node[pin={[pin distance=0.35cm]90: $\fluxmax_c$}] at (axis description cs:0.58,-0.05) {};
\end{axis}
\end{tikzpicture}%
		\label{img:LWRvsARSupply}}
	\subfloat[Limiting case $\rholim$.]{ 
		\tikzsetnextfilename{Limitingsupply}
%
%
\begin{tikzpicture}
\setlength\fwidth{0.4\textwidth}
\begin{axis}[%
scaled ticks=false,
width=0.951\fwidth,
height=0.75\fwidth,
at={(0\fwidth,0\fwidth)},
scale only axis,
xmin=0,
xmax=600,
xtick={0,300,377.976314968462, 600},
xticklabels={0,$\frac{\rhomax}{2}$,$\sigmalim$,$\rhomax$},
xlabel style={font=\color{white!15!black}},
xlabel={},
ymin=0,
ymax=40000,
ytick={15000},
yticklabels={\small $\fmax$},
ylabel style={font=\color{white!15!black}},
ylabel={},
axis background/.style={fill=white},
legend style={at={(0.03,0.8)}, font=\small, anchor=south west, legend cell align=left, align=left, draw=white!15!black},
legend columns ={2},
]
\addplot [color=black, line width=1.0pt]
table[row sep=crcr]{%
	0	15000\\
	301	14999.8333333333\\
	302.1	14999.265\\
	303.299999999999	14998.185\\
	304.6	14996.4733333333\\
	306	14994\\
	307.5	14990.625\\
	309.200000000001	14985.8933333333\\
	311	14979.8333333333\\
	313	14971.8333333333\\
	315.1	14961.9983333333\\
	317.4	14949.54\\
	319.9	14933.9983333333\\
	322.5	14915.625\\
	325.299999999999	14893.3183333333\\
	328.299999999999	14866.5183333333\\
	331.4	14835.6733333333\\
	334.700000000001	14799.3183333333\\
	338.200000000001	14756.7933333333\\
	341.799999999999	14708.7933333333\\
	345.6	14653.44\\
	349.6	14589.9733333333\\
	353.700000000001	14519.385\\
	358	14439.3333333333\\
	362.5	14348.9583333333\\
	367.1	14249.5983333333\\
	371.9	14138.3983333333\\
	376.9	14014.3983333333\\
	382	13879.3333333333\\
	387.299999999999	13729.785\\
	392.799999999999	13564.6933333333\\
	398.4	13386.24\\
	404.200000000001	13190.3933333333\\
	410.200000000001	12975.9933333333\\
	416.299999999999	12745.7183333333\\
	422.6	12494.8733333333\\
	429.1	12222.1983333333\\
	435.799999999999	11926.3933333333\\
	442.6	11610.8733333333\\
	449.6	11269.9733333333\\
	456.799999999999	10902.2933333333\\
	464.1	10511.865\\
	471.6	10092.24\\
	479.299999999999	9641.91833333333\\
	487.1	9165.59833333333\\
	495.1	8655.99833333333\\
	503.299999999999	8111.51833333334\\
	511.700000000001	7530.51833333333\\
	520.200000000001	6918.66\\
	528.9	6267.465\\
	537.799999999999	5575.19333333334\\
	546.799999999999	4848.29333333334\\
	556	4077.33333333334\\
	565.4	3260.47333333334\\
	574.9	2404.99833333334\\
	584.6	1500.47333333333\\
	594.5	544.958333333332\\
	600	0\\
};
\addlegendentry{$\LWRsupply(\rho)$}

\addplot [color=black, dashdotted, line width=1.0pt]
table[row sep=crcr]{%
	0	15000\\
	379	14999.8348122091\\
	380.1	14999.2883832417\\
	381.299999999999	14998.2551255803\\
	382.6	14996.6193742248\\
	384	14994.2551483421\\
	385.5	14991.0259887653\\
	387.200000000001	14986.4923234671\\
	389	14980.675484475\\
	391	14972.9802855948\\
	393.1	14963.4974946873\\
	395.4	14951.4537280585\\
	397.799999999999	14937.0277827356\\
	400.4	14919.2445641914\\
	403.1	14898.3897919532\\
	406	14873.2621273269\\
	409.1	14843.254503646\\
	412.299999999999	14808.8421721044\\
	415.700000000001	14768.4234415082\\
	419.200000000001	14722.6326080512\\
	422.9	14669.5712955396\\
	426.700000000001	14610.0535601673\\
	430.700000000001	14541.8553457403\\
	434.799999999999	14465.9919034526\\
	439.1	14379.8830621102\\
	443.5	14284.7675129072\\
	448.1	14177.6770046495\\
	452.799999999999	14060.0968135311\\
	457.6	13931.4110828854\\
	462.6	13788.0194190183\\
	467.700000000001	13631.8306539573\\
	473	13458.7737981748\\
	478.4	13271.0634411984\\
	483.9	13067.9261005279\\
	489.6	12844.5284341361\\
	495.4	12603.6037840503\\
	501.299999999999	12344.2753594372\\
	507.4	12060.864457436\\
	513.6	11756.6837809075\\
	519.9	11430.743370685\\
	526.4	11076.4301347411\\
	533	10697.7011651033\\
	539.700000000001	10293.4427942714\\
	546.6	9856.01213271816\\
	553.6	9390.08046997092\\
	560.700000000001	8894.39921019634\\
	567.9	8367.68106422776\\
	575.200000000001	7808.59916206516\\
	582.700000000001	7207.59386418122\\
	590.299999999999	6570.71079676994\\
	598	5896.46970149798\\
	600	5716.52366928448\\
};
\addlegendentry{$S^\ARZ(\rholim, \rho)$}

\end{axis}
\end{tikzpicture}%
		\label{img:ARSupply_limitingcase}}

	\caption{Supply functions.}
\end{figure}

%

With $\rholim$, we denote the density for which we have
\begin{equation}\label{eq:rholim}
	\ARZsupply(\rholim,0) = \fmax,
\end{equation}
see Figure~\ref{img:ARSupply_limitingcase}. This is used, among other criteria, to distinguish whether the ARZ or LWR supply is applied. The $w$-value is denoted $\wlim = w(\rholim)$. Plugging the densities $\rho_\Leftroad$, $\rho_\Rightroad$ at the junction into the augmented supply function, we can distinguish the following cases:

\paragraph{Case 1}

The LWR supply $\LWRsupply(\rho_\Rightroad)$ exceeds the maximum of the ARZ supply, which is exemplarily depicted in Figure~\ref{img:ARSupply_Case1}. The maximum of the ARZ supply~\eqref{eq:derivedARZsupply} is $\ARZsupply(\rho_\Leftroad,0)$ for which $\tilde{\rho} = 0$ since $w_\Leftroad \leq \vmax$. The exceedance is only possible for $\rho_\Leftroad \leq \rholim$. In this case, the considered supply is the supply value from the ARZ model. This result is independent of the value $\tilde{\rho}$:
\begin{align*}
	\LWRsupply(\rho_\Rightroad) > \ARZsupply(\rho_\Leftroad, 0) \quad
	\Rightarrow \qquad \min \lbrace \LWRsupply(\rho_\Rightroad), \ARZsupply(\rho_\Leftroad,\rho_\Rightroad) \rbrace = \ARZsupply(\rho_\Leftroad,\rho_\Rightroad).
\end{align*}

\begin{figure}[tbh]
	\centering
		\subfloat[Case 1.]{ 
		\tikzsetnextfilename{Case1}
%
%
\begin{tikzpicture}
\setlength\fwidth{0.35\textwidth}
\begin{axis}[%
legend style={font=\small},
scaled ticks=false,
width=0.951\fwidth,
height=0.75\fwidth,
at={(0\fwidth,0\fwidth)},
scale only axis,
xmin=0,
xmax=700,
xtick={264.575131106459,350},
xticklabels={,$\rho_\Rightroad$},
ymin=0,
ymax=30000,
ytick={14583.3333333333},
yticklabels={{$\LWRsupply(\rho_\Rightroad)$}},
axis background/.style={fill=white},
legend style={legend cell align=left, align=left, draw=white!15!black}
]
\addplot [color=black, line width=1.0pt]
  table[row sep=crcr]{%
0	15000\\
301	14999.8333333333\\
302.200000000001	14999.1933333333\\
303.5	14997.9583333333\\
304.9	14995.9983333333\\
306.5	14992.9583333333\\
308.200000000001	14988.7933333333\\
310.1	14982.9983333333\\
312.200000000001	14975.1933333333\\
314.4	14965.44\\
316.799999999999	14952.96\\
319.4	14937.2733333333\\
322.200000000001	14917.86\\
325.200000000001	14894.16\\
328.4	14865.5733333333\\
331.799999999999	14831.46\\
335.4	14791.14\\
339.200000000001	14743.8933333333\\
343.200000000001	14688.96\\
347.4	14625.54\\
351.799999999999	14552.7933333333\\
356.4	14469.84\\
361.200000000001	14375.76\\
366.200000000001	14269.5933333333\\
371.4	14150.34\\
376.700000000001	14019.5183333333\\
382.200000000001	13873.86\\
387.9	13712.265\\
393.799999999999	13533.5933333333\\
399.9	13336.665\\
406.200000000001	13120.26\\
412.700000000001	12883.1183333333\\
419.4	12623.94\\
426.299999999999	12341.385\\
433.4	12034.0733333333\\
440.700000000001	11700.585\\
448.200000000001	11339.46\\
455.9	10949.1983333333\\
463.799999999999	10528.26\\
471.9	10075.065\\
480.200000000001	9587.99333333333\\
488.700000000001	9065.385\\
497.4	8505.54\\
506.299999999999	7906.71833333333\\
515.4	7267.14\\
524.700000000001	6584.985\\
534.200000000001	5858.39333333333\\
543.9	5085.465\\
553.799999999999	4264.26\\
563.9	3392.79833333334\\
574.200000000001	2469.06\\
584.700000000001	1490.985\\
595.4	456.473333333335\\
600	0\\
};
\addlegendentry{LWR supply}

\addplot [color=black, dashed, line width=1.0pt]
  table[row sep=crcr]{%
0	12031.456446159\\
352.299999999999	12031.2754629167\\
353.6	12030.60352\\
355	12029.3229166667\\
356.5	12027.30734375\\
358.1	12024.42056375\\
359.9	12020.25961125\\
361.799999999999	12014.81319\\
363.9	12007.5269279167\\
366.200000000001	11998.01201\\
368.700000000001	11985.8383745833\\
371.299999999999	11971.1405420833\\
374.1	11952.9724970833\\
377.1	11930.79402625\\
380.299999999999	11904.0190795833\\
383.700000000001	11872.0142704167\\
387.200000000001	11835.2068266667\\
390.9	11791.9974404167\\
394.799999999999	11741.6239733333\\
398.9	11683.2707404167\\
403.1	11617.69975125\\
407.5	11542.6497395833\\
412.1	11457.15783875\\
416.799999999999	11362.30144\\
421.700000000001	11255.2380120833\\
426.799999999999	11134.8421066667\\
432	11002.56\\
437.4	10854.88533\\
443	10690.5129166667\\
448.700000000001	10511.3153745833\\
454.6	10313.03287\\
460.6	10097.8643033333\\
466.799999999999	9861.00477333333\\
473.200000000001	9600.81789333333\\
479.700000000001	9320.03200375\\
486.4	9012.92714666667\\
493.200000000001	8682.63256\\
500.200000000001	8322.76944333333\\
507.299999999999	7936.89944208334\\
514.6	7517.93720333334\\
522	7069.91\\
529.5	6591.52119791667\\
537.200000000001	6074.55016\\
545	5523.80208333334\\
553	4930.22541666667\\
561.1	4299.17859291668\\
569.299999999999	3629.09061708334\\
577.700000000001	2909.57882875\\
586.200000000001	2146.92167666667\\
594.799999999999	1339.37064000001\\
603.6	475.086853333351\\
608.299999999999	-2.44080374999612\\
};
\addlegendentry{ARZ supply}

\addplot [color=black, line width=1.0pt, draw=none, mark=asterisk, mark options={solid, black}, forget plot]
  table[row sep=crcr]{%
350	14583.3333333333\\
};
\addplot [color=black, line width=1.0pt, draw=none, mark=asterisk, mark options={solid, black}, forget plot]
  table[row sep=crcr]{%
264.57513110646	12031.456446159\\
};

\addplot [color=red, dashdotted, line width=1.0pt]
table[row sep=crcr]{%
	0	12031.456446159\\
	700	12031.456446159\\
};

\node[pin={[pin distance=0.15cm]90:$\tilde{\rho}$}] at (axis description cs:0.3780,0) {};
\end{axis}
\end{tikzpicture}%
		\label{img:ARSupply_Case1}}
 \\
		\subfloat[Case 2(a).]{ 
		\tikzsetnextfilename{Case2a}
%
%
\begin{tikzpicture}
\setlength\fwidth{0.35\textwidth}
\begin{axis}[%
legend style={font=\small},
scaled ticks = false,
width=0.951\fwidth,
height=0.75\fwidth,
at={(0\fwidth,0\fwidth)},
scale only axis,
xmin=0,
xmax=1000,
xtick={450,519.615242270663},
xticklabels={$\rho_\Rightroad$},
ymin=0,
ymax=30000,
ytick={11250,15000},
yticklabels={$\LWRsupply(\rho_\Rightroad)$, $\fmax$},
axis background/.style={fill=white},
legend style={legend cell align=left, align=left, draw=white!15!black}
]
\addplot [color=black, line width=1.0pt]
  table[row sep=crcr]{%
0	15000\\
301.299999999999	14999.7183333333\\
302.799999999999	14998.6933333333\\
304.5	14996.625\\
306.299999999999	14993.385\\
308.4	14988.24\\
310.700000000001	14980.9183333333\\
313.200000000001	14970.96\\
316	14957.3333333333\\
319.1	14939.1983333333\\
322.4	14916.3733333333\\
326	14887.3333333333\\
329.9	14850.9983333333\\
334	14807.3333333333\\
338.4	14754.24\\
343.1	14690.3983333333\\
348.1	14614.3983333333\\
353.4	14524.74\\
358.9	14421.7983333333\\
364.700000000001	14302.3183333333\\
370.799999999999	14164.56\\
377.200000000001	14006.6933333333\\
383.799999999999	13829.5933333333\\
390.700000000001	13628.9183333333\\
397.9	13402.5983333333\\
405.4	13148.4733333333\\
413.200000000001	12864.2933333333\\
421.200000000001	12551.76\\
429.5	12204.9583333333\\
438.1	11821.3983333333\\
447	11398.5\\
456.200000000001	10933.5933333333\\
465.700000000001	10423.9183333333\\
475.4	9872.47333333333\\
485.4	9271.14\\
495.700000000001	8616.91833333334\\
506.299999999999	7906.71833333333\\
517.200000000001	7137.36\\
528.299999999999	6313.185\\
539.700000000001	5423.985\\
551.4	4466.34\\
563.4	3436.74\\
575.700000000001	2331.58499999999\\
588.299999999999	1147.185\\
600	0\\
};
\addlegendentry{LWR supply}

\addplot [color=black, dashed, line width=1.0pt]
  table[row sep=crcr]{%
0	16137.4306091976\\
388.6	16137.1568811111\\
390.1	16136.1608748611\\
391.799999999999	16134.14769\\
393.6	16130.98752\\
395.700000000001	16125.95715375\\
398	16118.7788888889\\
400.5	16108.98609375\\
403.299999999999	16095.5411198611\\
406.299999999999	16078.2113823611\\
409.6	16055.6282311111\\
413.1	16027.61387625\\
416.9	15992.4223876389\\
420.9	15949.95814875\\
425.200000000001	15898.0484711111\\
429.799999999999	15835.2622788889\\
434.6	15761.6647588889\\
439.700000000001	15674.3223926389\\
445	15573.4548611111\\
450.5	15457.7656076389\\
456.299999999999	15323.46089625\\
462.299999999999	15171.06217125\\
468.6	14996.11377\\
475.1	14799.3643401389\\
481.799999999999	14579.05119\\
488.799999999999	14329.6173511111\\
496	14052.2311111111\\
503.4	13744.80843\\
511	13405.1623611111\\
518.9	13026.0324626389\\
527	12609.2801388889\\
535.299999999999	12152.3441698611\\
543.799999999999	11652.5433788889\\
552.5	11107.0724826389\\
561.4	10512.9978411111\\
570.5	9867.25310763889\\
579.9	9158.87536125001\\
589.5	8391.34828125\\
599.299999999999	7561.12754763889\\
609.299999999999	6664.51578375001\\
619.5	5697.65765625\\
629.9	4656.53487513889\\
640.5	3536.96109375001\\
651.200000000001	2345.99864888888\\
662.1	1068.88596375\\
670.9	-9.95261513888181\\
};
\addlegendentry{ARZ supply}

\addplot [color=black, line width=1.0pt, draw=none, mark=asterisk, mark options={solid, black}, forget plot]
  table[row sep=crcr]{%
450	11250\\
};
\addplot [color=black, line width=1.0pt, draw=none, mark=asterisk, mark options={solid, black}, forget plot]
  table[row sep=crcr]{%
519.615242270664	12990.3810567666\\
};
\node[pin={[pin distance=0.15cm]90:$\tilde{\rho}$}] at (axis description cs:0.519615242270663,0) {};

\addplot [color=red, dashdotted, line width=1.0pt]
table[row sep=crcr]{%
	0	12990.3810567666\\
	1000	12990.3810567666\\
};
\end{axis}

\end{tikzpicture}%
		\label{img:ARSupply_Case2a}}
		\subfloat[Case 2(b).]{ 
		\tikzsetnextfilename{Case2b}
%
%
\begin{tikzpicture}
\setlength\fwidth{0.35\textwidth}
\begin{axis}[%
legend style={font=\small},
scaled ticks=false,
width=0.951\fwidth,
height=0.75\fwidth,
at={(0\fwidth,0\fwidth)},
scale only axis,
xmin=0,
xmax=1000,
xtick={427.200187265877,450},
xticklabels={,$\rho_\Rightroad$},
ymin=0,
ymax=30000,
ytick={11250,15000},
yticklabels={$\LWRsupply(\rho_\Rightroad)$,$\fmax$},
axis background/.style={fill=white},
legend style={legend cell align=left, align=left, draw=white!15!black}
]
\addplot [color=black, line width=1.0pt]
  table[row sep=crcr]{%
0	15000\\
301.299999999999	14999.7183333333\\
302.799999999999	14998.6933333333\\
304.5	14996.625\\
306.299999999999	14993.385\\
308.4	14988.24\\
310.700000000001	14980.9183333333\\
313.200000000001	14970.96\\
316	14957.3333333333\\
319.1	14939.1983333333\\
322.4	14916.3733333333\\
326	14887.3333333333\\
329.9	14850.9983333333\\
334	14807.3333333333\\
338.4	14754.24\\
343.1	14690.3983333333\\
348.1	14614.3983333333\\
353.4	14524.74\\
358.9	14421.7983333333\\
364.700000000001	14302.3183333333\\
370.799999999999	14164.56\\
377.200000000001	14006.6933333333\\
383.799999999999	13829.5933333333\\
390.700000000001	13628.9183333333\\
397.9	13402.5983333333\\
405.4	13148.4733333333\\
413.200000000001	12864.2933333333\\
421.200000000001	12551.76\\
429.5	12204.9583333333\\
438.1	11821.3983333333\\
447	11398.5\\
456.200000000001	10933.5933333333\\
465.700000000001	10423.9183333333\\
475.4	9872.47333333333\\
485.4	9271.14\\
495.700000000001	8616.91833333334\\
506.299999999999	7906.71833333333\\
517.200000000001	7137.36\\
528.299999999999	6313.185\\
539.700000000001	5423.985\\
551.4	4466.34\\
563.4	3436.74\\
575.700000000001	2331.58499999999\\
588.299999999999	1147.185\\
600	0\\
};
\addlegendentry{LWR supply}

\addplot [color=black, dashed, line width=1.0pt]
  table[row sep=crcr]{%
0	11667.4952705211\\
349	11667.2154166667\\
350.6	11666.19747\\
352.299999999999	11664.29629625\\
354.200000000001	11661.16721\\
356.299999999999	11656.4688129167\\
358.700000000001	11649.49541625\\
361.299999999999	11639.99758375\\
364.1	11627.49253875\\
367.200000000001	11610.87216\\
370.5	11589.9475520833\\
374.1	11563.2849970833\\
377.9	11530.7572029167\\
382	11490.56\\
386.299999999999	11442.6518545833\\
390.9	11384.8099404167\\
395.700000000001	11317.1029870833\\
400.799999999999	11236.8377066667\\
406.1	11144.2181970833\\
411.700000000001	11036.02505375\\
417.5	10912.6236979167\\
423.5	10772.64890625\\
429.799999999999	10611.99839\\
436.299999999999	10431.3458129167\\
443	10229.0545833333\\
450	10000\\
457.200000000001	9745.22816\\
464.6	9462.79470333333\\
472.200000000001	9150.65624333333\\
480	8806.66666666666\\
488.1	8423.68335541667\\
496.4	8003.55648\\
504.9	7543.76629875\\
513.6	7041.67285333334\\
522.5	6494.51171875\\
531.6	5899.38965333334\\
540.9	5253.28014875\\
550.4	4553.01888\\
560.1	3795.29905541668\\
570	2976.66666666667\\
580.1	2093.51563875001\\
590.4	1142.08288\\
600.9	118.443232083326\\
602.1	-2.04128624999066\\
};
\addlegendentry{ARZ supply}

\addplot [color=black, line width=1.0pt, draw=none, mark=asterisk, mark options={solid, black}, forget plot]
  table[row sep=crcr]{%
450	11250\\
};
\addplot [color=black, line width=1.0pt, draw=none, mark=asterisk, mark options={solid, black}, forget plot]
  table[row sep=crcr]{%
427.200187265877	10680.0046816469\\
};

\addplot [color=red, dashdotted, line width=1.0pt]
table[row sep=crcr]{%
	0	10680.0046816469\\
	1000	10680.0046816469\\
};

\node[pin={[pin distance=0.15cm]90:$\tilde{\rho}$}] at (axis description cs:0.427200187265877,0) {};
\end{axis}
\end{tikzpicture}%
		\label{img:ARSupply_Case2b}}
	\caption{Supply functions. The red line marks the value of $\ARZsupply$.}
\end{figure}


\paragraph{Case 2}

The LWR supply for $\rho_\Rightroad$ exceeds the ARZ supply for some values, but it is below the maximum of the ARZ supply,
\begin{equation*} 
	\LWRsupply(\rho_\Rightroad) \leq \ARZsupply(\rho_\Leftroad, 0), 
\end{equation*}
Then, one has to distinguish further depending on $\tilde{\rho}$:
\paragraph{(a)} The ARZ supply evaluated at $\tilde{\rho}$ exceeds the LWR supply at $\rho_\Rightroad$, see Figure~\ref{img:ARSupply_Case2a}. The ALWR model uses the same supply function as the LWR model in this case. Therefore, we have
\begin{equation*}
	\min \lbrace \LWRsupply(\rho_\Rightroad), \ARZsupply(\rho_\Leftroad, \rho_\Rightroad) \rbrace = \LWRsupply(\rho_\Rightroad).
\end{equation*}

\paragraph{(b)} The contrary is true and we have
\begin{equation*}
	\min \lbrace \LWRsupply(\rho_\Rightroad), \ARZsupply(\rho_\Leftroad, \rho_\Rightroad) \rbrace =  \ARZsupply(\rho_\Leftroad, \rho_\Rightroad).
\end{equation*}
The ALWR model therefore uses the supply function of the ARZ model, see Figure~\ref{img:ARSupply_Case2b}.

\begin{table}[htbp]
	\renewcommand{\arraystretch}{1.2}
	\centering
	\caption{Classification of the supply function.}
	\resizebox{\textwidth}{!}{
		\begin{tabular}{|c|ccccc|c|c|}
			\hline
			& \multicolumn{1}{r}{} & \multicolumn{1}{c}{} & \multicolumn{1}{c}{} & \multicolumn{1}{c}{} &       &\multicolumn{1}{c|}{Supply} & Area \\
			\hline
			\hline
			\multirow{7}[14]{*}{$\sumofincomingdemands > \fmax$} & \multicolumn{1}{c|}{\multirow{6}[12]{*}{$\rho_\Leftroad >\rholim$}} & \multicolumn{4}{c|}{$\rho_\Rightroad \leq \nicefrac{\rhomax}{2}$}      &  ARZ & I\\
			\cline{3-8}
			~       & \multicolumn{1}{c|}{} & \multicolumn{1}{c|}{\multirow{5}[10]{*}{$\rho_\Rightroad > \frac{\rhomax}{2}$}} & \multicolumn{1}{c|}{\multirow{2}[4]{*}{$\rho_\Rightroad < \fluxmax_\Leftroad$}} & \multicolumn{2}{c|}{$\ARZsupply(\rho_\Leftroad,0) \geq \LWRsupply(\rho_\Rightroad)$}    & LWR & II \\
			\cline{5-8}         & \multicolumn{1}{c|}{} & \multicolumn{1}{c|}{} & \multicolumn{1}{c|}{} & \multicolumn{2}{c|}{$\ARZsupply(\rho_\Leftroad,0) < \LWRsupply(\rho_\Rightroad)$}    & ARZ & III \\
			\cline{4-8}         & \multicolumn{1}{c|}{} & \multicolumn{1}{c|}{} & \multicolumn{1}{c|}{\multirow{3}[6]{*}{$\rho_\Rightroad \geq \fluxmax_\Leftroad$}} & \multicolumn{2}{c|}{$\rho_\Leftroad \leq \rho_\Rightroad$}    & LWR & IV\\
			\cline{5-8}         & \multicolumn{1}{c|}{} & \multicolumn{1}{c|}{} & \multicolumn{1}{c|}{} & \multicolumn{1}{c|}{\multirow{2}[4]{*}{$\rho_\Leftroad > \rho_\Rightroad$}} & $\ARZsupply(\rho_\Leftroad,0) \geq \LWRsupply(\rho_\Rightroad)$   & ARZ & V \\
			\cline{6-8}        & \multicolumn{1}{c|}{} & \multicolumn{1}{c|}{} & \multicolumn{1}{c|}{} & \multicolumn{1}{c|}{} & $\ARZsupply(\rho_\Leftroad,0) < \LWRsupply(\rho_\Rightroad)$   &  ARZ & VI\\
			\cline{2-8}       & \multicolumn{5}{c|}{$\rho_\Leftroad \leq \rholim$}        & LWR & VII\\
			\hline
		\end{tabular}%
	}
	\label{tab:Riemannproblem_cases}%
	\renewcommand{\arraystretch}{1}
\end{table}%

Based on the previous discussion of the augmented supply function, we investigate whether the ARZ supply or the LWR supply is lower for different combinations of initial states $(\rho_\Leftroad,\rho_\Rightroad) \in [0, \rhomax]^2$. The summarized cases presented in Table~\ref{tab:Riemannproblem_cases} lead to a partitioning of the $\rho_\Leftroad$-$\rho_\Rightroad$-plane, which is shown in Figure~\ref{img:rho_l-rho_r-plane}. Note that the figure is only valid for $\sumofincomingdemands = \demandonramp + \LWRdemand(\rho_\Leftroad) > \fmax$. The ARZ supply is applied in the areas I, III, V and VI while the LWR supply is applied in the areas II, IV and VII. 

In the following, we prove this partitioning. Let us start with some notation to ease computations. We denote the maximum of the flux curve associated with the density $\rho_\Leftroad$ by
\begin{equation}\label{eq:sigmalFromRhol}
	\fluxmax_\Leftroad = \fluxmax(w_\Leftroad) \overset{\eqref{eq:sonicpoint}}{=} \rhomax \sqrt{ \frac{2 w_\Leftroad}{3 \vmax}}.
\end{equation}
We have that 
\begin{equation*} 
	\sigma(\rhomax) =  \sqrt{\frac{1}{3}} \rhomax \leq \sigma_\Leftroad \leq \sigma(0) = \sqrt{\frac{2}{3}} \rhomax  .
\end{equation*}
Plugging the density dependence of $w_\Leftroad$ given by~\eqref{calc:functionwofrho} in~\eqref{eq:sigmalFromRhol}, we denote the curve of maxima 
\begin{equation}
	\curveofmaxima(\rho_\Leftroad) \overset{\eqref{calc:functionwofrho}}{=} 
	\rhomax \sqrt{\eindrittel \left( \frac{\rhomax - \rho_\Leftroad}{\rhomax}\right)^2 + \eindrittel}. \label{eq:curveofmaxima}
\end{equation}
Therefore, the curve $\curveofmaxima(\rho_\Leftroad)$ gives an expression for the curve of maxima which is decreasing in $\rho_\Leftroad$.
The maximum of the derived supply function $\ARZsupply(\rho_\Leftroad, \cdot)$ is
\begin{equation} \label{eq:ARZSupply_fluxmax}
	\ARZsupply(\rho_\Leftroad, 0) = \max_{\rho_\Rightroad \in [0, \rhomax]} \ARZsupply(\rho_\Leftroad, \rho_\Rightroad)
	= \fluxmax_\Leftroad (w_\Leftroad - \pressure(\fluxmax_\Leftroad)) = \rhomax \left( \frac{2 w_\Leftroad}{3 \vmax}\right)^\einhalb \frac{2}{3} w_\Leftroad, 
\end{equation}
since for $\rho_\Rightroad = 0$, it follows that $\tilde{\rho} = 0$. From~\eqref{calc:functionwofrho}, we directly compute the density corresponding to a given $w$-value
\begin{equation}
	\frac{\rho_\Leftroad}{\rhomax} = 1 - \sqrt{2 \frac{w_\Leftroad}{\vmax} -1}. \label{calc:rhofromw}
\end{equation}
Then, setting 
\begin{equation*}
	\ARZsupply(\rholim,0) = \fmax = \frac{\rhomax \vmax}{4},
\end{equation*}
and using equation~\eqref{eq:ARZSupply_fluxmax} on the left hand side, we can compute $\wlim$ and the associated density $\rholim$ 
\begin{equation} \label{eq:wlim_rholim}
	\wlim = \frac{3}{2} \vmax \left( \frac{1}{4} \right)^{\frac{2}{3}},\qquad
	\frac{\rholim}{\rhomax} = 1- \sqrt{3 \left( \frac{1  }{4}  \right)^{\frac{2}{3}}-1}.
\end{equation}

Exemplarily, we briefly discuss case III in Table~\ref{tab:Riemannproblem_cases} and the corresponding region in Figure~\ref{img:rho_l-rho_r-plane}. All other cases are explained in the Appendix. Let $\rho_\Leftroad$ and $\rho_\Rightroad$ be given with
\begin{equation*}
	\rho_\Leftroad > \rholim, \qquad \rho_\Rightroad > \frac{\rhomax}{2}, \qquad \rho_\Rightroad < \fluxmax_\Leftroad, \qquad \ARZsupply(\rho_\Leftroad,0) < \LWRsupply(\rho_\Rightroad).
\end{equation*}

With~\eqref{eq:curveofmaxima}, the condition $\rho_\Rightroad < \fluxmax_\Leftroad$ is equivalent to
\begin{equation} \label{eq:line_upIII}
	\rho_\Rightroad < \curveofmaxima(\rho_\Leftroad).
\end{equation}

Next, we make use of $\rho_\Leftroad  > \rholim $, $\rho_\Rightroad > \frac{\rhomax}{2}$ and analyze the condition 
\begin{equation} \label{eq:maxARZbiggerLWRsupply}
	\ARZsupply(\rho_\Leftroad, 0) < \LWRsupply(\rho_\Rightroad).
\end{equation} 
Applying~\eqref{eq:ARZSupply_fluxmax} on the left-hand side and inserting $\LWRsupply(\rho_\Rightroad) = f(\rho_\Rightroad)$ on the right-hand side yields a bound on $\rho_\Rightroad$
\begin{equation} \label{eq:line_upVI}
	\rho_\Rightroad < \rhomax \left( \einhalb + \sqrt{\frac{1}{4} - \left( \eindrittel \left( \frac{\rhomax - \rho_\Leftroad}{\rhomax}\right)^2 + \eindrittel \right)^{\frac{3}{2}}} \right):= \curveofequalsupplyatmaxarz(\rho_\Leftroad).
\end{equation}

The area for admissible values of $\rho_\Leftroad$ and $\rho_\Rightroad$ is marked as III in Figure \ref{img:rho_l-rho_r-plane}. Since we have $ \ARZsupply(\rho_\Leftroad,0) < \LWRsupply(\rho_\Rightroad)$, the ARZ model is applied.

\begin{figure}
	\centering
	\tikzsetnextfilename{stateplane}
%
%
\begin{tikzpicture}
\setlength\fwidth{0.45\textwidth}
\begin{axis}[%
width=0.951\fwidth,
height=0.75\fwidth,
at={(0\fwidth,0\fwidth)},
scale only axis,
xmin=0,
xmax=600,
xtick={       0, 338.0873,      600},
xticklabels={0, $\rho_{\lim}$, $\rhomax$},
xlabel={$\rho_\Leftroad$},
ymin=0,
ymax=600,
ytick={  300, 600},
yticklabels={ $\frac{\rhomax}{2}$, $\rhomax$},
ylabel={$\rho_\Rightroad$},
axis background/.style={fill=white},
legend style={legend cell align=left, align=left, draw=white!15!black}
]
\addplot [name path=A, color=blue, solid, line width=1.0pt]
  table[row sep=crcr]{%
338.087258745563	377.976314968462\\
359.787258745563	373.140796063618\\
381.087258745563	368.746881696886\\
401.987258745563	364.787173430891\\
422.487258745563	361.25280774349\\
442.687258745563	358.11883435576\\
462.587258745563	355.378793898876\\
482.287258745563	353.013828555216\\
501.787258745563	351.020295018541\\
521.087258745563	349.393388953049\\
540.287258745563	348.121440241821\\
559.387258745563	347.202820050087\\
578.387258745563	346.634827344537\\
597.287258745563	346.413702079423\\
599.987258745563	346.410161591881\\
};

\addplot [name path=B, color=blue, dotted, line width=1.0pt]
  table[row sep=crcr]{%
338.087258745563	300\\
338.387258745563	307.033173243888\\
339.087258745563	312.83080186002\\
340.087258745563	318.125376046725\\
341.487258745563	323.595871089149\\
343.187258745563	328.844290625477\\
345.287258745563	334.191920178693\\
347.687258745563	339.375556191786\\
350.487258745563	344.610486675576\\
353.687258745563	349.856583625351\\
357.287258745563	355.086033989889\\
361.287258745563	360.278059859757\\
365.687258745563	365.416095519318\\
370.487258745563	370.486197231076\\
375.787258745563	375.567618033282\\
381.487258745563	380.541614634536\\
387.587258745563	385.40040676027\\
394.187258745563	390.2057344478\\
401.187258745563	394.86808931475\\
408.687258745563	399.439085153056\\
416.587258745563	403.843247125292\\
424.987258745563	408.122694192952\\
433.787258745563	412.212885598506\\
442.987258745563	416.106745564117\\
452.687258745563	419.832949556642\\
462.787258745563	423.339091427951\\
473.287258745563	426.616202209886\\
484.187258745563	429.654321867667\\
495.487258745563	432.442425014455\\
507.087258745563	434.948207706945\\
519.087258745563	437.184117502379\\
531.387258745563	439.120581233934\\
543.887258745563	440.737518637352\\
556.587258745563	442.032390734389\\
569.487258745563	443.000103319449\\
582.487258745563	443.629430492398\\
595.587258745563	443.9178462062\\
599.987258745563	443.93737405685\\
};

\addplot [ name path=C,color=black, line width=1.0pt]
  table[row sep=crcr]{%
370.820393249937	370.820393249937\\
599.920393249937	599.920393249937\\
};

\addplot [name path=D,color=black, line width=1.0pt]
  table[row sep=crcr]{%
338.087258745563	0\\
338.087258745563	600\\
};

\addplot [name path=E,color=black, line width=1.0pt]
  table[row sep=crcr]{%
338.087258745563	300\\
600	300\\
};

\node[] at (470,150) {I};
\node[] at (347,358) {\footnotesize II};
\node[] at (420,330) {III};
\node[] at (400,500) {IV};
\node[] at (550,480) {V};
\node[] at (500,385) {VI};
\node[] at (150,300) {VII};

\path[name path=lim] (axis cs:338.0873,0) -- (axis cs:600,0);

\addplot[gray!40] fill between[of=C and E];
\addplot[gray!40] fill between[of=B and E];
\addplot[gray!40] fill between[of=B and lim];
\end{axis}

\end{tikzpicture}%
	\caption{Partitioning of the $\rho_\Leftroad$-$\rho_\Rightroad$-plane. Areas for the LWR supply are in white and areas for the ARZ supply are in gray. The curves $\curveofmaxima$~\eqref{eq:curveofmaxima} and $\curveofequalsupplyatmaxarz$ \eqref{eq:line_upVI} for $\rho_\Leftroad \geq \rholim$ are displayed in blue (solid and dotted).}
	\label{img:rho_l-rho_r-plane}
\end{figure}


\section{Numerical schemes and computational results} \label{sec:NumericalTreatment}

Having presented a discussion of initial states, we introduce the numerical schemes to compute solutions to the first and second order models.

\subsection{Numerical scheme for the LWR and the combined model} \label{sec:discretization_firstorder}

We use the Godunov scheme to approximate the densities on the roads for the setting in Figure~\ref{img:onetoonewithOR}, see~\cite{GoaGoeKol2016}. We consider the time horizon $[0,T]$ and introduce a grid in time and space with step sizes $\dt$, $\dx$. In the numerical examples, we set $\dt \vmax \leq \dx$. We denote the number of time discretizations by $T = \Nt \dt$, $\Nt \in \mathbb{N}$ and the number of spatial discretizations $\Nxroad \in \mathbb{N}$ on road $\road \in \setofroads$ with $\lengthofroad_\road = \Nxroad \dx$. Each road is discretized with cells $\Cell{\road,j} = (x_{\road,j-\niceeinhalb}, x_{\road, j+\niceeinhalb})$, $j = 1, \dots, \Nxroad$. The discretized density values $\rho_{\road, j}^s$, $s=0, \dots, \Nt$ are the cell averages
\begin{equation*}
	\rho_{\road, j}^s \approx \frac{1}{\dx} \int_{\Cell{\road,j}} \rho_\road (x,t^s) \intx.
\end{equation*}
The Godunov scheme reads
\begin{align} \label{eq:LWR_Godunov}
	\begin{split}
		\rho_{\road, 1}^{s+1} &=  \rho_{\road, 1}^{s} - \dtbydx \left(\Godunov\left(\rho_{\road,1}^s, \rho_{\road, 2}^s \right) - \massflux_{\road, in}^s \right), \\
		\rho_{\road, j}^{s+1} &=  \rho_{\road, j}^{s} - \dtbydx \left(\Godunov\left(\rho_{\road,j}^s, \rho_{\road, j+1}^s \right) - \Godunov\left(\rho_{\road,j-1}^s, \rho_{\road, j}^s \right) \right), \qquad j=2, \dots, \Nxroad -1, \\
		\rho_{\road,\Nxroad}^{s+1} &= \rho_{\road,\Nxroad}^s - \dtbydx \left( \massflux_{\road, out}^s - \Godunov \left(\rho_{\road, \Nxroad -1}^s,\rho_{\road, \Nxroad}^s \right)\right),
	\end{split}
\end{align}
where the Godunov flux is given by the minimum of supply and demand
\begin{equation*}
	\Godunov \left(\rho_{\road,j}^s, \rho_{\road, j+1}^s \right) = \min \set{ \LWRdemand(\rho_{\road,j}^s), \LWRsupply(\rho_{\road,j+1}^s)}.
\end{equation*}

We note that the above Godunov scheme for the computation of the density values $\rho_{\road,2}^s, \dots, \rho_{\road,\Nxroad-1}^s$ is equivalent to the cell transmission method (CTM), see~\cite{Dag1994,FanSunPic2017}. Figure~\ref{img:CTM_roads} visualizes the fluxes at the cell interfaces. The inflow $\massflux_{\road, in}^s \geq 0$ and outflow $\massflux_{\road, out}^s \geq 0$ in~\eqref{eq:LWR_Godunov} are mass fluxes at road interfaces and are determined with coupling and boundary conditions described below.

\subsubsection*{1-to-1 junction with onramp}

\begin{figure}[tbh]
	\centering 
	\subfloat[Godunov scheme within the roads.]{ 
		\includegraphics[width=0.9\textwidth]{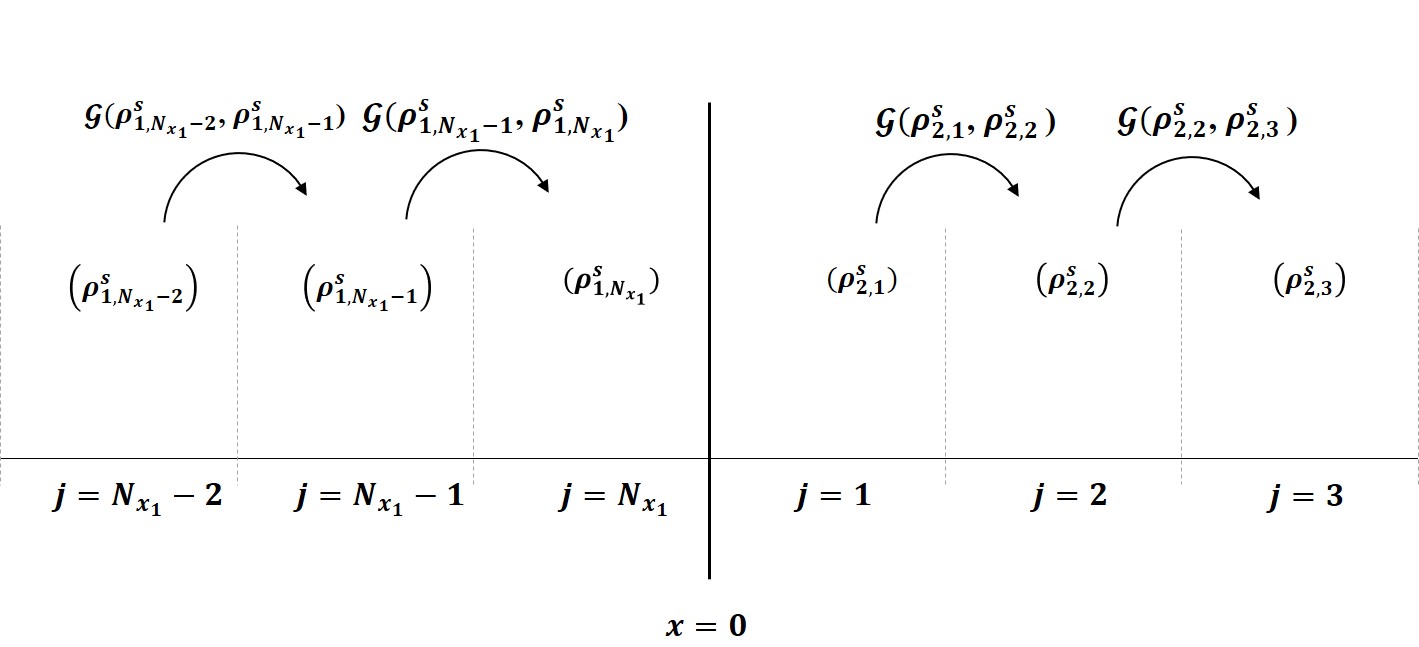}
		\label{img:CTM_roads}}
	\vspace{0.2cm}
	\subfloat[Calculation of fluxes at the junction.]{ 
		\includegraphics[width=0.9\textwidth]{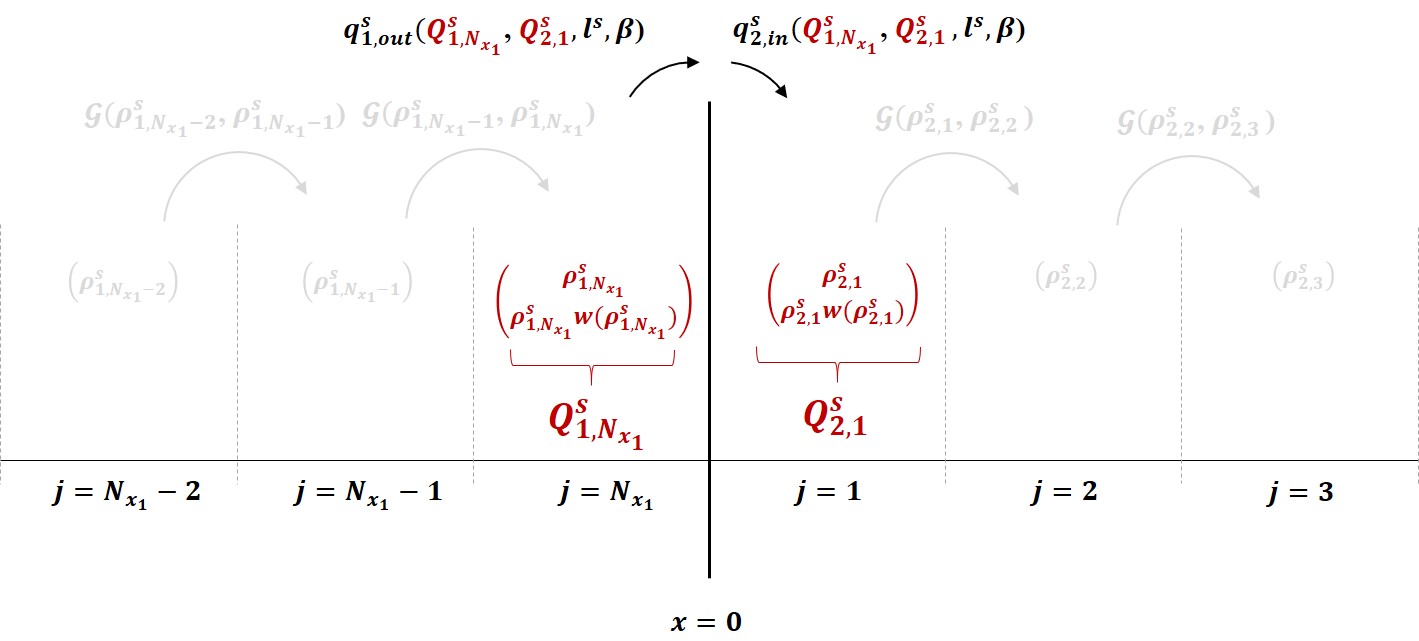}
		\label{img:CTM_junctions}}
	\caption{Illustration of the numerical scheme.}
\end{figure}

We consider a 1-to-1 junction with an additional onramp as shown in Figure~\ref{img:onetoonewithOR}. We denote the incoming road with index $\Leftroad$ and the outgoing road with index $\Rightroad$. As in~\cite{GoaGoeKol2016}, we assume that the demand of the onramp can be controlled and apply the demand function
\begin{align}\label{eq:controlledonrampdemand}
	&\demandonramp(l^s,t^s) = \control(t^s) \min \Big \lbrace \onrampinflow(t^s) + \frac{\queuelength^s}{\dt}, \fmax_\onramp \Big \rbrace,
\end{align}
for the onramp. Here, $\control(t) \in [0,1]$ denotes a time-dependent metering rate which can be used to control the flow from the onramp. Moreover, $\fmax_\onramp$ is the maximum flux allowed to enter the outgoing road from the onramp.  To define a unique solution, we assign priority parameters $\beta$ and $(1-\beta)$ to road $\road = \Leftroad$ and the onramp, respectively as shown in Figure~\ref{img:onetoonewithOR}, and compute
\begin{subequations} \label{eq:numflux_11withor}
\begin{align}
	\massflux_{\Leftroad,out}^s &= \min \bigg \{ \LWRdemand_\Leftroad(\rho_{\Leftroad,\Nxmyroad{\Leftroad}}^s), \max \Big \{ \beta \ALWRsupply,\ALWRsupply - \demandonramp(\queuelength^s,t^s) \Big \} \bigg \}, \\
	\massflux_{\onramp}^s &= \min \bigg \{ \demandonramp(\queuelength^s,t^s), \max \Big \{ (1-\beta) \ALWRsupply, \ALWRsupply - \LWRdemand( \rho_{\Leftroad, \Nxmyroad{\Leftroad}}^s) \Big \} \bigg \}, \\
	\massflux_{\Rightroad, in}^s &= \massflux_{\Leftroad,out}^s + \massflux_\onramp^s, 
\end{align}
\end{subequations}
where
$\ALWRsupply =  \ALWRsupply(\rho_{\Leftroad, \Nxmyroad{\Leftroad}}^s, \rho_{\Rightroad,1}^s, \queuelength^s)$ using~\eqref{eq:ARZSupplyexplicit} and~\eqref{eq:explicittilderho} to evaluate the ALWR supply~\eqref{eq:ALWRsupply}. Similarly to the computation of the mass flux in a 2CTM for the second order model , which will be studied in section~\ref{sec:discretization_secondorder}, the computation of the mass fluxes $q_{1,out}^s$, $q_{\onramp}^s$ and $q_{2,in}^s$ considers the $w$-value of the incoming road $e=1$ in the supply function. This value of the mass flux is then used as input for the boundary cells, see Figure~\ref{img:CTM_junctions}. Especially important is that the $w$-value of the incoming road is derived from the density. Therefore, not the whole second order dynamics have to be tracked. We compute first order with the CTM approach and derive the second order property $w$ from the density. 

Based on the computed fluxes, the evolution of the queue at the onramp reads
\begin{align*}
	\queuelength^{s+1}=  \queuelength^{s} + \dt \left( \onrampinflow(t^s) - \massflux_\onramp^s  \right).
\end{align*}
In general, we start with empty queues, i.e., $\queuelength^0 = 0$.

\subsubsection*{Onramp at origin} 

Additionally, we have to provide boundary data for the inflow at the beginning of the incoming road $\road = \Leftroad$. We consider an onramp at a junction with only one outgoing road, where we denote the outgoing road with the index $\Leftroad$. The demand function of the onramp is given by
\begin{align} \label{eq:onrampdemand}
	\demandonramp(l^s,t^s) =  \begin{cases}
		\fmax_\onramp & \text{if } l(t^s)> 0,\\
		\minimum{ \onrampinflow(t^s) + \frac{\queuelength^s}{\dt}, \fmax_\onramp} & \text{otherwise,} 
	\end{cases}
\end{align}
where $\onrampinflow(t^s)$ is the flow of cars arriving at the onramp at time $t^s$ and $\fmax_\onramp$ the maximum flux allowed to exit the onramp. The inflow into the system is given by 
\begin{align*}
	\massflux_{1, in}^s = \min \{ \LWRdemand(\rho_{1,1}^s), \demandonramp(l^s,t^s) \}.  
\end{align*}
Based on the computed inflow, the evolution of the queue at the onramp is given by
\begin{align*}
	\queuelength^{s+1}= \queuelength^s + \dt \left( \onrampinflow(t^s) - \massflux_{1, in}^s  \right).
\end{align*}

\subsubsection*{Junction with outflow} At a junction with only one incoming road (denoted by index 1), we consider absorbing boundary conditions 
\begin{align*}
	\massflux_{1, out}^s = \min \{ \LWRdemand(\rho_{1, \Nxmyroad{\Leftroad}}^s), \LWRsupply(\rho_{1, \Nxmyroad{\Leftroad}}^s) \}.  
\end{align*}

\begin{Remark}  \label{rem:LWRvsALWRnumerics}
	In comparison to the LWR onramp model in~\cite{GoaGoeKol2016}, only the supply function $\ALWRsupply$ on the outgoing road of the 1-to-1 junction with onramp has been adapted in the augmented model. Therefore, replacing $\ALWRsupply$ by $\LWRsupply(\rho_{\Rightroad,1}^s)$ in~\eqref{eq:numflux_11withor} gives the numerical discretization of the LWR model.
\end{Remark}

\subsection{Capacity drop in comparison to the LWR model} \label{sec:RP_ALWRvsLWR}

In this section, we consider the evolution of the left- and right-hand state $\rho_\Leftroad$ and $\rho_\Rightroad$ for the Riemann problem~\eqref{eq:LWR_RP_11withOR} at the 1-to-1 junction with onramp, cf. Figure~\ref{img:onetoonewithOR}, within the ALWR model. To ensure the activation condition for the minimum in the supply function~\eqref{eq:ALWRsupply}, we set $\onrampinflow(t) = \fmax_\onramp = \fmax$. For the particular setting, solutions of Riemann problems are discussed and phase space trajectories are compared to the LWR model.

We obtain the numerical solution of the ALWR model with the scheme described in section~\ref{sec:discretization_firstorder}. The only difference to the LWR network approach presented in~\cite{GoaGoeKol2016} is the modelling of the supply at the junction of the network, where we impose second-order-like conditions in the supply function, cf. Section \ref{sec:capdropandALWRmodel}. The numerical solution of the LWR model is obtained as explained in remark~\ref{rem:LWRvsALWRnumerics}. In comparison to the LWR model and as extension of the analysis in Section~\ref{sec:augmentedsupply}, the most exciting point in the ALWR model here is that the evolution of the states at the junction may lead to a change in the ``active'' model. 

\begin{figure}[tbh]
	\centering 
	\subfloat[$\rho_\Leftroad \leq \nicefrac{\rhomax}{2}$ and $\rho_\Rightroad \geq \nicefrac{\rhomax}{2}$.]{ 
		\tikzsetnextfilename{trajectories1}
%
%
\begin{tikzpicture}
\setlength\fwidth{0.35\textwidth}
\begin{axis}[%
width=0.951\fwidth,
height=0.75\fwidth,
at={(0\fwidth,0\fwidth)},
scale only axis,
xmin=0,
xmax=600,
ticks=none,
xlabel={$\rho_\Leftroad$},
ymin=0,
ymax=600,
ylabel={$\rho_\Rightroad$},
axis background/.style={fill=white},
axis x line*=bottom,
axis y line*=left,
legend style={at={(0.03,0.75)}, font=\small, anchor=south west, align=left, draw=white!15!black}
]

\addplot [color=black, line width=1.0pt]
table[row sep=crcr]{%
	150	350\\
	150	350\\
};
\addlegendentry{LWR model}

\addplot [color=black, dotted, line width=1.0pt]
table[row sep=crcr]{%
	150	350\\
	150	350\\
};
\addlegendentry{ALWR model}

\addplot [name path=A1, color=black, dashdotted, line width=1.0pt]
table[row sep=crcr]{%
	338.087258745563	300\\
	338.387258745563	307.033173243888\\
	339.087258745563	312.83080186002\\
	340.087258745563	318.125376046725\\
	341.487258745563	323.595871089149\\
	343.187258745563	328.844290625477\\
	345.287258745563	334.191920178693\\
	347.687258745563	339.375556191786\\
	350.487258745563	344.610486675576\\
	353.687258745563	349.856583625351\\
	357.287258745563	355.086033989889\\
	361.287258745563	360.278059859757\\
	365.687258745563	365.416095519318\\
	370.487258745563	370.486197231076\\
	370.787258745563	370.787248289118\\
};

\addplot [name path=A2,color=black, dashdotted, line width=1.0pt]
table[row sep=crcr]{%
	370.820393249937	370.820393249937\\
	599.920393249937	599.920393249937\\
};

\addplot [name path=A3, color=black, dashdotted, line width=1.0pt]
table[row sep=crcr]{%
	338.087258745563	0\\
	338.087258745563	300\\
};

\addplot [color=black, line width=1.0pt]
table[row sep=crcr]{%
	150	350\\
	456.12494995996	350\\
	456.12494995996	350\\
};

\addplot [color=black, line width=1.0pt]
table[row sep=crcr]{%
	150	350\\
	515.058131676066	350\\
	515.058131676066	350\\
};

\addplot [color=black, line width=1.0pt]
table[row sep=crcr]{%
	150	350\\
	585.043856274785	350\\
	585.043856274785	350\\
};

\addplot [color=black, dotted, line width=1.0pt]
table[row sep=crcr]{%
	150	350\\
	356.25	350\\
	359.846619193743	349.371174408343\\
	364.120398954611	347.839468060519\\
	369.183958931772	345.254721424305\\
	375.162469745617	341.450040339177\\
	382.192455690068	336.243392413242\\
	390.419285537639	329.440952616481\\
	399.992954824527	320.842726900631\\
	411.061779672749	310.250960436335\\
	423.763721987615	297.481704016513\\
	438.215282809912	282.379622920118\\
	454.498258009881	264.835655986825\\
	470.362290188079	244.806475909032\\
	479.253012741302	223.533143411531\\
	483.783776694208	205.833774897785\\
	485.968045292423	192.641376514641\\
	486.991039428499	183.633588016803\\
	487.779237793178	172.253906627687\\
	487.862471881552	169.399804280662\\
	487.862471881552	169.399804280662\\
};

\addplot [color=black, dotted, line width=1.0pt]
table[row sep=crcr]{%
	150	350\\
	387.5	350\\
	431.068229142062	342.030208382543\\
	478.833732300144	325.665868733046\\
	514.228669317263	302.437648670809\\
	528.225008265539	275.631894916106\\
	532.435173337927	247.756050111979\\
	533.554986325542	219.978441954477\\
	533.913917438338	183.088435095237\\
	533.938388930222	160.521185728673\\
	533.938388930222	160.521185728673\\
};

\addplot [color=black, dotted, line width=1.0pt]
table[row sep=crcr]{%
	150	350\\
	443.75	350\\
	543.508677251531	331.579894151361\\
	582.980332244028	302.755412613223\\
	587.956273284932	272.531528255175\\
	588.204262871485	242.238043653875\\
	588.214827803673	156.201983118657\\
	588.214827803673	156.201983118657\\
};

\addplot [color=black, draw=none, mark=diamond*, mark options={solid, black},mark size=3pt]
table[row sep=crcr]{%
	150	350\\
};
\node at (axis cs:150,390){$\beta_1$};

\addplot [color=black, draw=none, mark=diamond*, mark options={solid, black},mark size=3pt]
table[row sep=crcr]{%
	456.12494995996	350\\
};
\node at (axis cs:456,390){ $\beta_2$};

\addplot [color=black, draw=none, mark=diamond*, mark options={solid, black},mark size=3pt]
table[row sep=crcr]{%
	515.058131676066	350\\
};
\node at (axis cs:516,390){$\beta_3$};

\addplot [color=black, draw=none, mark=diamond*, mark options={solid, black},mark size=3pt]
table[row sep=crcr]{%
	585.043856274785	350\\
};
\node at (axis cs:585,390){$\beta_4$};

\addplot [color=black, draw=none, mark=diamond*, mark options={solid, black},mark size=3pt]
table[row sep=crcr]{%
	487.862471881552	169.399804280662\\
};
\node at (axis cs:487,130){ $\beta_2$};

\addplot [color=black, draw=none, mark=diamond*, mark options={solid, black},mark size=3pt]
table[row sep=crcr]{%
	533.938388930222	160.521185728673\\
};
\node at (axis cs:533,120){ $\beta_3$};

\addplot [color=black, draw=none, mark=diamond*, mark options={solid, black},mark size=3pt]
table[row sep=crcr]{%
	588.214827803673	156.201983118657\\
};
\node at (axis cs:585,116){ $\beta_4$};

\path[name path=lim] (axis cs:338.0873,0) -- (axis cs:600,0);
\addplot[gray!40] fill between[of=A1 and lim];
\addplot[gray!40] fill between[of=A2 and lim];
\addplot[gray!40] fill between[of=A3 and lim];
\end{axis}
\end{tikzpicture}%
		\label{img:trajectories_ex1}
	}
	\subfloat[$\rho_\Leftroad \leq \nicefrac{\rhomax}{2}$ and $\rho_\Rightroad \leq \nicefrac{\rhomax}{2}$.]{ 
		\tikzsetnextfilename{trajectories2}
%
%
\begin{tikzpicture}
\setlength\fwidth{0.35\textwidth}
\begin{axis}[%
width=0.951\fwidth,
height=0.75\fwidth,
at={(0\fwidth,0\fwidth)},
scale only axis,
xmin=0,
xmax=600,
xlabel={$\rho_\Leftroad$},
ymin=0,
ymax=600,
ticks=none,
ylabel={$\rho_\Rightroad$},
axis background/.style={fill=white},
axis x line*=bottom,
axis y line*=left,
legend style={at={(0.03,0.75)}, font=\small, anchor=south west, align=left, draw=white!15!black}
]

\addplot [color=black, line width=1.0pt]
table[row sep=crcr]{%
	200	100\\
	200	299.940067966175\\
};
\addlegendentry{LWR model}

\addplot [color=black, dotted, line width=1.0pt]
table[row sep=crcr]{%
	200	100\\
	200	299.940067966175\\
};
\addlegendentry{ALWR model}

\addplot [name path=A1, color=black, dashdotted, line width=1.0pt]
table[row sep=crcr]{%
	338.087258745563	300\\
	338.387258745563	307.033173243888\\
	339.087258745563	312.83080186002\\
	340.087258745563	318.125376046725\\
	341.487258745563	323.595871089149\\
	343.187258745563	328.844290625477\\
	345.287258745563	334.191920178693\\
	347.687258745563	339.375556191786\\
	350.487258745563	344.610486675576\\
	353.687258745563	349.856583625351\\
	357.287258745563	355.086033989889\\
	361.287258745563	360.278059859757\\
	365.687258745563	365.416095519318\\
	370.487258745563	370.486197231076\\
	370.787258745563	370.787248289118\\
};

\addplot [name path=A2, color=black, dashdotted, line width=1.0pt]
table[row sep=crcr]{%
	370.820393249937	370.820393249937\\
	599.920393249937	599.920393249937\\
};

\addplot [name path=A3, color=black, dashdotted, line width=1.0pt]
table[row sep=crcr]{%
	338.087258745563	0\\
	338.087258745563	300\\
};

\addplot [color=black, line width=1.0pt]
table[row sep=crcr]{%
	200	100\\
	228.333333333333	166.666666666667\\
	256.666666666667	196.296296296296\\
	285	214.220393232739\\
	313.333333333333	226.483961461315\\
	341.666666666667	235.491641332018\\
	370	242.427188562079\\
	398.333333333333	247.95156959019\\
	426.666666666667	252.466634770398\\
	444.925925925926	256.232336120483\\
	454.920052583448	259.4250167896\\
	459.919681429351	262.168898893807\\
	462.295840581909	264.554219245319\\
	463.395940798262	266.6482248675\\
	463.898885016013	268.502126374982\\
	464.231096049232	271.640128181649\\
	464.316435419196	278.91748846025\\
	464.31676725155	299.940067966175\\
};

\addplot [color=black, line width=1.0pt]
table[row sep=crcr]{%
	200	100\\
	258.333333333333	166.666666666667\\
	316.666666666667	196.296296296296\\
	375	214.220393232739\\
	433.333333333333	226.483961461316\\
	478.703703703704	235.491641332018\\
	500.478680841335	242.427188562079\\
	508.492511721532	247.95156959019\\
	511.04396598161	252.466634770398\\
	511.811373352865	256.232336120483\\
	512.104461041857	262.168898893807\\
	512.132034355964	286.3324037626\\
	512.132034355964	299.940067966176\\
};

\addplot [color=black, line width=1.0pt]
table[row sep=crcr]{%
	200	100\\
	318.333333333333	166.666666666667\\
	436.666666666667	196.296296296296\\
	540.537037037037	214.220393232739\\
	579.106926726109	226.483961461316\\
	584.272465815286	235.491641332018\\
	584.587741114114	242.427188562079\\
	584.604989415154	299.940067966176\\
};

\addplot [color=black, dotted, line width=1.0pt]
table[row sep=crcr]{%
	200	100\\
	228.333333333333	166.666666666667\\
	256.666666666667	196.296296296296\\
	285	214.220393232739\\
	313.333333333333	226.483961461315\\
	341.666666666667	235.491641332018\\
	370.683561836166	241.45067165327\\
	404.84572594915	238.837143622572\\
	442.535539248977	229.342172837981\\
	469.353020310492	215.265747579827\\
	484.907125930433	201.011025229811\\
	492.591675885005	189.243874743242\\
	496.027051863861	180.753551623753\\
	497.485674936493	175.163127598106\\
	498.338921557625	169.637408426283\\
	498.510414387076	166.870714716169\\
	498.510414387076	166.870714716169\\
};

\addplot [color=black, dotted, line width=1.0pt]
table[row sep=crcr]{%
	200	100\\
	258.333333333333	166.666666666667\\
	316.666666666667	196.296296296296\\
	375	214.220393232739\\
	438.001214397965	217.148199332052\\
	492.098702452549	206.912807428377\\
	520.037169408676	192.471544271056\\
	530.063235677919	180.302173313715\\
	532.932439110182	172.012741269576\\
	533.682772706233	166.954900773627\\
	533.938388930222	160.521185728673\\
	533.938388930222	160.521185728673\\
};

\addplot [color=black, dotted, line width=1.0pt]
table[row sep=crcr]{%
	200	100\\
	318.333333333333	166.666666666667\\
	436.666666666667	196.296296296296\\
	542.68300321718	192.760731431313\\
	582.819176600034	178.982097966191\\
	587.946821996541	169.003087476871\\
	588.214384797981	159.896603741362\\
	588.214827803673	156.201983118657\\
	588.214827803673	156.201983118657\\
};

\addplot [color=black, draw=none, mark=diamond*, mark options={solid, black},mark size=3pt]
table[row sep=crcr]{%
	200	100\\
};

\addplot [color=black, draw=none, mark=diamond*, mark options={solid, black},mark size=3pt]
table[row sep=crcr]{%
	464.31676725155	299.940067966175\\
};
\node at (axis cs:464,340){$\beta_2$};

\addplot [color=black, draw=none, mark=diamond*, mark options={solid, black},mark size=3pt]
table[row sep=crcr]{%
	512.132034355964	299.940067966176\\
};
\node at (axis cs:512,340){$\beta_3$};

\addplot [color=black, draw=none, mark=diamond*, mark options={solid, black},mark size=3pt]
table[row sep=crcr]{%
	584.604989415154	299.940067966176\\
};
\node at (axis cs:584,340){$\beta_4$};

\addplot [color=black, draw=none, mark=diamond*, mark options={solid, black},mark size=3pt]
table[row sep=crcr]{%
	498.510414387076	166.870714716169\\
};
\node at (axis cs:490,126){$\beta_2$};

\addplot [color=black, draw=none, mark=diamond*, mark options={solid, black},mark size=3pt]
table[row sep=crcr]{%
	533.938388930222	160.521185728673\\
};
\node at (axis cs:533,120){$\beta_3$};

\addplot [color=black, draw=none, mark=diamond*, mark options={solid, black},mark size=3pt]
table[row sep=crcr]{%
	588.214827803673	156.201983118657\\
};
\node at (axis cs:585,116){$\beta_4$};

\addplot [color=black, draw=none, mark=diamond*, mark options={solid, black},mark size=3pt]
table[row sep=crcr]{%
	200	300\\
};
\node at (axis cs:200,340){$\beta_1$};

\path[name path=lim] (axis cs:338.0873,0) -- (axis cs:600,0);
\addplot[gray!40] fill between[of=A1 and lim];
\addplot[gray!40] fill between[of=A2 and lim];
\addplot[gray!40] fill between[of=A3 and lim];

\end{axis}
\end{tikzpicture}%
		\label{img:trajectories_ex2}
	}
	\hfill
	\subfloat[$\rho_\Leftroad \geq \nicefrac{\rhomax}{2}$ and $\rho_\Rightroad \leq \nicefrac{\rhomax}{2}$.]{ 
		\tikzsetnextfilename{trajectories3}
%
%
\begin{tikzpicture}
\setlength\fwidth{0.35\textwidth}
\begin{axis}[%
width=0.951\fwidth,
height=0.75\fwidth,
at={(0\fwidth,0\fwidth)},
scale only axis,
xmin=0,
xmax=600,
ticks=none,
xlabel={$\rho_\Leftroad$},
ymin=0,
ymax=600,
ytick={ },
ylabel={$\rho_\Rightroad$},
axis background/.style={fill=white},
axis x line*=bottom,
axis y line*=left,
legend style={at={(0.03,0.75)}, font=\small, anchor=south west, align=left, draw=white!15!black}
]

\addplot [color=black, line width=1.0pt]
table[row sep=crcr]{%
	550	100\\
	460.833333333333	166.666666666667\\
	432.721064814815	196.296296296296\\
	418.362929738851	214.220393232739\\
	410.013291178244	226.483961461316\\
	404.841750785129	235.491641332018\\
	401.52209627231	242.427188562079\\
	399.344202886437	247.95156959019\\
	397.895418474534	252.466634770398\\
	396.266150312195	259.4250167896\\
	395.518134140819	264.554219245319\\
	395.075361984373	270.155653113144\\
	394.882776674421	278.119566151233\\
	394.868329805052	298.819732472478\\
};
\addlegendentry{LWR model}

\addplot [color=black, dotted, line width=1.0pt]
table[row sep=crcr]{%
	550	100\\
	490.825875898644	133.341619371877\\
	471.008409709176	150.885099096739\\
	461.058123230475	161.511069467498\\
	455.431883286887	168.357903540947\\
	449.982541056596	175.988205631197\\
	447.303087978376	180.498336610852\\
	446.32498499776	182.569163454517\\
	446.32498499776	182.569163454517\\
};
\addlegendentry{ALWR model}

\addplot [name path=A1,color=black, dashdotted, line width=1.0pt]
table[row sep=crcr]{%
	338.087258745563	300\\
	338.387258745563	307.033173243888\\
	339.087258745563	312.83080186002\\
	340.087258745563	318.125376046725\\
	341.487258745563	323.595871089149\\
	343.187258745563	328.844290625477\\
	345.287258745563	334.191920178693\\
	347.687258745563	339.375556191786\\
	350.487258745563	344.610486675576\\
	353.687258745563	349.856583625351\\
	357.287258745563	355.086033989889\\
	361.287258745563	360.278059859757\\
	365.687258745563	365.416095519318\\
	370.487258745563	370.486197231076\\
	370.787258745563	370.787248289118\\
};

\addplot [name path=A2,color=black, dashdotted, line width=1.0pt]
table[row sep=crcr]{%
	370.820393249937	370.820393249937\\
	599.920393249937	599.920393249937\\
};

\addplot [name path=A3,color=black, dashdotted, line width=1.0pt]
table[row sep=crcr]{%
	338.087258745563	0\\
	338.087258745563	300\\
};

\addplot [color=black, line width=1.0pt]
table[row sep=crcr]{%
	550	100\\
	483.333333333333	166.666666666667\\
	464.814814814815	196.296296296296\\
	457.041609510745	214.220393232739\\
	453.438164314536	226.483961461316\\
	451.699380534179	235.491641332018\\
	450.844877110089	242.427188562079\\
	450.210328678744	252.466634770398\\
	450.013128276017	264.554219245319\\
	450	298.819732472478\\
};

\addplot [color=black, line width=1.0pt]
table[row sep=crcr]{%
	550	100\\
	520.833333333333	166.666666666667\\
	514.554398148148	196.296296296296\\
	512.831748540291	214.220393232739\\
	512.336159895762	226.483961461316\\
	512.137155060681	247.95156959019\\
	512.132034355964	298.819732472478\\
};

\addplot [color=black, line width=1.0pt]
table[row sep=crcr]{%
	550	100\\
	580.833333333333	166.666666666667\\
	584.387731481481	196.296296296296\\
	584.604413040165	226.483961461316\\
	584.604989415154	298.819732472478\\
};

\addplot [color=black, dotted, line width=1.0pt]
table[row sep=crcr]{%
	550	100\\
	508.327118804425	133.341619371877\\
	496.34079906671	149.170208901804\\
	491.582601869578	157.764494128846\\
	489.530990716099	162.654000130597\\
	488.205836265196	167.132035531982\\
	487.862471881552	169.399804280662\\
	487.862471881552	169.399804280662\\
};

\addplot [color=black, dotted, line width=1.0pt]
table[row sep=crcr]{%
	550	100\\
	537.495856980728	133.341619371877\\
	534.811325790859	146.988101197561\\
	533.993444977608	156.847813742719\\
	533.938388930222	160.521185728673\\
	533.938388930222	160.521185728673\\
};

\addplot [color=black, dotted, line width=1.0pt]
table[row sep=crcr]{%
	550	100\\
	584.165838062812	133.341619371877\\
	588.023061904543	145.22401531154\\
	588.21482780283	155.998326282247\\
	588.214827803673	156.201983118657\\
	588.214827803673	156.201983118657\\
};

\addplot [color=black, draw=none, mark=diamond*, mark options={solid, black},mark size=3pt]
table[row sep=crcr]{%
	550	100\\
};

\addplot [color=black, draw=none, mark=diamond*, mark options={solid, black},mark size=3pt]
table[row sep=crcr]{%
	394.868329805051	298.819732472478\\
};
\node at (axis cs:394,340){$\beta_1$};

\addplot [color=black, draw=none, mark=diamond*, mark options={solid, black},mark size=3pt]
table[row sep=crcr]{%
	450	298.819732472478\\
};
\node at (axis cs:450,340){$\beta_2$};

\addplot [color=black, draw=none, mark=diamond*, mark options={solid, black},mark size=3pt]
table[row sep=crcr]{%
	512.132034355964	298.819732472478\\
};
\node at (axis cs:512,340){$\beta_3$};

\addplot [color=black, draw=none, mark=diamond*, mark options={solid, black},mark size=3pt]
table[row sep=crcr]{%
	584.604989415154	298.819732472478\\
};
\node at (axis cs:584,340){$\beta_4$};

\addplot [color=black, draw=none, mark=diamond*, mark options={solid, black},mark size=3pt]
table[row sep=crcr]{%
	446.32498499776	182.569163454517\\
};
\node at (axis cs:400,200){$\beta_1$};

\addplot [color=black, draw=none, mark=diamond*, mark options={solid, black},mark size=3pt]
table[row sep=crcr]{%
	487.862471881552	169.399804280662\\
};
\node at (axis cs:487,200){$\beta_2$};

\addplot [color=black, draw=none, mark=diamond*, mark options={solid, black},mark size=3pt]
table[row sep=crcr]{%
	533.938388930222	160.521185728673\\
};
\node at (axis cs:533,200){$\beta_3$};

\addplot [color=black, draw=none, mark=diamond*, mark options={solid, black},mark size=3pt]
table[row sep=crcr]{%
	588.214827803673	156.201983118657\\
};
\node at (axis cs:585,116){$\beta_4$};

\path[name path=lim] (axis cs:338.0873,0) -- (axis cs:600,0);
\addplot[gray!40] fill between[of=A1 and lim];
\addplot[gray!40] fill between[of=A2 and lim];
\addplot[gray!40] fill between[of=A3 and lim];

\end{axis}
\end{tikzpicture}%
		\label{img:trajectories_ex3}
	}
	\caption{Comparison of the LWR and ALWR model with priority parameters $\beta_1= 0.9$, $\beta_2 = 0.75$, $\beta_3 = 0.5$ and $\beta_4 = 0.1$.}
	\label{img:LWR_NumTreatment}
\end{figure}

Due to the construction of the augmented supply function within the ALWR model as minimum of the supply functions of the LWR and the ARZ model, the solution to the Riemann problem in the ALWR model is restricted by the solution in the LWR model in the sense that the (final) supply of the outgoing road is at most as large as for the same Riemann problem in the LWR model. This means that for $\rho_\Rightroad \in [0,\nicefrac{\rhomax}{2}]$, the arising densities on the outgoing road can never exceed $\nicefrac{\rhomax}{2}$. Analogously, for  $\rho_\Rightroad \in [\nicefrac{\rhomax}{2}, \rhomax]$, the density on the outgoing road can never exceed $\rho_\Rightroad$. We obtain weak entropy solutions on the incoming and outgoing road, see~\eqref{eq:LWR_weakentropysolution}. When the current densities at the left-hand and right-hand side of the junction are in a region where the LWR supply is applied, the solution is identical to the LWR model. But as soon as the combination of the density values is in the region where the ARZ supply is applied, the supply drops below the LWR supply and the models differ.

Figure~\ref{img:trajectories_ex1} shows the phase space trajectory for a Riemann problem with $\rho_\Leftroad\in [0, \rholim]$ and $\rho_\Rightroad \in [\nicefrac{\rhomax}{2}, \rhomax]$ in the LWR and the ALWR model. Different priority parameters $\beta$ are applied at the onramp. For the choice of the priority parameter $\beta= \beta_1$, there is no restriction on the incoming road and the density at the outgoing road remains constant. Decreasing priority parameters of the first road at some point lead to congestion at the end of the first road (as it is the case for $\beta \in \{\beta_2, \beta_3, \beta_4\}$). When the phase path enters the region where the ARZ supply is applied, the phase paths of the two models diverge. In the LWR model, only the density at the incoming road increases further. In contrast, in the ALWR model, there is a drop in the density at the outgoing road due to the congestion at the end of the incoming road. The markers at the end of each trajectory indicate the final states that are reached in the respective models for the respective priority parameters.

Figure~\ref{img:trajectories_ex2} shows the phase space trajectory for a Riemann problem with $\rho_\Rightroad \in [0, \nicefrac{\rhomax}{2}]$. The density on the outgoing road increases up to $\nicefrac{\rhomax}{2}$ in the LWR model (since the accumulated demand is above $\fmax$). In the ALWR model we see a different behavior. The density at the outgoing road increases at the beginning, but starts to drop down within the region where the ARZ supply is applied. Figure~\ref{img:trajectories_ex3} examplarily shows some trajectories with initial data in the region where the ARZ supply is applied. The fluxes in the final situation for the examples considered before are given in Table~\ref{tab:fluxvalues_ex}. A capacity drop in the solution for the ALWR model can be seen in all of the cases where the model differs from the LWR model. Note that here, the final fluxes $f_2(0,T)$ at the junction match for $\beta_2, \beta_3, \beta_4$ since the examples share the same priority parameters as well as the desired inflow from the onramp.

%
%
%

\begin{table}[tbhp]
	\centering
	\caption{Outflow $f_2(0,T)$ in the final state at the junction with onramp (flux values in relation to $f^\max$).}

\begin{tabular}{|c|c|c|c|c|c|c|}
	\toprule
	~& \multicolumn{2}{|c|}{Figure~\ref{img:trajectories_ex1}} & \multicolumn{2}{|c|}{Figure~\ref{img:trajectories_ex2}} & \multicolumn{2}{|c|}{Figure~\ref{img:trajectories_ex3}} \\
	Priority Parameter & LWR& ALWR & LWR& ALWR & LWR& ALWR\\
	\hline
	\hline
	$\beta_1 = 0.90$& 0.97 & 0.97 & 1 & 1 & 1&0.84\\
	$\beta_2= 0.75$& 0.97 & 0.81 & 1 & 0.81 & 1& 0.81\\
	$\beta_3 = 0.50$& 0.97 & 0.78 & 1 & 0.78 & 1&0.78\\
	$\beta_4 = 0.10$& 0.97 & 0.77 & 1 & 0.77 & 1& 0.77\\
	\bottomrule
\end{tabular}%
\label{tab:fluxvalues_ex}%
\end{table}%

\subsection{Comparison to second order models} \label{sec:comparison_ALWR_ARZ}

\subsubsection{Numerical scheme for the ARZ and the Greenberg model} \label{sec:discretization_secondorder}

Within the ARZ and the Greenberg model, the traffic state $\staterhow_\road = (\rho_\road, \rho_\road w_\road)$, consisting of the density $\rho_\road$ and the generalized momentum $\rho_\road w_\road$, is computed on each road. Let $\constrafficstate_{\road,j}^s$ denote the average value of the function $\staterhow_\road(x,t)$ on the interval $\Cell{e,j}$ at time $t^s = s \dt$, i.e.,
\begin{equation*}
	\constrafficstate_{\road,j}^s =  \frac{1}{\Delta x } \int_{\Cell{\road,j}} \staterhow_{\road,j}(x,t^s) \intx,
\end{equation*}
As in~\cite{KolGoeGoa2017}, we apply a splitting approach consisting of a Godunov scheme for the flux term and an implicit Euler scheme for the relaxation term
\begin{align} \label{eq:ARZ_Godunov}
	\begin{split}
		\constrafficstate_{\road, 1}^{s+1} &=  \constrafficstate_{\road, 1}^{s} - \dtbydx \left(\Godunov\left(\constrafficstate_{\road,1}^s, \constrafficstate_{\road, 2}^s \right) - \colvector{\massflux_{\road, in}^s \\ \momflux_{\road, in}^s} \right) + \dt \RHSARZ(\constrafficstate_{\road, 1}^{s+1}), \\
		\constrafficstate_{\road, j}^{s+1} &=  \constrafficstate_{\road, j}^{s} - \dtbydx \left(\Godunov\left(\constrafficstate_{\road,j}^s, \constrafficstate_{\road, j+1}^s \right) - \Godunov\left(\constrafficstate_{\road,j-1}^s, \constrafficstate_{\road, j}^s \right) \right)  \\
		&+ \dt \RHSARZ(\constrafficstate_{\road, j}^{s+1}), \qquad j=2, \dots, \Nxroad -1, \\
		\constrafficstate_{\road,\Nxroad}^{s+1} &= \constrafficstate_{\road,\Nxroad}^s - \dtbydx \left( \colvector{\massflux_{\road, out}^s \\ \momflux_{\road,out}^s} - \Godunov \left(\constrafficstate_{\road, \Nxroad -1}^s,\constrafficstate_{\road, \Nxroad}^s \right)\right) + \dt \RHSARZ(\constrafficstate_{\road,\Nxroad}^{s+1}),
	\end{split}
\end{align}
where there is no relaxation term for the ARZ model and for the Greenberg model, we set $\RHSARZ(\constrafficstate_{\road,j}^s) = (0,  \nicefrac{- \rho_{\road,j}^s(v_{\road,j}^s - \LWRvelocity(\rho_{\road,j}^s))}{\relaxationtime})^T$ equal to the right-hand side of the relaxed system~\eqref{eq:Greenberg_network}.  The vectors $(\massflux_{\road, in}^s, \momflux_{\road, in}^s)^T$ and $(\massflux_{\road, out}^s, \momflux_{\road, out}^s)^T$ in~\eqref{eq:ARZ_Godunov} denote mass and momentum flux at the road interfaces. The Godunov flux for the second order system  formulated with the ARZ demand and supply functions~\eqref{eq:ARZDemand}-\eqref{eq:ARZSupplyimplicit} is given by 
\begin{align*}
	\Godunov\left(\constrafficstate_{\road,j}^s, \constrafficstate_{\road, j+1}^s \right) &= \colvector{1 \\ w_{\road, j}^s} \min \set{ \ARZdemand \left( \rho_{\road,j}^s, w_{\road,j}^s \right), \ARZsupply \left(\tilde{\rho}_{\road,j+1}^s,  w_{\road,j}^s \right)},
\end{align*}
where the intermediate state $\tilde{\constrafficstate}_{\road,j+1}^s = (\tilde{\rho}_{\road,j+1}^s, \tilde{\rho}_{\road,j+1}^s w_{\road,j}^s)^T$ used to evaluate the supply is such that 
$$
\tilde{\rho}_{\road,j+1}^s  = \pressure^{-1} \left(\max\{w_{\road,j}^s-v_{\road,j+1}^s,0\} \right).$$

In the following, we explain how the flux terms for the inflow and the outflow are given by coupling and boundary conditions.
\subsubsection*{Onramp at origin }

We model a junction with an onramp at the beginning of the network similar as in the first order model. For the onramp with queue length $\queuelength^s$ at time $t^s$, we use the demand function~\eqref{eq:onrampdemand} to determine the mass flow $\massflux_{1, in}^s$. However, due to the second order system, we have to provide also boundary data for the momentum flow. We follow the idea introduced by~\cite{KolGoeGoa2017} to compute the $w$-value at the boundary. We compute
\begin{equation*}
	\rho_{-}^s = \frac{\rhomax}{2} - \sqrt{\left(\frac{\rhomax}{2}\right)^2 - \frac{\rhomax \demandonramp(l^s,t^s)}{\vmax}}.
\end{equation*}
Let $w_{-}^s = w(\rho_{-}^s) = \LWRvelocity(\rho_{-}^s) + \pressure(\rho_{-}^s)$, then we can model the boundary fluxes by
\begin{equation*}
	\colvector{\massflux_{1, in}^s \\ \momflux_{1, in}^s} = \colvector{1 \\ w_{-}^s} \min \Big \lbrace \demandonramp(l^s,t^s), \ARZsupply(\tilde{\rho}_{1,0}^s, w_{-}^s) \Big \rbrace,
\end{equation*}
with $\tilde{\rho}_{1,0}^s  = \pressure^{-1} \left(\max\{w_{-}^s-v_{1,1}^s,0\} \right)$.

\subsubsection*{1-to-1 junction with onramp}

At the 1-to-1 junction with onramp, we denote the incoming road $\road = \Leftroad$ and outgoing road $\road =\Rightroad$. We assume that the onramp can be controlled by the time--dependent metering rate $\control(t)$ and compute the controlled onramp demand with equation~\eqref{eq:controlledonrampdemand}. The fluxes at the road interfaces are given by
\begin{align*}
	\colvector{\massflux_{\Leftroad, out}^s \\ \momflux_{\Leftroad,out}^s} &= \colvector{1 \\ w_{\Leftroad, \Nxmyroad{\Leftroad}}^s} \min \bigg \lbrace  \Demand_\Leftroad, \max \Big \lbrace \beta \Supply_\Rightroad, \Supply_\Rightroad - \demandonramp(l^s,t^s) \Big \rbrace \bigg \rbrace, \\
	\colvector{\massflux_{\onramp, out}^s \\ \momflux_{\onramp,out}^s} &= \colvector{1 \\ w_{\Leftroad, \Nxmyroad{\Leftroad}}^s} \min \bigg \lbrace \demandonramp(l^s,t^s), \max \Big \lbrace \Big (1-\beta) \Supply_\Rightroad, \Supply_\Rightroad - \Demand_\Leftroad \rbrace \bigg \rbrace, \\	
	\colvector{\massflux_{\Rightroad, in} \\ \momflux_{\Rightroad, in}} &= \colvector{1 \\ w_{\Leftroad,\Nxmyroad{\Leftroad}}^s} ( \massflux_{\Leftroad, out}^s + \massflux_{\onramp, out}^s),	
\end{align*}
where
\begin{equation*}
	\Demand_\Leftroad =\ARZdemand(\rho_{\Leftroad,\Nxmyroad{\Leftroad}}^s, w_{\Leftroad, \Nxmyroad{\Leftroad}}^s), \qquad 
	\Supply_\Rightroad = \ARZsupply(\tilde{\rho}_{\Rightroad,1}^s, w_{\Leftroad, \Nxmyroad{\Leftroad}}^s), \qquad \tilde{\rho}_{\Rightroad,1}^s = \pressure^{-1}\left(\max \lbrace w_{\Leftroad, \Nxmyroad{\Leftroad}}^s - v_{\Rightroad,1}^s,0 \rbrace \right)
\end{equation*}

\subsubsection*{Junction with outflow }

At a junction with only one incoming road (denoted by index 1), we consider absorbing boundary conditions. We have
\begin{align*}
	\colvector{\massflux_{1,out}^s \\ \momflux_{1,out}^s} &= \colvector{1 \\ w_{N_{x_\Leftroad} -1}^s} \min \Big \{ \ARZdemand \left(\rho_{N_{x_\Leftroad} -1}^s, w_{N_{x_\Leftroad} -1}^s \right),\ARZsupply \left(\rho_{N_{x_\Leftroad} -1}^s, w_{N_{x_\Leftroad} -1}^s \right) \Big \}.
\end{align*}

Analogously to Godunov scheme for the first order models, the Godunov approximation for the second order models is mathematically equivalent to a cell transmission model (2CTM), see~\cite{FanSunPic2017}, using the supply function of the ARZ model and the two quantities $\rho$, $w$.

\subsubsection{Capacity drop in comparison to second order models} \label{sec:RP_ARZvsALWR}
We consider the network depicted in Figure~\ref{img:onetoonewithOR} with the parameters given in Table~\ref{tab:testNetwork}, which are inspired by the onramp scenario in~\cite[Section 4.3]{KolGoeGoa2017}. We set $\dx = 0.25 \, \km$, $\dt = 2 \cdot 10^{-3}\, \text{h} $ and the desired inflow from the onramp to $\onrampinflow = 4000\, \ch$ and the maximum flux allowed to exit the onramp equal to the maximum flux of the outgoing road $\fmax_\onramp = 4500\, \ch$. The priority is $\beta = 0.5$ and the incoming road therefore experiences congestion. We compute the time evolution of the (half-) Riemann problem~\eqref{eq:LWR_RP_11withOR} with $\rho_\Leftroad^0 = 140$, $\rho_\Rightroad^0 = 90$ within the LWR and ALWR model. Within the ARZ and the Greenberg model, we compute the time evolution of the (half-) Riemann problem~\eqref{eq:ARZ_RP_11withOR} with initial states $(\rho_\Leftroad^0, w_\Leftroad^0)$, $(\rho_\Rightroad^0, w_\Rightroad^0)$ where $w_\road^0 = \LWRvelocity(\rho_\road^0) + \pressure(\rho_\road^0)$, $\road = \Leftroad, \Rightroad$.

The density at the end of the incoming road increases in all models, see Figure~\ref{img:11OR_allmodels_densities}. The density in the Greenberg model increases similar to the ARZ model but is relaxed towards the ALWR flux as time progresses. The discharge flow of the junction in the LWR model stays at the level of the maximum flow even as further congestion on the incoming road sets on, compare Figure~\ref{img:11OR_allmodels_fluxes}. In contrast, all of the other models predict a decrease in the discharge flow at the junction. At $t=0$, the flux $(\rho v)_\Rightroad(0,0)$ is identical for the Greenberg and the ARZ model. However, the flux in the Greenberg model is relaxed until a state on the LWR fundamental diagram is reached. The ALWR model mimics the flux of the second order model at the junction. We see a capacity drop since the discharge flow is lower than the maximum flux due to congestion on the incoming road.

\begin{table}[hbt]
	\centering
	\caption{Properties of the roads in Figure~\ref{img:onetoonewithOR}.}
	\label{tab:testNetwork}
	\begin{tabular}{c|c|c|c|c}
		road & length $[\km]$ & $\rhomax$ $[\ckm]$ & $\vmax$ $[\kmh]$ & initial density $\rho^0$ $[\ckm]$\\
		\hline
		\hline
		road $\Leftroad$  & 4 & 180 & 100 & 140\\
		road $\Rightroad$  & 2 & 180 & 100 & 90\\
	\end{tabular}
\end{table}

\begin{figure}[tbh]
	\centering 
	\tikzsetnextfilename{flowcomparison_legend}
	\begin{tikzpicture}
		\begin{axis}[height=2.5cm, axis lines=none, legend columns=3, xmin=0, xmax=1, ymin=0, ymax=1] 
			\addplot[mygray, line width=1.0] coordinates {(-1,-1)};
			\addplot[myblue, line width=1.0] coordinates {(-1,-1)};
			\addplot[myred, line width=1.0] coordinates {(-1,-1)};
			\addplot[black,dashed, line width=1.0] coordinates {(-1,-1)};
			\addplot[black,dashdotted, line width=1.0] coordinates {(-1,-1)};
			\addplot[black, dotted, line width=1.0] coordinates {(-1,-1)};
			\addlegendentry{LWR };
			\addlegendentry{ALWR };
			\addlegendentry{ARZ};
			\addlegendentry{Greenberg $\relaxationtime = 0.005 \,\text{h} $ };
			\addlegendentry{Greenberg $\relaxationtime = 0.01 \,\text{h}$};
			\addlegendentry{Greenberg $\relaxationtime = 0.02 \,\text{h}$};
		\end{axis}
	\end{tikzpicture}
	\vspace{0.25cm}
	
	\tikzsetnextfilename{flowcomparison_densities}
	\subfloat[Density at the end of the incoming road.]{ 
		\tikzsetnextfilename{ALWRARComparison_densities} 
%
%
\begin{tikzpicture}
\setlength\fwidth{0.35\textwidth}
\begin{axis}[%
width=0.951\fwidth,
height=0.75\fwidth,
at={(0\fwidth,0\fwidth)},
scale only axis,
xmin=0.00,
xmax=0.1,
xtick={0, 0.05, 0.1},
scaled ticks=false,
xlabel={$t$},
ylabel={$\rho_\Leftroad(0,t)$},
ymin=140,
ymax=165,
axis background/.style={fill=white},
		legend style={at={(0.02,0.02)}, anchor=south west, font=\small, legend columns = 2, draw=white!15!black}
]
\addplot [color=mygray,line width=1.0]
table[row sep=crcr]{%
	0	140\\
	0.00200000000000955	146.888888888889\\
	0.00399999999999068	150.505130315501\\
	0.00600000000000023	152.234593451075\\
	0.00800000000000978	153.020617353206\\
	0.00999999999999091	153.369069746187\\
	0.0120000000000005	153.521785299533\\
	0.01400000000001	153.5883754878\\
	0.0159999999999911	153.617346611456\\
	0.0180000000000007	153.629938656424\\
	0.0200000000000102	153.635409352334\\
	0.0219999999999914	153.637785692605\\
	0.0240000000000009	153.638817835458\\
	0.0260000000000105	153.639266122149\\
	0.0279999999999916	153.639460821857\\
	0.0320000000000107	153.639582109512\\
	0.0420000000000016	153.639609871047\\
	0.102000000000004	153.639610306789\\
};

\addplot [color=myblue,line width=1.0]
table[row sep=crcr]{%
	0	140\\
	0.00200000000000955	149.993512747028\\
	0.00399999999999068	155.55897152603\\
	0.00600000000000023	158.215467264387\\
	0.00800000000000978	159.37195915567\\
	0.00999999999999091	159.852939869022\\
	0.0120000000000005	160.048962203112\\
	0.01400000000001	160.128174002215\\
	0.0159999999999911	160.160072016282\\
	0.0180000000000007	160.172899047082\\
	0.0200000000000102	160.178054209259\\
	0.0219999999999914	160.180125587746\\
	0.0240000000000009	160.180957805109\\
	0.0260000000000105	160.181292152612\\
	0.0279999999999916	160.181426476411\\
	0.0320000000000107	160.181502120415\\
	0.0480000000000018	160.181516669187\\
	0.102000000000004	160.181516679066\\
};

\addplot [color=black, dashed,line width=1.0]
table[row sep=crcr]{%
	0	140\\
	0.00200000000000955	149.993512747028\\
	0.00399999999999068	156.250898460134\\
	0.00600000000000023	159.433048782631\\
	0.00800000000000978	160.739814586218\\
	0.00999999999999091	161.124056319014\\
	0.0120000000000005	161.121716245825\\
	0.01400000000001	160.985437036446\\
	0.0159999999999911	160.822334943786\\
	0.0180000000000007	160.67306701722\\
	0.0200000000000102	160.549748259465\\
	0.0219999999999914	160.452988675145\\
	0.0240000000000009	160.379317545276\\
	0.0260000000000105	160.324305053575\\
	0.0279999999999916	160.283783119993\\
	0.0300000000000011	160.254242753935\\
	0.0320000000000107	160.232888771658\\
	0.0339999999999918	160.217564840356\\
	0.0360000000000014	160.206641344408\\
	0.0380000000000109	160.198904236283\\
	0.039999999999992	160.193458690431\\
	0.0420000000000016	160.18965077479\\
	0.0440000000000111	160.187006044577\\
	0.0459999999999923	160.185182494459\\
	0.0480000000000018	160.18393507292\\
	0.0500000000000114	160.183089223234\\
	0.0519999999999925	160.182521330986\\
	0.054000000000002	160.182144381998\\
	0.0560000000000116	160.181897509902\\
	0.0579999999999927	160.181738423168\\
	0.0620000000000118	160.181576123441\\
	0.0680000000000121	160.181507983543\\
	0.0879999999999939	160.181513617545\\
	0.102000000000004	160.181516385636\\
};

\addplot [color=black, dashdotted,line width=1.0]
table[row sep=crcr]{%
	0	140\\
	0.00200000000000955	149.993512747028\\
	0.00399999999999068	156.312318284977\\
	0.00600000000000023	159.627363378478\\
	0.00800000000000978	161.082917520519\\
	0.00999999999999091	161.588372804822\\
	0.0120000000000005	161.665772453606\\
	0.01400000000001	161.569973544721\\
	0.0159999999999911	161.415620780426\\
	0.0180000000000007	161.251627738458\\
	0.0200000000000102	161.097518125826\\
	0.0219999999999914	160.959990457378\\
	0.0240000000000009	160.840291628653\\
	0.0260000000000105	160.737468528507\\
	0.0279999999999916	160.649788978536\\
	0.0300000000000011	160.575351627305\\
	0.0320000000000107	160.512336908366\\
	0.0339999999999918	160.459099666825\\
	0.0360000000000014	160.414192789637\\
	0.0380000000000109	160.376361766997\\
	0.039999999999992	160.34452802184\\
	0.0420000000000016	160.317768907496\\
	0.0440000000000111	160.295297786347\\
	0.0459999999999923	160.276445573182\\
	0.0480000000000018	160.260644212722\\
	0.0500000000000114	160.247412155113\\
	0.0519999999999925	160.23634172055\\
	0.054000000000002	160.227088178403\\
	0.0560000000000116	160.219360349177\\
	0.0579999999999927	160.21291254236\\
	0.0600000000000023	160.20753765709\\
	0.0620000000000118	160.203061289681\\
	0.063999999999993	160.199336709644\\
	0.0660000000000025	160.196240582644\\
	0.0680000000000121	160.1936693343\\
	0.0699999999999932	160.191536062693\\
	0.0720000000000027	160.189767919838\\
	0.0740000000000123	160.188303893356\\
	0.0759999999999934	160.187092929158\\
	0.078000000000003	160.186092344318\\
	0.0800000000000125	160.1852664866\\
	0.0819999999999936	160.184585603374\\
	0.0840000000000032	160.184024888111\\
	0.0860000000000127	160.183563677296\\
	0.0879999999999939	160.183184774655\\
	0.092000000000013	160.18261912705\\
	0.0960000000000036	160.182240291877\\
	0.0999999999999943	160.181988021597\\
	0.102000000000004	160.181895892929\\
};

\addplot [color=black, dotted,line width=1.0]
table[row sep=crcr]{%
	0	140\\
	0.00200000000000955	149.993512747028\\
	0.00399999999999068	156.351034961535\\
	0.00600000000000023	159.758047873485\\
	0.00800000000000978	161.330975742085\\
	0.00999999999999091	161.951049190707\\
	0.0120000000000005	162.126597913747\\
	0.01400000000001	162.108571533224\\
	0.0159999999999911	162.01201144476\\
	0.0180000000000007	161.887808237752\\
	0.0219999999999914	161.631646918905\\
	0.0240000000000009	161.512497690144\\
	0.0260000000000105	161.401664951189\\
	0.0279999999999916	161.299263236004\\
	0.0300000000000011	161.204972282789\\
	0.0320000000000107	161.11830590681\\
	0.0339999999999918	161.038729466543\\
	0.0360000000000014	160.965710426844\\
	0.0380000000000109	160.898739312649\\
	0.039999999999992	160.837337566701\\
	0.0420000000000016	160.781059644501\\
	0.0440000000000111	160.729492611027\\
	0.0459999999999923	160.68225469531\\
	0.0480000000000018	160.638993452598\\
	0.0500000000000114	160.599383823185\\
	0.0519999999999925	160.563126215104\\
	0.054000000000002	160.529944665047\\
	0.0560000000000116	160.499585098984\\
	0.0579999999999927	160.471813699018\\
	0.0600000000000023	160.446415376285\\
	0.0620000000000118	160.423192346662\\
	0.063999999999993	160.401962804692\\
	0.0660000000000025	160.382559690633\\
	0.0680000000000121	160.364829545341\\
	0.0699999999999932	160.348631447769\\
	0.0720000000000027	160.333836029962\\
	0.0740000000000123	160.320324564609\\
	0.0759999999999934	160.307988120443\\
	0.078000000000003	160.29672678096\\
	0.0800000000000125	160.28644892219\\
	0.0819999999999936	160.277070545451\\
	0.0840000000000032	160.268514661271\\
	0.0860000000000127	160.260710720856\\
	0.0879999999999939	160.253594091727\\
	0.0900000000000034	160.247105574324\\
	0.092000000000013	160.241190956608\\
	0.0939999999999941	160.235800603858\\
	0.0960000000000036	160.230889081056\\
	0.0980000000000132	160.226414805418\\
	0.0999999999999943	160.222339726806\\
	0.102000000000004	160.218629033889\\
};

\addplot [color=myred,line width=1.0]
table[row sep=crcr]{%
	0	140\\
	0.00200000000000955	149.993512747028\\
	0.00399999999999068	156.39711566735\\
	0.00600000000000023	159.921732678862\\
	0.00800000000000978	161.660268900617\\
	0.00999999999999091	162.463816190603\\
	0.0120000000000005	162.823007004426\\
	0.01400000000001	162.981056859167\\
	0.0159999999999911	163.050108602131\\
	0.0180000000000007	163.08018256126\\
	0.0200000000000102	163.093262607758\\
	0.0219999999999914	163.098948094244\\
	0.0240000000000009	163.101418752666\\
	0.0260000000000105	163.102492268712\\
	0.0279999999999916	163.102958694951\\
	0.0300000000000011	163.103161345651\\
	0.0339999999999918	163.10328764487\\
	0.0440000000000111	163.103316621272\\
	0.063999999999993	163.103355727031\\
	0.0699999999999932	163.103463202121\\
	0.0740000000000123	163.103636584027\\
	0.078000000000003	163.103969292045\\
	0.0800000000000125	163.104228003162\\
	0.0819999999999936	163.104571944374\\
	0.0840000000000032	163.105023658188\\
	0.0860000000000127	163.105610188118\\
	0.0879999999999939	163.106363666603\\
	0.0900000000000034	163.107321928103\\
	0.092000000000013	163.108529141424\\
	0.0939999999999941	163.110036454628\\
	0.0960000000000036	163.111902645271\\
	0.0980000000000132	163.114194768158\\
	0.0999999999999943	163.11698879244\\
	0.102000000000004	163.120370219566\\
};

\end{axis}
\end{tikzpicture}%
		\label{img:11OR_allmodels_densities}}
	\tikzsetnextfilename{flowcomparison_fluxes}
	\subfloat[Mass flux into the outgoing road.]{ 
		\tikzsetnextfilename{ALWRARComparison_fluxes} 
%
%
\begin{tikzpicture}
	\setlength\fwidth{0.35\textwidth}
	\begin{axis}[%
		scaled x ticks = false,
		width=0.951\fwidth,
		height=0.75\fwidth,
		at={(0\fwidth,0\fwidth)},
		scale only axis,
		xmin=0.0,
		xmax=0.1,
		xtick={0, 0.05, 0.1},
		ymin=3300,
		ymax=4600,
		xlabel={$t$},
		ylabel={$(\rho v)_\Rightroad (0,t)$},
		axis background/.style={fill=white},
		legend style={at={(0.02,0.02)}, anchor=south west, font=\small, legend columns = 2, draw=white!15!black}
		]
\addplot [mygray, line width=1.0]
table[row sep=crcr]{%
0	4500\\
0.101999999999862	4500\\
};

\addplot [myblue, line width=1.0]
table[row sep=crcr]{%
0	3723.84403546519\\
0.00199999999995271	3609.50022555202\\
0.00399999999990541	3560.34412368753\\
0.00599999999985812	3540.48816706962\\
0.00799999999981083	3532.56735822217\\
0.0100000000002183	3529.40196276056\\
0.012000000000171	3528.13359940983\\
0.0140000000001237	3527.62461994433\\
0.0160000000000764	3527.42023637785\\
0.0180000000000291	3527.33814204254\\
0.0199999999999818	3527.30516360303\\
0.0219999999999345	3527.29191509055\\
0.0239999999998872	3527.28659263261\\
0.0259999999998399	3527.2844543721\\
0.0279999999997926	3527.28359533821\\
0.0300000000002001	3527.28325022585\\
0.0320000000001528	3527.2831115787\\
0.0360000000000582	3527.28303350038\\
0.0520000000001346	3527.28301848344\\
0.101999999999862	3527.28301847324\\
};

\addplot [color=black, dashed,line width=1.0]
  table[row sep=crcr]{%
	0	3723.84403546519\\
	0.00199999999995271	3691.05317390699\\
	0.00399999999990541	3655.12826183422\\
	0.00599999999985812	3624.20675494823\\
	0.00799999999981083	3599.58331354141\\
	0.0100000000002183	3580.69569797176\\
	0.012000000000171	3566.49791514204\\
	0.0140000000001237	3555.93892383086\\
	0.0160000000000764	3548.13504059983\\
	0.0180000000000291	3542.39440623199\\
	0.0199999999999818	3538.18925838015\\
	0.0219999999999345	3535.12126317757\\
	0.0239999999998872	3532.89180232808\\
	0.0259999999998399	3531.27820758797\\
	0.0279999999997926	3530.11518114481\\
	0.0300000000002001	3529.28050453709\\
	0.0320000000001528	3528.68415979317\\
	0.0340000000001055	3528.2600999473\\
	0.0360000000000582	3527.96005338022\\
	0.0380000000000109	3527.74887922786\\
	0.0399999999999636	3527.60109941411\\
	0.0419999999999163	3527.49831917133\\
	0.043999999999869	3527.4273157136\\
	0.0459999999998217	3527.3786274011\\
	0.0479999999997744	3527.34551636389\\
	0.0500000000001819	3527.32320872901\\
	0.0520000000001346	3527.30834040598\\
	0.0540000000000873	3527.29855449161\\
	0.05600000000004	3527.29221006793\\
	0.0579999999999927	3527.28817250944\\
	0.0599999999999454	3527.28566318657\\
	0.0619999999998981	3527.28415226702\\
	0.0639999999998508	3527.2832826511\\
	0.0659999999998035	3527.28281629514\\
	0.068000000000211	3527.28259655636\\
	0.0700000000001637	3527.28252194541\\
	0.0740000000000691	3527.28257459297\\
	0.0859999999997854	3527.28290221571\\
	0.0960000000000036	3527.28299456959\\
	0.101999999999862	3527.28301061808\\
};

\addplot [color=black, dashdotted,line width=1.0]
table[row sep=crcr]{%
0	3723.84403546519\\
0.00199999999995271	3704.70428330942\\
0.00399999999990541	3681.67307430079\\
0.00599999999985812	3659.61726721708\\
0.00799999999981083	3639.87794539779\\
0.0100000000002183	3622.72183953857\\
0.012000000000171	3608.02009211111\\
0.0140000000001237	3595.50079328547\\
0.0160000000000764	3584.86924508464\\
0.0180000000000291	3575.8541202755\\
0.0199999999999818	3568.21781071348\\
0.0219999999999345	3561.75529227506\\
0.0239999999998872	3556.29066504918\\
0.0259999999998399	3551.67349854648\\
0.0279999999997926	3547.77539443538\\
0.0300000000002001	3544.48690122495\\
0.0320000000001528	3541.714813629\\
0.0340000000001055	3539.37983340759\\
0.0360000000000582	3537.41455199077\\
0.0380000000000109	3535.76171491094\\
0.0399999999999636	3534.372731303\\
0.0419999999999163	3533.20639558874\\
0.043999999999869	3532.22779237717\\
0.0459999999998217	3531.40735927986\\
0.0479999999997744	3530.72008563729\\
0.0500000000001819	3530.14482807108\\
0.0520000000001346	3529.66372634292\\
0.0540000000000873	3529.26170524646\\
0.05600000000004	3528.92605021668\\
0.0579999999999927	3528.64604604412\\
0.0599999999999454	3528.41266956048\\
0.0619999999998981	3528.21832844272\\
0.0639999999998508	3528.05663939198\\
0.0659999999998035	3527.92223990084\\
0.068000000000211	3527.81062864928\\
0.0700000000001637	3527.71803028193\\
0.0720000000001164	3527.64128093239\\
0.0740000000000691	3527.57773138792\\
0.0760000000000218	3527.5251652409\\
0.0779999999999745	3527.48172976241\\
0.0799999999999272	3527.44587756705\\
0.0819999999998799	3527.41631742355\\
0.0839999999998327	3527.39197281072\\
0.0859999999997854	3527.3719470273\\
0.0880000000001928	3527.35549384323\\
0.0900000000001455	3527.34199283243\\
0.0920000000000982	3527.33092865744\\
0.0940000000000509	3527.32187368729\\
0.0960000000000036	3527.31447342466\\
0.0979999999999563	3527.30843429847\\
0.0999999999999091	3527.30351344697\\
0.101999999999862	3527.2995101741\\
};

\addplot [color=black, dotted,line width=1.0]
table[row sep=crcr]{%
0	3723.84403546519\\
0.00199999999995271	3713.40009501799\\
0.00799999999981083	3673.56791768605\\
0.0100000000002183	3661.42048707774\\
0.012000000000171	3650.19369512124\\
0.0140000000001237	3639.86631025276\\
0.0160000000000764	3630.38259934576\\
0.0180000000000291	3621.67951399864\\
0.0199999999999818	3613.69565517068\\
0.0219999999999345	3606.37341951631\\
0.0239999999998872	3599.65935391853\\
0.0259999999998399	3593.50408515876\\
0.0279999999997926	3587.86210667416\\
0.0300000000002001	3582.69151518493\\
0.0320000000001528	3577.95374318475\\
0.0340000000001055	3573.61330417945\\
0.0360000000000582	3569.63755455244\\
0.0380000000000109	3565.99647234539\\
0.0399999999999636	3562.66245235033\\
0.0419999999999163	3559.61011658337\\
0.043999999999869	3556.81613911216\\
0.0459999999998217	3554.25908422185\\
0.0479999999997744	3551.91925695806\\
0.0500000000001819	3549.77856514304\\
0.0520000000001346	3547.82039202068\\
0.0540000000000873	3546.02947874294\\
0.05600000000004	3544.39181596454\\
0.0579999999999927	3542.89454386417\\
0.0599999999999454	3541.52585995818\\
0.0619999999998981	3540.27493411819\\
0.0639999999998508	3539.13183024625\\
0.0659999999998035	3538.08743410055\\
0.068000000000211	3537.1333868021\\
0.0700000000001637	3536.262023587\\
0.0720000000001164	3535.46631740142\\
0.0740000000000691	3534.73982696628\\
0.0760000000000218	3534.0766489667\\
0.0779999999999745	3533.47137404711\\
0.0799999999999272	3532.91904631732\\
0.0819999999998799	3532.41512609669\\
0.0839999999998327	3531.95545564491\\
0.0859999999997854	3531.53622764642\\
0.0880000000001928	3531.15395623365\\
0.0900000000001455	3530.80545035059\\
0.0920000000000982	3530.48778927311\\
0.0940000000000509	3530.1983001167\\
0.0960000000000036	3529.93453717519\\
0.0979999999999563	3529.69426294574\\
0.0999999999999091	3529.47543070672\\
0.101999999999862	3529.27616852509\\
};

\addplot [myred, line width=1.0]
table[row sep=crcr]{%
0	3723.84403546519\\
0.0500000000001819	3723.84406899043\\
0.0540000000000873	3723.84417380007\\
0.05600000000004	3723.84429674612\\
0.0579999999999927	3723.84450921961\\
0.0599999999999454	3723.84486385504\\
0.0619999999998981	3723.8454376402\\
0.0639999999998508	3723.84634037941\\
0.0659999999998035	3723.84772515423\\
0.068000000000211	3723.84980101017\\
0.0700000000001637	3723.85284807619\\
0.0720000000001164	3723.85723529378\\
0.0740000000000691	3723.86344089405\\
0.0760000000000218	3723.87207571582\\
0.0779999999999745	3723.88390940624\\
0.0799999999999272	3723.89989948917\\
0.0819999999998799	3723.92122322703\\
0.0839999999998327	3723.94931214075\\
0.0859999999997854	3723.98588899087\\
0.0880000000001928	3724.0330069627\\
0.0900000000001455	3724.09309074094\\
0.0920000000000982	3724.16897910592\\
0.0940000000000509	3724.26396863488\\
0.0960000000000036	3724.38185805003\\
0.0979999999999563	3724.52699271957\\
0.0999999999999091	3724.7043087907\\
0.101999999999862	3724.91937641395\\
};

\end{axis}
\end{tikzpicture}%
		\label{img:11OR_allmodels_fluxes}}
	\caption{Densities and fluxes.}
\end{figure}

\subsubsection{The ramp metering control problem} \label{sec:RampMetering_ALWR}

We consider an onramp with ramp metering as introduced in Section~\ref{sec:discretization_firstorder} and~\ref{sec:discretization_secondorder} with a piecewise constant control function $\control(t)$ on intervals of 15 minutes. Our aim is to compare the optimal control of the ALWR model with optimal control results of the Greenberg model. We consider the relaxation time $\relaxationtime = 0.005\,\text{h}$ for the Greenberg model~\eqref{eq:Greenberg_network} and the parameters from Table~\ref{tab:testNetwork} but set the initial densities to $\rho_\road^0 = 50 \ckm$, $\road = \Leftroad, \Rightroad$ such that both roads have equal traffic volume at $t=0$. The priority parameter $\beta$ at the onramp equals 0.5 and we set the maximum inflow from the onramp to $\fmax_\onramp= 2000 \, \ch$. Additionally, we consider an onramp at the origin \emph{in} with $\fmax_\onramp = 4500 \, \ch$. The time horizon is $T=3$ hours. The applied discretization parameters are $\Delta x = 0.25\, \km$ and $\Delta t = 2\cdot 10^{-3} \, \text{h}$ and we apply the discretization schemes described in section~\ref{sec:discretization_firstorder} and~\ref{sec:discretization_secondorder}. 

For the control scenario, we apply the inflow conditions shown in Figure~\ref{img:RampMetering_Inflows}. At the beginning of the time horizon, we consider a first rush-hour with moderate demand at the onramp. Later on, we consider a second rush-hour, with increased onramp demand. For the given scenario, we are interested in minimizing the total travel time
\begin{equation}\label{eq:objective}
	\sum\limits_{ \road = \Leftroad}^{\Rightroad} \int\limits_0^T \int\limits_0^{\lengthofroad_\road} \rho_\road(x,t) \, \intx \, \intt \ +  \sum\limits_{ \road = \Leftroad}^{\Rightroad} \int\limits_0^T \queuelength_\road(t) \, \intt,
\end{equation}
where $\queuelength_\Leftroad$ denotes the queue at \emph{in} and $\queuelength_\Rightroad$ the queue at \emph{ramp}. To solve this optimization task, we apply a first-discretize-then-optimize approach. Thus, for given control decisions (time-dependent piecewise constant metering rates), the discretization schemes are always used to evaluate the objective function~\eqref{eq:objective}. The Matlab solver \emph{fmincon}\footnote{https://de.mathworks.com/help/optim/ug/fmincon.html} is used for the optimization of the control decisions. 

\begin{figure}[tbhp]
	\centering 
	\subfloat[Desired inflows at \emph{in} and \emph{ramp}.]{ 
		\tikzsetnextfilename{RampMetering_Inflows} 
%
%
\begin{tikzpicture}
\setlength\fwidth{0.35\textwidth}
\begin{axis}[%
width=0.951\fwidth,
height=0.75\fwidth,
at={(0\fwidth,0\fwidth)},
scale only axis,
xmin=0,
xmax=3,
ymin=0,
ymax=5000,
xlabel={$t$},
ylabel={inflow},
axis background/.style={fill=white},
legend style={legend pos=north west, legend cell align=left, align=left, draw=white!15!black, legend columns=2}
]
\addplot [color=black,line width=1.0]
  table[row sep=crcr]{%
0	3611.11111111111\\
2	3611.11111111111\\
2.00199999999995	3200\\
3	3200\\
};
\addlegendentry{flow \emph{in}}

\addplot [color=black, dashed,line width=1.0]
  table[row sep=crcr]{%
0	500\\
0.25	500\\
0.251999999999953	0\\
0.5	0\\
0.501999999999953	1500\\
0.948000000000093	1500\\
0.950000000000045	0\\
3	0\\
};
\addlegendentry{flow \emph{ramp}}

\end{axis}
\end{tikzpicture}%
		\label{img:RampMetering_Inflows}}
	\tikzsetnextfilename{RampMetering_controls}
	\subfloat[Optimal control $\control(t)$ at \emph{ramp}.]{ 
		\tikzsetnextfilename{RampMetering_controls} 
%
%
\begin{tikzpicture}
\setlength\fwidth{0.35\textwidth}
\begin{axis}[%
width=0.951\fwidth,
height=0.75\fwidth,
at={(0\fwidth,0\fwidth)},
scale only axis,
xmin=0,
xmax=3,
ymin=0,
ymax=1,
xlabel={$t$},
ylabel={$\control(t)$},
axis background/.style={fill=white},
axis x line*=bottom,
axis y line*=left,
legend style={legend pos=south east, legend cell align=left, align=left, draw=white!15!black}
]
\addplot[const plot, color=black, dashed,line width=1.0] table[row sep=crcr] {%
0	1\\
0.5	0.470575327258505\\
0.75	0.444145268048465\\
1	0.430790723813216\\
1.25	0.999997329429308\\
1.5	1\\
2.75	1\\
};
\addlegendentry{opt. control ALWR}

\addplot[const plot, color=black, dotted,line width=1.0] table[row sep=crcr] {%
0	1\\
0.5	0.497300800155455\\
0.75	0.494275846745833\\
1	0.486699483298891\\
1.25	0.999937572626162\\
1.5	0.999999191491467\\
2.75	0.999999191491467\\
};
\addlegendentry{opt. control Greenberg}

\end{axis}
\end{tikzpicture}%
		\label{img:RampMetering_controls}}
	\caption{Inflow and controls.}
	\label{img:RampMetering_InflowandControls}
\end{figure}

\begin{table}[hbt]
	\centering
	\caption{Optimization results (total travel time) for the network in Figure~\ref{img:onetoonewithOR}.}
	\label{tab:optResultsOnramp}
	\begin{tabular}{c|c|c|c}
		& Greenberg & ALWR & LWR \\
		\hline
		\hline
		no control    & 1236.5 & 1316.0 & 1019.3 \\
		optimal control ALWR & 996.6 & 1019.6 & 1019.6 \\
		optimal control Greenberg & \phantom{1}982.3 & 1322.9 & 1019.5 \\
	\end{tabular}
\end{table}

Table~\ref{tab:optResultsOnramp} shows the total travel times for the different models with and without optimization ($\control(t) = 1 ~\forall t$). The resulting queues are shown in Figure~\ref{fig:queuesOnrampScenario}, where we show the Greenberg model to demonstrate the benefit of the optimized control computed with the cheaper ALWR model. 

\begin{figure}[H]
	\centering
	\tikzsetnextfilename{control_queuelength_legend}
	\begin{tikzpicture}
		\begin{axis}[height=2.5cm, axis lines=none, legend columns=3, xmin=0, xmax=1, ymin=0, ymax=1] 
			\addplot[black,solid, line width=1.0] coordinates {(-1,-1)};
			\addplot[black,dashed, line width=1.0] coordinates {(-1,-1)};
			\addplot[black,dotted,thick,line width=1.0] coordinates {(-1,-1)};
			\addlegendentry{\ no control\quad \ };
			\addlegendentry{\ optimal control ALWR};
			\addlegendentry{\ optimal control Greenberg\quad \ };
		\end{axis}
	\end{tikzpicture}

		\subfloat[Queue at \emph{in}.]{ 
		\tikzsetnextfilename{RampMetering_BVTQueue_in} 
%
%
\begin{tikzpicture}
\setlength\fwidth{0.32\textwidth}
\begin{axis}[%
width=0.951\fwidth,
height=0.75\fwidth,
at={(0\fwidth,0\fwidth)},
scale only axis,
xmin=0,
xmax=3,
ymin=0,
ymax=300,
xlabel={$t$},
ylabel={queue $\queuelength_\Leftroad(t)$ at \emph{in} },
axis background/.style={fill=white},
axis x line*=bottom,
axis y line*=left,
legend style={legend cell align=left, align=left, draw=white!15!black}
]
\addplot [color=black,line width=1.0]
  table[row sep=crcr]{%
0	0\\
0.873999999999995	0\\
0.876000000000005	0.16709444996124\\
0.882000000000005	12.1210867529871\\
0.894000000000005	34.1861346406244\\
0.936000000000007	99.9976288450187\\
1.00800000000001	209.942953647962\\
1.018	220.246408356009\\
1.02799999999999	227.212940429532\\
1.036	230.807356293928\\
1.04400000000001	232.998540782248\\
1.05000000000001	233.884403356467\\
1.054	234.166654960052\\
1.05600000000001	234.22424131185\\
1.05799999999999	234.229701745176\\
1.06	234.185512263398\\
1.06399999999999	233.957363784055\\
1.06999999999999	233.296743935787\\
1.078	231.896799897496\\
1.08799999999999	229.445523475204\\
1.102	224.965617940991\\
1.12	217.878276784493\\
1.148	205.235847508266\\
1.19800000000001	180.858341057406\\
1.27199999999999	143.010792174745\\
1.34999999999999	101.424162277507\\
1.47399999999999	32.6990794062406\\
1.53399999999999	0\\
3	0\\
};

\addplot [color=black, dashed,line width=1.0]
  table[row sep=crcr]{%
0	0\\
3	0\\
};

\addplot [color=black, dotted,line width=1.0]
  table[row sep=crcr]{%
0	0\\
3	0\\
};

\end{axis}
\end{tikzpicture}%
		\label{img:RampMetering_BVTQueue_in}}
	\qquad
	\subfloat[Queue at \emph{ramp}.]{ 
	\tikzsetnextfilename{RampMetering_BVTQueue_ramp} 
%
%
\begin{tikzpicture}
\setlength\fwidth{0.32\textwidth}
\begin{axis}[%
width=0.951\fwidth,
height=0.75\fwidth,
at={(0\fwidth,0\fwidth)},
scale only axis,
xmin=0,
xmax=3,
ymin=0,
ymax=300,
xlabel={$t$},
ylabel={queue $\queuelength_\Rightroad(t)$ at \emph{ramp} },
axis background/.style={fill=white},
axis x line*=bottom,
axis y line*=left,
legend style={legend cell align=left, align=left, draw=white!15!black}
]
\addplot [color=black,line width=1.0]
  table[row sep=crcr]{%
0	0\\
3	0\\
};

\addplot [color=black, dashed,line width=1.0]
  table[row sep=crcr]{%
0	0\\
0.50200000000001	0\\
0.507999999999981	3.82367140015646\\
0.767999999999972	150.075983357295\\
0.951999999999998	262.63052471546\\
1.05599999999998	171.744017915727\\
1.25	4.59721707619957\\
1.25400000000002	1.59493896489948e-06\\
3	0\\
};

\addplot [color=black, dotted,line width=1.0]
  table[row sep=crcr]{%
0	0\\
0.50200000000001	0\\
0.50800000000001	3.52969119828998\\
0.891999999999996	198.461763447234\\
0.951999999999998	229.148661837735\\
1.11000000000001	74.624294224379\\
1.18600000000001	1.33087053801779\\
1.19	0.350654336304018\\
1.19399999999999	0.092389499997438\\
1.19800000000001	0.0243425471356034\\
1.202	0.00641371152636339\\
1.20599999999999	0.00168986816845518\\
1.214	0.000117311241552898\\
1.27799999999999	0\\
3	0\\
};

\end{axis}
\end{tikzpicture}%
	\label{img:RampMetering_BVTQueue_ramp}}
	
	\caption{Queues in the Greenberg model with and without optimization.}
	\label{fig:queuesOnrampScenario}
\end{figure}

Running a simulation with the Greenberg model using the optimal control of the ALWR model leads to a total travel time of 996.6, which is about 1.5\% away from the optimal solution found by optimizing the Greenberg model, see Table~\ref{tab:optResultsOnramp}. Actually, the optimal metering rates for the two models do not differ much, see Figure~\ref{img:RampMetering_controls}. The key in improving the travel times is the reduction of the flow from the onramp in the time frame of high onramp demand, compare Figure~\ref{img:RampMetering_InflowandControls}. Both controls do not reduce the onramp demand during the first rush-hour, but reduce it during the second rush-hour to ensure that the incoming road stays in free flow. Without control, the queue at the onramp stays empty whereas more than 200 cars accumulate in the queue at the origin, see Figure~\ref{fig:queuesOnrampScenario}. When the optimal control of the Greenberg or the ALWR model is used, the queue at the origin is reduced to zero, while the cars accumulate in the queue at the onramp during the rush-hour time. Ramp metering leads to an increased outflow at the end of the outgoing road early within the time horizon, see Figure~\ref{img:RampMetering_Outflows}. Note that an optimal control strategy cannot be recognized by an optimal control approach based on the LWR model since full inflow, i.e., $\control(t) = 1 ~\forall t$ is optimal for the LWR model. However, this is clearly not optimal for the Greenberg model, see Table~\ref{tab:optResultsOnramp}.

\begin{figure}[tbh]
	
	\centering
	\begin{tikzpicture}
		\begin{axis}[height=2.5cm, axis lines=none, legend columns=3, xmin=0, xmax=1, ymin=0, ymax=1] 
			\addplot[black,solid, line width=1.0] coordinates {(-1,-1)};
			\addplot[black,dashed, line width=1.0] coordinates {(-1,-1)};
			\addplot[black,dotted,thick,line width=1.0] coordinates {(-1,-1)};
			\addlegendentry{\ no control\quad \ };
			\addlegendentry{\ optimal control ALWR};
			\addlegendentry{\ optimal control Greenberg\quad \ };
		\end{axis}
	\end{tikzpicture}

\tikzsetnextfilename{RampMetering_Outflow_legend}
\subfloat[ALWR.]{ 
	\tikzsetnextfilename{RampMetering_ALWROutflow} 
	\input{./img/RampMetering_ALWROutflow.tex}
	\label{img:RampMetering_ALWROutflow}}
		\qquad 
	\subfloat[Greenberg.]{ 
		\tikzsetnextfilename{RampMetering_BVTOutflow} 
		\input{./img/RampMetering_BVTOutflow.tex}
		\label{img:RampMetering_BVTOutflow}}
	\caption{Flow at the end of the outgoing road \emph{out} with and without optimization.}
	\label{img:RampMetering_Outflows}
	
\end{figure}


\section{Conclusion}

We have established a new model for the 1-to-1 junction with onramp which couples the LWR model to boundary conditions of the second order traffic model. The main difference to earlier presented versions of the LWR model is the shape of the supply function at the junction, which allows for formation of congestion when the incoming road is congested. A discussion of the new supply function was presented and we have shown that the fluxes at the junctions are only dependent on the densities of the adjacent roads. Numerical studies have shown that the ALWR model is able to capture the capacity drop phenomenon. Even though the junction model is a modification of the most simple traffic model for traffic flows, we achieved promising results and were able to capture at least some aspects of second order traffic models. Optimization results showed that the ALWR model is a suitable surrogate for the Greenberg model. In the considered scenario, the ALWR model optimum is close to the optimum of the Greenberg model. Note that similar considerations can be done analogously for exponents $\gamma > 1$ in the pressure function which is postponed to future work.

\section*{Acknowledgment}
This work was supported by the DFG grant No.\ GO 1920/7-1 and by DAAD (Project-ID 57445223). 


\section{Appendix}

We begin with some general observations which are useful to derive the partitioning of the $\rho_\Leftroad$-$\rho_\Rightroad$-plane in Figure~\ref{img:rho_l-rho_r-plane}. For fixed $\rho_\Leftroad$, the value of $\tilde{\rho}$ given by~\eqref{eq:explicittilderho} increases with $\rho_\Rightroad$
\begin{equation}\label{eq:tilderhoincreaseswithrhor}
	\partial_{\rho_\Rightroad} \tilde{\rho} \geq 0.
\end{equation}
Moreover, due to~\eqref{eq:tilderhoincreaseswithrhor}, the partial derivative for the density on the outgoing road of the derived ARZ supply~\eqref{eq:derivedARZsupply} satisfies 
\begin{equation} \label{eq:partial_rhor_SAR}
	\partial_{\rho_\Rightroad} \ARZsupply(\rho_\Leftroad, \rho_\Rightroad) \leq 0.
\end{equation}
Thus, the larger $\rho_\Rightroad$, the larger $\tilde{\rho}$ and the lower the ARZ supply. Furthermore, for $\rho_\Leftroad = \rho_\Rightroad$, we have $\tilde{\rho} = \rho_\Leftroad = \rho_\Rightroad$. 
Now, we have a closer look at $\tilde{\rho}$ for $\rho_\Rightroad \geq \rho_\Leftroad$. We show that 
\begin{equation}\label{eq:tilderho_bigger_rhor}
	\tilde{\rho} = \pressure^{-1}\left(\pressure(\rho_\Leftroad) + \LWRvelocity_\Leftroad - \LWRvelocity_\Rightroad \right) \geq \rho_\Rightroad.
\end{equation}
This is equivalent to
\begin{align*}
	\pressure(\rho_\Leftroad) + \LWRvelocity_\Leftroad - \LWRvelocity_\Rightroad &\geq \pressure(\rho_\Rightroad) \\
	\vmax \left( \frac{\rho_\Rightroad - \rho_\Leftroad}{\rhomax} + \frac{\rho_\Leftroad^2}{2 \left(\rhomax\right)^2}\right) &\geq \frac{\vmax}{2} \left( \frac{\rho_\Rightroad}{{\rhomax}}\right)^2 \\
	\frac{\rho_\Rightroad - \rho_\Leftroad}{\rhomax} &\geq \frac{\rho_\Rightroad^2- \rho_\Leftroad^2}{2{\left( \rhomax \right)}^2} \\
	2 \rhomax &\geq \rho_\Leftroad + \rho_\Rightroad, 
\end{align*}
which is satisfied for any combination $(\rho_\Leftroad, \rho_\Rightroad) \in [0, \rhomax]^2$. In the following, we prove the partitioning of the $\rho_\Leftroad$-$\rho_\Rightroad$-plane.


\paragraph{I} 

The analysis for 
\begin{equation*}
	\rho_\Leftroad > \rholim, \qquad 	\rho_\Rightroad \leq \frac{\rhomax}{2},
\end{equation*}
is straightforward. The area of admissible combinations $(\rho_\Leftroad,\rho_\Rightroad)$ is shown in area I of Figure \ref{img:rho_l-rho_r-plane}. By the definition of $\rholim$, cf. equation~\eqref{eq:rholim}, it holds that 
\begin{align*}
	\LWRsupply(\rho_\Rightroad) = \fmax > \ARZsupply(\rho_\Leftroad, 0).
\end{align*} 
Therefore, $\LWRsupply(\rho_\Rightroad) > \ARZsupply(\rho_\Leftroad, \rho_\Rightroad)$ holds and the ARZ supply is applied.


\paragraph{VII}
We analyze more closely the case
\begin{equation*}
	\rho_\Leftroad \leq \rholim,
\end{equation*} 
which is the area VII in Figure~\ref{img:rho_l-rho_r-plane}. By the definition of $\rholim$, cf. equation~\eqref{eq:rholim}, we have that $\ARZsupply(\rho_\Leftroad,0) \geq \fmax$. \\

\begin{itemize}
	\item[a)] Assume first that $\rho_\Leftroad \geq \rho_\Rightroad$. Then 
	\begin{equation*}
		\tilde{\rho} = \pressure^{-1} \left(\max \lbrace 0, \LWRvelocity_\Leftroad - \LWRvelocity_\Rightroad + \pressure(\rho_\Leftroad) \rbrace \right) \leq \rho_\Leftroad < \curveofmaxima(\rho_\Leftroad),
	\end{equation*}
	and this leads to
	\begin{equation*}
		\ARZsupply(\rho_\Leftroad, \rho_\Rightroad) = \ARZsupply(\rho_\Leftroad, 0) \geq \fmax \geq \LWRsupply(\rho_\Rightroad).
	\end{equation*}
	\item[b)] Assume now $\rho_\Rightroad > \rho_\Leftroad$ and $\rho_\Rightroad \geq \nicefrac{\rhomax}{2}$.
	\begin{itemize}
		\item[b1)] Assume $\tilde{\rho} \geq \fluxmax_\Leftroad$. Then we have
		\begin{equation*}
			\ARZsupply(\rho_\Leftroad, \rho_\Rightroad) = \tilde{\rho} (w_\Leftroad -(w_\Leftroad - \LWRvelocity_\Rightroad)) = \tilde{\rho}  \LWRvelocity_\Rightroad \geq \rho_\Rightroad \LWRvelocity_\Rightroad = \LWRsupply(\rho_\Rightroad),
		\end{equation*}
		where we made use of the relation of $\tilde{\rho}$ and $\rho_\Rightroad$ in equation~\eqref{eq:tilderho_bigger_rhor}.
		\item[b2)] If instead we have $\tilde{\rho} < \fluxmax_\Leftroad$, then
		\begin{equation*}
			\ARZsupply(\rho_\Leftroad, \rho_\Rightroad) = \ARZsupply(\rho_\Leftroad,0) \geq \LWRsupply(\rho_\Rightroad),
		\end{equation*}
		due to the assumption $\rho_\Leftroad \leq \rholim$. 
	\end{itemize}
\item[c)] Assume $\rho_\Rightroad > \rho_\Leftroad$ and $\rho_\Rightroad < \nicefrac{\rhomax}{2}$.  From a) and b), we deduce
\begin{equation*}
	\ARZsupply\left(\rho_\Leftroad, \frac{\rhomax}{2}\right) \geq \fmax, \qquad \forall \rho_\Leftroad \in [0, \rholim].
\end{equation*}

Moreover since~\eqref{eq:partial_rhor_SAR} holds true, we know that 
\begin{equation*}
	\ARZsupply(\rho_\Leftroad, \rho_\Rightroad) \geq \fmax = \LWRsupply(\rho_2), \qquad \forall \rho_\Leftroad \in [0, \rholim],~ \rho_\Rightroad \in  \left(\rho_\Leftroad, \frac{\rhomax}{2} \right).
\end{equation*}
\end{itemize}

\paragraph{II} 
Let $(\rho_\Leftroad,\rho_\Rightroad)$ be given with 
\begin{equation*}
	\rho_\Leftroad > \rholim, \qquad \rho_\Rightroad > \frac{\rhomax}{2}, \qquad \rho_\Rightroad < \fluxmax_\Leftroad, \qquad \ARZsupply(\rho_\Leftroad,0) \geq \LWRsupply(\rho_\Rightroad).
\end{equation*}
Because $\rho_\Rightroad < \fluxmax_\Leftroad$, we get the inequality $\rho_\Rightroad <  \curveofmaxima(\rho_\Leftroad)$ from~\eqref{eq:line_upIII}. Moreover due to $\ARZsupply(\rho_\Leftroad,0) \geq \LWRsupply(\rho_\Rightroad)$, compare~\eqref{eq:maxARZbiggerLWRsupply}, we know from~\eqref{eq:line_upVI} that $\rho_\Rightroad \geq \curveofequalsupplyatmaxarz(\rho_\Leftroad)$. This gives the area II in the $\rho_\Leftroad$-$\rho_\Rightroad$-plane in Figure~\ref{img:rho_l-rho_r-plane}. For given $(\rho_\Leftroad,\rho_\Rightroad)$ in area II, we have to proof that the LWR model is applied. We can proceed similarly as for region VII:

For $\rho_\Leftroad \geq \rho_\Rightroad$, it holds that $\tilde{\rho} \leq \rho_\Leftroad \leq \curveofmaxima(\rho_\Leftroad)$ and $\ARZsupply(\rho_\Leftroad, \rho_\Rightroad) = \ARZsupply(\rho_\Leftroad,0) \geq \LWRsupply(\rho_\Rightroad)$. 

 For $\rho_\Rightroad > \rho_\Leftroad \geq \nicefrac{\rhomax}{2}$, we distinguish again. If $\tilde{\rho} \geq \fluxmax_\Leftroad$, then $\ARZsupply(\rho_\Leftroad, \rho_\Rightroad) = \tilde{\rho} \LWRvelocity_\Rightroad \geq \rho_\Rightroad \LWRvelocity_\Rightroad$.
If instead $\tilde{\rho}< \fluxmax_\Leftroad$, then $\ARZsupply(\rho_\Leftroad, \rho_\Rightroad) = \ARZsupply(\rho_\Leftroad,0) \geq \LWRsupply(\rho_\Rightroad)$.

\paragraph{IV}

Let $(\rho_\Leftroad, \rho_\Rightroad)$ satisfy the assumptions
\begin{equation*}
	\rho_\Leftroad > \rholim, \qquad \rho_\Rightroad > \frac{\rhomax}{2}, \qquad  \rho_\Rightroad \geq \sigma_\Leftroad, \qquad \rho_\Leftroad \leq \rho_\Rightroad.
\end{equation*}
The area for admissible values of $\rho_\Leftroad$ and $\rho_\Rightroad$ is marked in Figure \ref{img:rho_l-rho_r-plane} as area IV. We know from equation~\eqref{eq:tilderho_bigger_rhor} and the assumption here, that $\tilde{\rho} \geq \rho_\Rightroad \geq \fluxmax_\Leftroad$. Therefore,
\begin{equation*}
	\ARZsupply(\rho_\Leftroad, \rho_\Rightroad) = \tilde{\rho} \LWRvelocity_\Rightroad \geq \LWRsupply(\rho_\Rightroad).
\end{equation*}


\paragraph{V}

Let 
\begin{equation*}
	\rho_\Leftroad > \rholim, \qquad \rho_\Rightroad > \frac{\rhomax}{2}, \qquad \rho_\Rightroad \geq \fluxmax_\Leftroad, \qquad \rho_\Leftroad > \rho_\Rightroad, \qquad \ARZsupply(\rho_\Leftroad,0)  \geq \LWRsupply(\rho_\Rightroad).
\end{equation*}

For the region V, it holds that $w_\Leftroad \geq w(\rhomax) =\nicefrac{\vmax}{2}$ and $\LWRvelocity_\Rightroad<\nicefrac{\vmax}{2}$. Thus,
\begin{equation*}
	w_\Leftroad - \LWRvelocity_\Rightroad > 0.
\end{equation*}
Similarly to the proof of~\eqref{eq:tilderho_bigger_rhor}, one can show 
\begin{equation} \label{eq:tilderho_smaller_rhor}
	\tilde{\rho} = \pressure^{-1}(\pressure(\rho_\Leftroad) + \LWRvelocity_\Leftroad - \LWRvelocity_\Rightroad) \leq \rho_\Rightroad. 
\end{equation}

Moreover, we prove
\begin{align}
	\tilde{\rho} &> \fluxmax_\Leftroad \notag \\
	\pressure^{-1}(w_\Leftroad - \LWRvelocity_\Rightroad) &> \fluxmax_\Leftroad \notag \\
	w_\Leftroad - \LWRvelocity_\Rightroad &> \pressure(\fluxmax_\Leftroad) = \frac{w_\Leftroad}{3} \notag \\
	\frac{2}{3} w_\Leftroad &> \LWRvelocity_\Rightroad \notag \\
	\frac{2}{3} \left( \vmax \frac{\rhomax - \rho_\Leftroad}{\rhomax} + \frac{\vmax}{2} \left(\frac{\rho_\Leftroad}{\rhomax}\right)^2\right) &> \vmax \frac{\rhomax - \rho_\Rightroad}{\rhomax} \notag \\
	\rho_\Rightroad &> \rhomax - \frac{2}{3} (\rhomax -\rho_\Leftroad) - \frac{1}{3} \frac{\rho_\Leftroad^2}{\rhomax} =: \tilderhoequalssigmal(\rho_\Leftroad). \label{eq:tilderhoequalssigmal}
\end{align}

\begin{figure}[H]
	\centering
	\tikzsetnextfilename{regionII}
%
%
\begin{tikzpicture}
	\setlength\fwidth{0.5\textwidth}
	\begin{axis}[%
		width=0.951\fwidth,
		height=0.75\fwidth,
		at={(0\fwidth,0\fwidth)},
		scale only axis,
		xmin=338.0873,
		xmax=600,
		xtick={ 338.0873},
		xticklabels={$\rho_{\lim}$},
		xlabel={$\rho_\Leftroad$},
		ymin=300,
		ymax=600,
		ytick={  300},
		yticklabels={ $\frac{\rhomax}{2}$},
		ylabel={$\rho_\Rightroad$},
		axis background/.style={fill=white},
		legend style={at={(0.35,0.95)}, legend cell align=left, align=left, draw=white!15!black}
		]
		\addplot [color=black,dotted, line width=1.0pt]
	  table[row sep=crcr]{%
		299	349.666111111111\\
		317	355.506111111111\\
		335	360.986111111111\\
		353	366.106111111111\\
		371	370.866111111111\\
		389	375.266111111111\\
		407	379.306111111111\\
		425	382.986111111111\\
		443	386.306111111111\\
		461	389.266111111111\\
		479	391.866111111111\\
		497	394.106111111111\\
		515	395.986111111111\\
		533	397.506111111111\\
		551	398.666111111111\\
		569	399.466111111111\\
		587	399.906111111111\\
		600	400\\
	};
		\addlegendentry{$\tilderhoequalssigmal(\rho_\Leftroad)$}
		
				\addplot [color=black,solid, line width=1.0pt]
		table[row sep=crcr]{%
			300	300\\
			370.820393249937 370.820393249937 \\
		};
		\addlegendentry{$\rho_\Leftroad$}
		
		\addplot [name path=A, color=blue, line width=1.0pt]
		table[row sep=crcr]{%
			338.087258745563	377.976314968462\\
			359.787258745563	373.140796063618\\
			381.087258745563	368.746881696886\\
			401.987258745563	364.787173430891\\
			422.487258745563	361.25280774349\\
			442.687258745563	358.11883435576\\
			462.587258745563	355.378793898876\\
			482.287258745563	353.013828555216\\
			501.787258745563	351.020295018541\\
			521.087258745563	349.393388953049\\
			540.287258745563	348.121440241821\\
			559.387258745563	347.202820050087\\
			578.387258745563	346.634827344537\\
			597.287258745563	346.413702079423\\
			599.987258745563	346.410161591881\\
		};

		\addplot [name path=B, color=blue, dotted, line width=1.0pt]
		table[row sep=crcr]{%
			338.087258745563	300\\
			338.387258745563	307.033173243888\\
			339.087258745563	312.83080186002\\
			340.087258745563	318.125376046725\\
			341.487258745563	323.595871089149\\
			343.187258745563	328.844290625477\\
			345.287258745563	334.191920178693\\
			347.687258745563	339.375556191786\\
			350.487258745563	344.610486675576\\
			353.687258745563	349.856583625351\\
			357.287258745563	355.086033989889\\
			361.287258745563	360.278059859757\\
			365.687258745563	365.416095519318\\
			370.487258745563	370.486197231076\\
			375.787258745563	375.567618033282\\
			381.487258745563	380.541614634536\\
			387.587258745563	385.40040676027\\
			394.187258745563	390.2057344478\\
			401.187258745563	394.86808931475\\
			408.687258745563	399.439085153056\\
			416.587258745563	403.843247125292\\
			424.987258745563	408.122694192952\\
			433.787258745563	412.212885598506\\
			442.987258745563	416.106745564117\\
			452.687258745563	419.832949556642\\
			462.787258745563	423.339091427951\\
			473.287258745563	426.616202209886\\
			484.187258745563	429.654321867667\\
			495.487258745563	432.442425014455\\
			507.087258745563	434.948207706945\\
			519.087258745563	437.184117502379\\
			531.387258745563	439.120581233934\\
			543.887258745563	440.737518637352\\
			556.587258745563	442.032390734389\\
			569.487258745563	443.000103319449\\
			582.487258745563	443.629430492398\\
			595.587258745563	443.9178462062\\
			599.987258745563	443.93737405685\\
		};

		\addplot [ name path=C,color=black, line width=1.0pt]
		table[row sep=crcr]{%
			370.820393249937	370.820393249937\\
			599.920393249937	599.920393249937\\
		};

		\addplot [name path=D,color=black, line width=1.0pt]
		table[row sep=crcr]{%
			338.087258745563	0\\
			338.087258745563	600\\
		};

		\addplot [name path=E,color=black, line width=1.0pt]
		table[row sep=crcr]{%
			338.087258745563	300\\
			600	300\\
		};

\node[] at (axis cs: 347,358) {II};
\node[] at (axis cs: 420,330) {III};
\node[]  at (axis cs:533,480){V};
\node[] at (axis cs: 500,385) {VI};

		\path[name path=lim] (axis cs:338.0873,0) -- (axis cs:600,0);

		\addplot[gray!40] fill between[of=C and E];
		\addplot[gray!40] fill between[of=B and E];
		\addplot[gray!40] fill between[of=B and lim];
		
	\end{axis}

\end{tikzpicture}%
	\caption{Zoom into the regions II, III, V and VI. The curves~\eqref{eq:curveofmaxima} and~\eqref{eq:line_upVI} are displayed in blue (solid and dotted).}
	\label{img:regionII}
\end{figure}

For $(\rho_\Leftroad, \rho_\Rightroad)$ in region V, we have $\rho_\Rightroad > \tilderhoequalssigmal(\rho_\Leftroad)$, see Figure~\ref{img:regionII}. Therefore $\tilde{\rho} > \fluxmax_\Leftroad$ and, using also~\eqref{eq:tilderho_smaller_rhor}, we get
\begin{equation*}
	\ARZsupply(\rho_\Leftroad, \rho_\Rightroad) = \tilde{\rho} \LWRvelocity_\Rightroad \leq \rho_\Rightroad \LWRvelocity_\Rightroad = \LWRsupply(\rho_\Rightroad).
\end{equation*}

\paragraph{VI}
Let 
\begin{equation*}
	\rho_\Leftroad > \rholim, \qquad \rho_\Rightroad > \frac{\rhomax}{2}, \qquad \rho_\Rightroad \geq \fluxmax_\Leftroad, \qquad \rho_\Leftroad > \rho_\Rightroad, \qquad  \ARZsupply(\rho_\Leftroad,0) < \LWRsupply(\rho_\Rightroad).
\end{equation*}
Since $\ARZsupply(\rho_\Leftroad, 0) < \LWRsupply(\rho_\Rightroad)$ is claimed, the ARZ supply is applied. Now, since $\rho_\Rightroad \geq \fluxmax_\Leftroad$, we get from~\eqref{eq:line_upIII}
\begin{equation*}
	\rho_\Rightroad \geq \curveofmaxima(\rho_\Leftroad),
\end{equation*}
which is the area above the lower curve limiting region VI in Figure~\ref{img:rho_l-rho_r-plane}. Next, by the condition $\ARZsupply(\rho_\Leftroad,0) < \LWRsupply(\rho_\Rightroad)$, we know from~\eqref{eq:line_upVI} that
\begin{equation*}
	\rho_\Rightroad < \curveofequalsupplyatmaxarz(\rho_\Leftroad),
\end{equation*}
which is below the upper limiting curve of region VI in Figure~\ref{img:rho_l-rho_r-plane}.


\bibliography{Literature}

\begin{thebibliography}{10}

\bibitem{AndDonRaz2015}
{\sc B.~Andreianov, C.~Donadello, U.~Razafison, and M.~D. Rosini}, {\em
  {Riemann problems with non-local point constraints and capacity drop}},
  Mathematical Biosciences and Engineering, 12 (2015), pp.~259--278.

\bibitem{AndDonRos2016}
{\sc B.~Andreianov, C.~Donadello, and M.~D. Rosini}, {\em {A second-order model
  for vehicular traffics with local point constraints on the flow}},
  Mathematical Models and Methods in Applied Sciences, 26 (2016), pp.~751--802.

\bibitem{AwKlaMat2002}
{\sc A.~Aw, A.~Klar, T.~Materne, and M.~Rascle}, {\em {Derivation of continuum
  traffic flow models from microscopic follow-the-leader models}}, SIAM Journal
  on Applied Mathematics, 63 (2002), pp.~259--278.

\bibitem{AwRas2000}
{\sc A.~Aw and M.~Rascle}, {\em {Resurrection of "Second Order" Models of
  Traffic Flow}}, SIAM Journal on Applied Mathematics, 60 (2000), pp.~916--938.

\bibitem{Ban1991}
{\sc J.~H. Banks}, {\em {Freeway bottlenecks : a basis for ramp metering?}},
  Transportation Research Record, 1320 (1991), pp.~83--90.

\bibitem{bressanhyperbolic}
{\sc A.~Bressan}, {\em {Hyperbolic Systems of Conservation Laws: The
  One-Dimensional Cauchy Problem}}, vol.~20, Oxford University Press Inc., New
  York, reprinted~ed., 2005.

\bibitem{BreCanGar2014}
{\sc A.~Bressan, S.~{\v{C}}ani{\'{c}}, M.~Garavello, M.~Herty, and B.~Piccoli},
  {\em {Flows on networks: recent results and perspectives}}, EMS Surveys in
  Mathematical Sciences, 1 (2014), pp.~47--111.

\bibitem{CasBer1999}
{\sc M.~J. Cassidy and R.~L. Bertini}, {\em Some traffic features at freeway
  bottlenecks}, Transportation Research Part B: Methodological, 33 (1999),
  pp.~25--42.

\bibitem{ChuRudCas2007}
{\sc K.~Chung, J.~Rudjanakanoknad, and M.~J. Cassidy}, {\em Relation between
  traffic density and capacity drop at three freeway bottlenecks},
  Transportation Research Part B: Methodological, 41 (2007), pp.~82--95.

\bibitem{CocGarPic2005}
{\sc G.~M. Coclite, M.~Garavello, and B.~Piccoli}, {\em {Traffic flow on a road
  network}}, SIAM Journal on Mathematical Analysis, 36 (2005), pp.~1862--1886.

\bibitem{Dag1994}
{\sc C.~F. Daganzo}, {\em {The cell transmission model: A dynamic
  representation of highway traffic consistent with the hydrodynamic theory}},
  Transportation Research Part B: Methodological, 28 (1994), pp.~269--287.

\bibitem{Dag1995}
\leavevmode\vrule height 2pt depth -1.6pt width 23pt, {\em {Requiem for
  second-order fluid approximations of traffic flow}}, Transportation Research
  Part B, 29 (1995), pp.~277--286.

\bibitem{SanDonPel2019}
{\sc E.~{Dal Santo}, C.~Donadello, S.~F. Pellegrino, and M.~D. Rosini}, {\em
  {Representation of capacity drop at a road merge via point constraints in a
  first order traffic model}}, ESAIM: Mathematical Modelling and Numerical
  Analysis, 53 (2019), pp.~1--34.

\bibitem{ReiDelSam2014}
{\sc M.~L. {Delle Monache}, J.~D. Reilly, S.~Samaranayake, W.~Krichene,
  P.~Goatin, and A.~M. Bayen}, {\em {A PDE-ODE model for a junction with ramp
  buffer}}, SIAM Journal on Applied Mathematics, 74 (2008), pp.~22--39.

\bibitem{FanSunPic2017}
{\sc S.~Fan, Y.~Sun, B.~Piccoli, B.~Seibold, and D.~B. Work}, {\em {A collapsed
  generalized Aw-Rascle-Zhang model and its model accuracy}}, arXiv:1702.03624,
   (2017).

\bibitem{GarPic2006}
{\sc M.~Garavello and B.~Piccoli}, {\em {Traffic flow on a road network using
  the Aw-Rascle model}}, Communications in Partial Differential Equations, 31
  (2006), pp.~243--275.

\bibitem{PicGar2006}
\leavevmode\vrule height 2pt depth -1.6pt width 23pt, {\em {Traffic Flow On
  Networks}}, vol.~1 of AIMS on Applied Mathematics, American Institute of
  Mathematical Sciences (AIMS), Springfield, MO, 2006.

\bibitem{Goa2006}
{\sc P.~Goatin}, {\em {The Aw-Rascle vehicular traffic flow model with phase
  transitions}}, Mathematical and Computer Modelling, 44 (2006), pp.~287--303.

\bibitem{GoaGoeKol2016}
{\sc P.~Goatin, S.~G{\"{o}}ttlich, and O.~Kolb}, {\em {Speed limit and ramp
  meter control for traffic flow networks}}, Engineering Optimization, 48
  (2016), pp.~1121--1144.

\bibitem{KolGoeGoa2017}
\leavevmode\vrule height 2pt depth -1.6pt width 23pt, {\em {Capacity drop and
  traffic control for a second order traffic model}}, Networks and
  Heterogeneous Media, 12 (2017), pp.~663--681.

\bibitem{Gre2001}
{\sc J.~M. Greenberg}, {\em {Extensions and amplifications of a traffic model
  of Aw and Rascle}}, SIAM Journal on Applied Mathematics, 62 (2002),
  pp.~729--745.

\bibitem{HalAgy1991}
{\sc F.~L. Hall and K.~Agyemang-Duah}, {\em {Freeway capacity drop and the
  definition of capacity}}, Transportation Research Record 1320, 1320 (2000),
  pp.~91--98.

\bibitem{HauBas2007}
{\sc B.~Haut and G.~Bastin}, {\em {A second order model of road junctions in
  fluid models of traffic networks}}, Networks and Heterogeneous Media, 2
  (2007), pp.~227--253.

\bibitem{HauBasChi2005}
{\sc B.~Haut, G.~Bastin, and Y.~Chitour}, {\em {A macroscopic traffic model for
  road networks with a representation of the capacity drop phenomenon at the
  junctions}}, IFAC Proceedings Volumes, 38 (2005), pp.~114--119.

\bibitem{Hel2001}
{\sc D.~Helbing}, {\em {Traffic and related self-driven many-particle
  systems}}, Reviews of Modern Physics, 73 (2001), pp.~1067--1141.

\bibitem{HerKla2003}
{\sc M.~Herty and A.~Klar}, {\em {Modeling, simulation, and optimization of
  traffic flow networks}}, SIAM Journal on Scientific Computing, 25 (2003),
  pp.~1066--1087.

\bibitem{MouHer2009}
{\sc M.~Herty and S.~Moutari}, {\em {A macro-kinetic hybrid model for traffic
  flow on road networks}}, Computational Methods in Applied Mathematics, 9
  (2009), pp.~238--252.

\bibitem{HerMouRas2006}
{\sc M.~Herty, S.~Moutari, and M.~Rascle}, {\em {Optimization criteria for
  modelling intersections of vehicular traffic flow}}, Networks and
  Heterogeneous Media, 1 (2006), pp.~275--294.

\bibitem{HerRas2006}
{\sc M.~Herty and M.~Rascle}, {\em {Coupling conditions for a class of
  second-order models for traffic flow}}, SIAM Journal on Mathematical
  Analysis, 38 (2006), pp.~595--616.

\bibitem{HolRis1995}
{\sc H.~Holden and N.~H. Risebro}, {\em {A mathematical model of traffic flow
  on a network of unidirectional roads}}, SIAM Journal on Mathematical
  Analysis, 26 (1995), pp.~999--1017.

\bibitem{KolCosGoa2018}
{\sc O.~Kolb, G.~Costeseque, P.~Goatin, and S.~G{\"{o}}ttlich}, {\em
  {Pareto-optimal coupling conditions for the Aw-Rascle-Zhang traffic flow
  model at junctions}}, SIAM Journal on Applied Mathematics, 78 (2018),
  pp.~1981--2002.

\bibitem{Kru1970}
{\sc S.~N. Kru{\v{z}}kov}, {\em {First order quasilinear equations in several
  independent variables}}, Mathematics of the USSR-Sbornik, 10 (1970).

\bibitem{Lax_maththeoryofshockwaves}
{\sc P.~D. Lax}, {\em {Hyperbolic Systems of Conservation Laws and the
  Mathematical Theory of Shock Waves}}, Society for Industrial and Applied
  Mathematics, Philadelphia, PA, 1973.

\bibitem{Leb2003}
{\sc J.~P. Lebacque}, {\em Two-phase bounded-acceleration traffic flow model:
  analytical solutions and applications}, Transportation Research Record, 1852
  (2003), pp.~220--230.

\bibitem{trafficandgranularflow_lebacque}
{\sc J.~P. Lebacque}, {\em {Intersection Modeling, Application to Macroscopic
  Network Traffic Flow Models and Traffic Management}}, in Traffic and Granular
  Flow '03, Springer-Verlag, Berlin, 2005, pp.~261--278.

\bibitem{LigWhi1955}
{\sc M.~J. Lighthill and G.~B. Whitham}, {\em {On kinematic waves. II. A theory
  of traffic flow on long crowded roads}}, Proceedings of the Royal Society A:
  Mathematical, Physical and Engineering Sciences, 229 (1955), pp.~317--345.

\bibitem{Pap2002}
{\sc M.~Papageorgiou and A.~Kotsialos}, {\em Freeway ramp metering: An
  overview}, IEEE transactions on intelligent transportation systems, 3 (2002),
  pp.~271--281.

\bibitem{ParBui2012}
{\sc C.~Parzani and C.~Buisson}, {\em {Second-order model and capacity drop at
  merge}}, Transportation Research Record: Journal of the Transportation
  Research Board, 2315 (2012), pp.~25--34.

\bibitem{Ras2002}
{\sc M.~Rascle}, {\em {An improved macroscopic model of traffic flow:
  derivation and links with the Lighthill-Whitham model}}, Mathematical and
  Computer Modelling, 35 (2002), pp.~581--590.

\bibitem{ReiSamDel2015}
{\sc J.~Reilly, S.~Samaranayake, M.~L. Delle~Monache, W.~Krichene, P.~Goatin,
  and A.~M. Bayen}, {\em Adjoint-based optimization on a network of discretized
  scalar conservation laws with applications to coordinated ramp metering},
  Journal of Optimization Theory and Applications, 167 (2015), pp.~733--760.

\bibitem{Ric1956}
{\sc P.~I. Richards}, {\em {Shock waves on the highway}}, Operations Research,
  4 (1956), pp.~42--51.

\bibitem{SieMau2006}
{\sc F.~Siebel and W.~Mauser}, {\em On the fundamental diagram of traffic
  flow}, SIAM Journal on Applied Mathematics, 66 (2006), pp.~1150--1162.

\bibitem{SieMauMou2009}
{\sc F.~Siebel, W.~Mauser, S.~Moutari, and M.~Rascle}, {\em Balanced vehicular
  traffic at a bottleneck}, Mathematical and Computer Modelling, 49 (2009),
  pp.~689--702.

\bibitem{TreKes2013}
{\sc M.~Treiber and A.~Kesting}, {\em {Traffic Flow Dynamics: Data, Models and
  Simulation}}, Springer-Verlag, Berlin, 1~ed., 2013.

\bibitem{Zha2002}
{\sc H.~M. Zhang}, {\em {A non-equilibrium traffic model devoid of gas-like
  behavior}}, Transportation Research Part B: Methodological, 36 (2002),
  pp.~275--290.

\end{thebibliography}
\end{document}